\documentclass[journal, final]{IEEEtran}

\usepackage{latexsym, amsfonts, amssymb, amsthm, setspace, verbatim,cite,mathrsfs,graphicx,bm,stfloats,url, tikz,xcolor, bm, enumerate, paralist, subfig}
\usepackage[cmex10]{amsmath}

\numberwithin{equation}{section}
\renewcommand{\theequation}{\arabic{section}.\arabic{equation}}
\newtheorem{thm}{Theorem}
\newtheorem{lem}{Lemma}

\newtheorem{applem}{Lemma}[section]

\newcommand{\be}{\begin{equation}}
\newcommand{\ee}{\end{equation}}
\newcommand{\ben}{\begin{equation*}}
\newcommand{\een}{\end{equation*}}
\newcommand{\mc}{\mathcal}
\newcommand{\mbf}{\mathbf}
\newcommand{\abs}[1]{\lvert#1\rvert}
\newcommand{\norm}[1]{\lVert#1\rVert}
\newcommand{\e}{\epsilon}
\newcommand{\expec}{\mathbb{E}}
\newcommand{\snr}{\textsf{snr}}
\newcommand{\reals}{\mathbb{R}}

\newcommand{\vem}{\varepsilon_M}
\newcommand{\mcb}{\mathcal{B}_{M,L}}

\newcommand{\mscrs}{\mathscr{S}}

\newcommand{\xtl}{x_{\textsf{L}}(\tau)}

\newcommand{\fullA}{\mathbf{A}}
\newcommand{\fullb}{\boldsymbol{\beta}}
\newcommand{\fully}{\mathbf{y}}
\newcommand{\fullw}{\mathbf{w}}
\newcommand{\fullz}{\mathbf{z}}
\newcommand{\fullh}{\mathbf{h}}
\newcommand{\fullq}{\mathbf{q}}
\newcommand{\fullbamp}{\mathbf{b}}
\newcommand{\fullm}{\mathbf{m}}
\newcommand{\fullH}{\mathbf{H}}
\newcommand{\fullQ}{\mathbf{Q}}
\newcommand{\fullB}{\mathbf{B}}
\newcommand{\fullM}{\mathbf{M}}
\newcommand{\fullL}{\boldsymbol\Lambda}
\newcommand{\fulla}{\boldsymbol \alpha}
\newcommand{\fullg}{\boldsymbol \gamma}
\newcommand{\fullC}{\mathbf{C}}
\newcommand{\sect}{\textsf{sec}}
\newcommand{\ind}{\textsf{ind}}
\newcommand{\proj}{\boldsymbol{\mathsf{P}}}
\newcommand{\fullD}{\boldsymbol \Delta}
\newcommand{\vecZ}{\mathbf{Z}}
\newcommand{\Qmat}{\pmb{\mathbb{Q}}}
\newcommand{\Mmat}{\pmb{\mathbb{M}}}
\DeclareMathOperator*{\argmin}{arg\,min}

\begin{document}

\title{The Error Probability of \\  Sparse Superposition   Codes with  \\ Approximate Message Passing Decoding}

\author{Cynthia~Rush,~\IEEEmembership{Member,~IEEE,}
        and~Ramji~Venkataramanan,~\IEEEmembership{Senior Member,~IEEE}
\thanks{This work was supported in part by EPSRC Grant EP/N013999/1, by an Isaac Newton Trust Research Grant, and by a Marie Curie Career Integration Grant under Grant Agreement 631489. This paper was presented in part at the 2017 IEEE International Symposium on Information Theory.}%
\thanks{C.~Rush is with the Department of Statistics, New York, NY 10027, Columbia University, USA (e-mail: cynthia.rush@columbia.edu).}%
\thanks{R.~Venkataramanan is with Department of Engineering, University of Cambridge, Cambridge CB2 1PZ, UK (e-mail: rv285@cam.ac.uk).}
%
%
}

\maketitle

\begin{abstract}
Sparse superposition codes, or sparse regression codes (SPARCs),  are a recent class of codes for reliable communication over the AWGN channel at rates approaching the channel capacity.  Approximate message passing (AMP) decoding, a computationally efficient technique for decoding SPARCs, has been proven to  be asymptotically capacity-achieving for the AWGN channel.  In this paper, we refine the asymptotic result by deriving a  large deviations bound on the probability of AMP decoding error. This bound gives insight into the error performance of the AMP decoder for large but finite problem sizes,  giving an error exponent as well as guidance on how the code parameters  should be chosen at finite block lengths. For an appropriate choice of code parameters, we show that for any fixed rate less than the channel capacity, the decoding error probability decays exponentially in  $n/(\log n)^{2T}$, where $T$, the number of AMP iterations required for successful decoding, is bounded in terms of the gap from capacity.
\end{abstract}


\section{Introduction} \label{sec:intro}
A long-standing goal in information theory is to construct efficient codes for the memoryless additive white Gaussian noise (AWGN) channel, with {provably} low probability of decoding error at rates close to the channel capacity.   The input-output relationship of the  real-valued AWGN channel is  given by \be
y = u + w,
\label{eq:model1}
\ee
where $u$ is the input symbol, $y$ is the output symbol, and $w
$ is independent Gaussian noise with zero mean and variance $\sigma^2$.  There is an average power constraint $P$ on the input: if $u_1, u_2, \ldots, u_n$ are transmitted over $n$ channel uses, it is required that $\tfrac{1}{n} \sum_{i=1}^n u_i^2 \leq P$.  Then the signal-to-noise ratio is given by $\snr = P/\sigma^2$ and the channel capacity is $\mc{C}: =\tfrac{1}{2} \log (1 + \snr).$

Sparse superposition codes, or sparse regression codes (SPARCs),  are a class of codes introduced by Barron and Joseph \cite{AntonyML,AntonyFast} for reliable communication  over the AWGN channel at rates close to $\mc{C}$.  In \cite{AntonyFast}, the authors proposed the first feasible SPARC decoder, called the `adaptive successive decoder', and showed for any fixed rate $R < \mc{C}$, the probability of decoding error decays to zero exponentially in $\tfrac{n}{\log n}$. Despite these strong theoretical guarantees,  the rates achieved by this decoder for practical block lengths are significantly less than $\mc{C}$. Subsequently, a adaptive soft-decision iterative decoder was proposed by Cho and Barron \cite{BarronC12}, with improved finite  length performance for rates closer to capacity. Theoretically, the decoding error probability of the adaptive soft-decision decoder was shown to decay exponentially   in $n/(\log n)^{2T}$,  where $T$ is the minimum number of iterations \cite{ChoB13, choThesis}.

Recently,   decoders for SPARCs based on Approximate Message Passing (AMP) techniques were proposed in \cite{barbKrzISIT14, RushGVIT17, BarbKrz17}.  AMP decoding has several attractive features, notably,  the absence of tuning parameters, its superior empirical performance at finite block lengths, and  its low complexity when implemented using implicitly defined Hadamard design matrices \cite{RushGVIT17, BarbKrz17}. Furthermore, its decoding performance in each iteration can be predicted using a deterministic scalar iteration called `state evolution'.

In this paper, we provide a non-asymptotic analysis of the AMP decoder proposed in \cite{RushGVIT17}. In \cite{RushGVIT17}, it was proved that the state evolution predictions for the AMP decoder are asymptotically accurate, and that for any fixed rate $R < \mc{C}$, the  probability of decoding error goes to zero with growing block length.
However this result did not specify the rate of decay of the probability of error.  In this paper, we refine the asymptotic result in \cite{RushGVIT17}, and derive a large deviations bound for the probability of error of the AMP decoder (Theorem \ref{thm:main_amp_perf}). This bound gives insight into the error performance of the AMP decoder for large but finite problem sizes,  giving an error exponent as well as guidance on how the code parameters  should be chosen at finite block lengths. 

The error probability bound for the AMP decoder is of the same order as the bound for  the Cho-Barron soft-decision decoder \cite{ChoB13, choThesis}: both bounds decay exponentially in $n/(\log n)^{2T}$,  where $T$ is the minimum number of iterations. However, the AMP decoder has slightly lower complexity and has been empirically found to have better error performance (see Remark 6 on p.\pageref{rem:twodec}).

In the rest of this section, we describe the sparse regression codebook, briefly review the AMP decoder, and then list the main contributions of this paper.


\subsection{The sparse regression codebook} \label{sec:sparc}
\begin{figure}[t]
\centering
\includegraphics[width=3.5in]{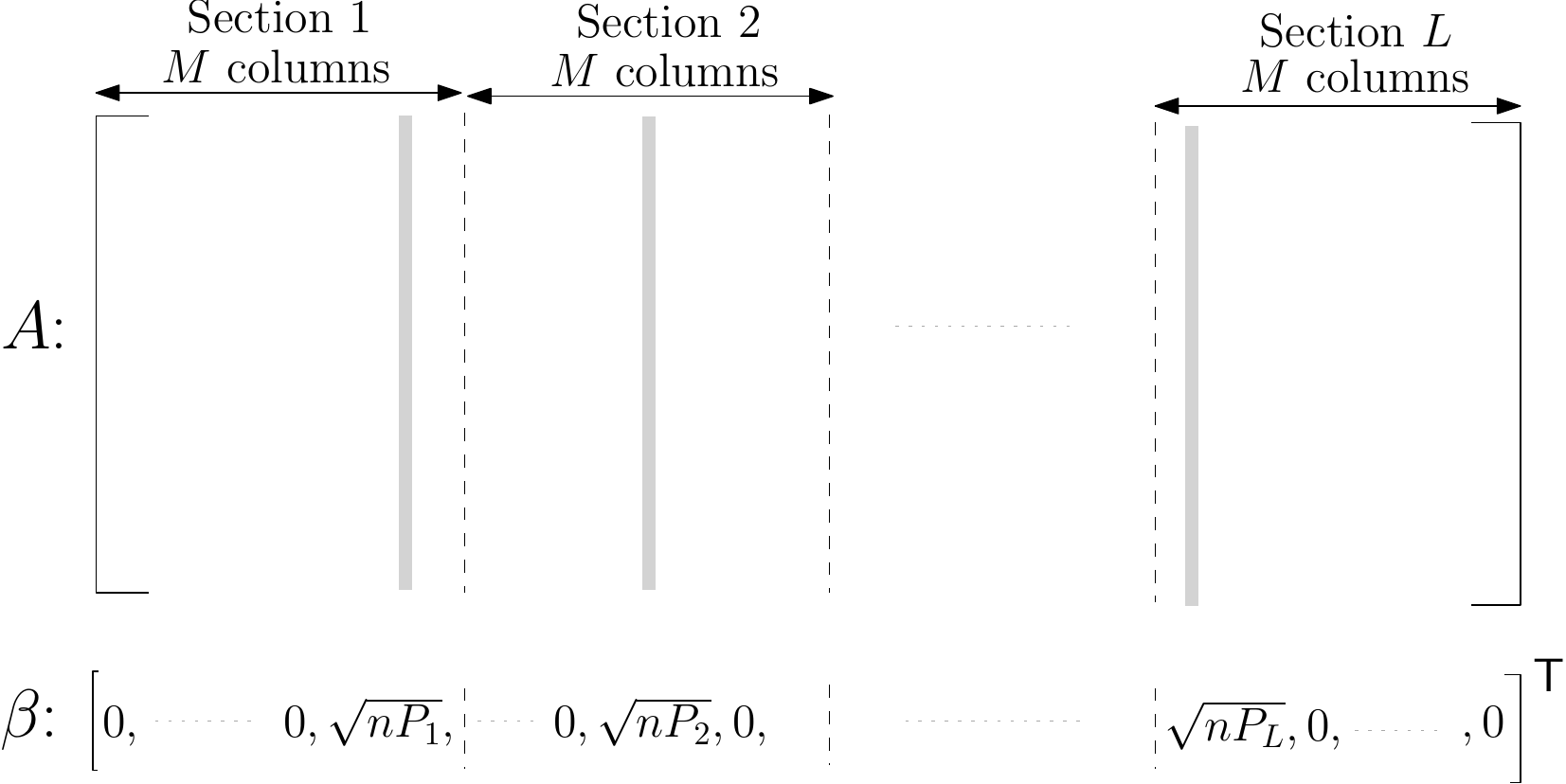}
\caption{\small{$\fullA$ is an $n \times ML$ matrix and $\fullb$ is a $ML \times 1$ vector. The positions of the non-zeros in $\fullb$  correspond to the gray columns of $\fullA$ which combine to form the codeword $\fullA \beta$.}}
\vspace{-7pt}
\label{fig:sparserd}
\end{figure}

A sparse regression code (SPARC) is defined in terms of a design matrix, or `dictionary', $\fullA$ of dimension $n \times ML$. The entries of $\fullA$   are i.i.d.\ $ \mathcal{N}(0,\tfrac{1}{n})$. Here $n$ is the block length, and $M$ and $L$ are integers whose values will be specified below in terms of $n$ and the rate $R$.  As shown in Fig.\ \ref{fig:sparserd}, we think of the matrix $\fullA$ as being composed of $L$ sections with $M$ columns each. 

Each codeword is a linear combination of $L$ columns, with one column selected from each of the $L$ sections. The codeword is formally expressed as  $\fullA \fullb$, where $\fullb$ is  an $ML \times 1$ vector $(\beta_1, \ldots, \beta_{ML})$ with the following property: there is exactly one non-zero $\beta_j$ for  $1 \leq j \leq M$, one non-zero $\beta_j$ for $M+1 \leq j \leq 2M$, and so forth.  The non-zero value of $\fullb$ in section $\ell$  is set to $\sqrt{n P_\ell}$, where $P_1, \ldots, P_L$ are positive constants that satisfy $\sum_{\ell=1}^L P_\ell = P$.  Denote the set of all $\fullb$'s that satisfy this property by $\mcb(P_1, \ldots,P_L)$.  For the main  result in this paper, we use  an exponentially decaying allocation of the form $P_\ell  \propto  e^{-2\mc{C}\ell/L},$ for $\ell \in \{1, 2, \ldots, L\}$.

Both the design matrix $\fullA$ and the power allocation are known to the encoder and the decoder before communication begins.

As each of the $L$ sections contains $M$ columns, the total number of codewords is $M^L$. To obtain a rate of $R$ nats/sample, we require
\be
M^L = e^{nR} \quad \text{ or } \quad L \log M = nR.
\label{eq:ml_nR}
\ee
(Throughout the paper, rate will be measured in nats unless otherwise mentioned.) An important case is when $M$ equals $L^\textsf{a}$, for some constant $\textsf{a} >0$. Then \eqref{eq:ml_nR} becomes $\textsf{a} L \log L = nR.$  In this case, $L= \Theta(\tfrac{n}{\log n})$, and the size of the design matrix $\fullA$ (given by $n \times ML = n \times L^{\textsf{a}+1}$) now grows polynomially in $n$.

Given a sequence of information bits, the encoder maps them to a message vector $\fullb_0 \in \mcb$ and generates the codeword  $\fullA \fullb_0 \in \reals^n$.  At the decoder, the task is to recover $\fullb_0$ from the channel output sequence
\be
  \fully = \fullA \fullb_0 +\fullw. \label{eq:AMPoutput}\ee
Assuming that the transmitted message vector is uniformly distributed over $\mcb$,   the maximum-likelihood  decoder minimizes the probability of the decoded message vector not being equal to the transmitted one.  The maximum-likelihood decoding rule for a SPARC is given by 
$$\widehat{\fullb}_{\textsf{ML}} = \argmin_{\fullb \in \mcb}\  \norm{\fully - \fullA \fullb},$$ 
where $\| \cdot \|$ denotes the  $\ell_2$-norm.
  This decoder was analyzed in \cite{AntonyML} and shown to have probability of error decaying exponentially in $n$ for any fixed $R < \mc{C}$. However, it is infeasible as the decoding complexity is exponential in $n$. This motivates the need for low-complexity SPARC decoding techniques such as AMP.

\subsection{Notation}
 For a positive integer $m$, we use $[m]$ to denote the set  $\{1, \dots, m \}$. 
Throughout the paper, we use boldface  to denote vectors or matrices, plain font for scalars, and  subscripts to denote entries of a vector or matrix. Bold lower case letters or Greek symbols are used for vectors, and bold upper case for matrices.   For example, $\textbf{x}$ denotes a vector, with $x_i$ being the $i^{th}$ element of $\textbf{x}$. Similarly,  $\textbf{X}$ is a matrix,and  its $(i,j)^{th}$ entry is denoted by $X_{i, j}$. 
The transpose of $\textbf{X}$ is denoted by $\textbf{X}^*$. 
The number of columns in the design matrix $\fullA$ is denoted by $N = ML,$ so $\fullA$ has dimensions $n \times N$.  

For length-$N$ vectors such as $\fullb$, we will need to refer to specific entries as well as sections.   Indices such as  $j$ will be used to denote specific entries, while the subscript $(\ell)$ will be used to denote the entire section $\ell \in [L]$.  Therefore $\beta_j$ denotes the $j$th entry of $\fullb$, for $j \in [ML]$, and $\fullb_{(\ell)}$ denotes the length-$M$ vector containing the entries in the $\ell$th section  of $\fullb$,   for $\ell \in [L]$.  For example, $\boldsymbol{\beta}_{(2)}= \{ \beta_{M+1}, \ldots, \beta_{2M} \}$ is the vector containing the entries in the second section of $\boldsymbol{\beta}$. 

We will use a mapping $\ind: [L] \rightarrow [ML]^M$ that maps a section index to the indices of the entries corresponding to that section.  For example $\ind(\ell) = [(\ell - 1)M + 1, (\ell - 1)M + 1, \ldots, \ell M]$ is the length-$M$ vector of indices contained in section $\ell$.  We define a complementary function $\sect: [ML] \rightarrow [L]$ that maps the index of an entry to the section containing it.  For example, $\sect(i) = \ell$ indicates that $i \in \ind(\ell)$, and $\ind(\sect(i))$ returns the length-$M$ vector of the indices for the section to which $i$ belongs.

In the analysis, we will treat the message as a random vector $\fullb$ which is uniformly distributed over 
$\mcb(P_1,\ldots,P_L)$. We will denote the true message vector by $\fullb_0$, noting that $\fullb_0$ is a \emph{realization} of the random vector $\fullb$.

 The indicator function of an event $\mc{A}$ is denoted by $\mathbf{1}(\mc{A})$.  The $t \times t$ identity matrix is denoted by $\mathsf{I}_t$ and we suppress the subscript if the dimensions are clear from context.  $\log$  and $\ln$ are both used to denote the natural logarithm.   We will use $k, K, \kappa, \kappa_0, \kappa_1, \ldots, \kappa_8$ to denote generic universal positive constants.


\subsection{The AMP channel decoder} \label{sec:amp_channel_decoder}

Approximate message passing refers to a class of iterative algorithms obtained via quadratic or Gaussian approximations of standard message passing algorithms such as belief propagation or min-sum. Approximations of loopy belief propagation were first used for CDMA multiuser detection  \cite{boutrosCaire02,tanaka05}.  AMP was then proposed in the context of compressed sensing \cite{DonMalMont09,MontChap11,BayMont11,Rangan11, krz12},  and has since been applied to other high-dimensional estimation problems represented by  dense factor graphs where standard message passing is infeasible, e.g. low-rank matrix estimation\cite{rangan2012iterative,deshpande2014information, lesieur2015phase,montRV17}.  

AMP techniques were used to develop efficient SPARC decoders in \cite{barbKrzISIT14, RushGVIT17, BarbKrz17}. The problem of  recovering the SPARC message vector $\fullb$ from the channel output sequence $\fully$ in \eqref{eq:AMPoutput} is similar to the compressed sensing recovery problem, with one key difference: in a SPARC we know that $\fullb \in \mcb$, and an effective decoder must take advantage of this structure. 

We now describe the AMP decoder from  \cite{RushGVIT17} and give some insight into its working.  
Given the received vector $\fully= \fullA \fullb_0 + \fullw$, the AMP decoder generates successive estimates of the message vector, denoted by $\{ \fullb^t \}_{t \geq 0}$, where $\fullb^t \in \mathbb{R}^{ML}$. We initialize the algorithm with $\fullb^0= \textbf{0}$, the all-zeros vector. For $t=0,1,\ldots$, the decoder computes
\begin{align}
\fullz^t & = \fully - \fullA\fullb^t + \frac{\fullz^{t-1}}{\tau^2_{t-1}}\Big( P - \frac{\norm{\fullb^t}^2}{n} \Big), \label{eq:amp1}\\
\beta^{t+1}_i & = \eta^t_{i}( \fullb^t + \fullA^*\fullz^t), \quad  \text{ for } i=1,\ldots,ML, \label{eq:amp2}
\end{align}
where  quantities with negative indices are set equal to zero. The constants $\{ \tau_{t} \}_{t \geq 0}$, and the estimation functions $\{\eta^t_i(\cdot)\}_{t \geq 0}$ are defined as follows. Define
\begin{align}
\tau^2_{0}  =\sigma^2+P,  \qquad  \tau^2_{t+1}  =  \sigma^2 + P(1-x_{t+1}), \ \  t \geq 0,
\label{eq:taut_def}
\end{align}
where $x_{t+1} = x(\tau_t)$, with 
\be
\begin{split}
&  x(\tau) := \sum_{\ell=1}^{L} \frac{P_\ell}{P} \\
 & \, \times  \expec\Bigg[
\frac{\exp\Big\{ \frac{\sqrt{n P_\ell}}{\tau} \, \Big(U^{\ell}_1  + \frac{\sqrt{n P_\ell}}{\tau}\Big)\Big\} }{\exp\Big\{ \frac{\sqrt{n P_\ell}}{\tau} \, \Big(U^{\ell}_1  + \frac{\sqrt{n P_\ell}}{\tau}\Big)\Big\}  + \sum_{j=2}^M 
\exp\Big\{ \frac{\sqrt{n P_\ell}}{\tau}U^{\ell}_j \Big\}} \Bigg]
\end{split}
\label{eq:xt_tau_def}
\ee
In \eqref{eq:xt_tau_def}, the expectation is over $\{ U^\ell_j\}$, which are i.i.d.\ $\mc{N}(0,1)$ random variables for $j\in [M]$ and $\ell \in [L]$.   Hence $x(\tau)$ is a deterministic function of $\tau$.   For consistency, we define $x_0=0$.
Recall that ${\sect(i)}$ returns a value in $[L]$ indicating the section to which index $i$ belongs and $\ind({\sect(i)})$ returns the vector of section indices for $i$'s section.  For $i\in [ML]$, define
\be
\eta^t_i(\mathbf{s}) = \sqrt{nP_{\sect(i)}}\,  \frac{\exp\{{s_i \sqrt{n P_{\sect(i)}}}/{\tau^2_t}\} }
{\sum_{j \in \ind(\sect(i))} \, \exp\{{s_j \sqrt{n P_{\sect(i)}}}/{\tau^2_t}\}}.
\label{eq:eta_def}
\ee
Notice that $\eta^t_i(\mathbf{s})$ depends on all the components of $s$ in the \emph{section} containing $i$, i.e.\ all components of $s$ belonging to $\ind(\sect(i))$.

   For any rate $R < \mc{C}$, the AMP decoder is run for a finite number of iterations $T$, where $T$ is specified later in Section \ref{subsec:del_vs_CR}. After  $T$ iterations, the maximum value in each section $\ell \in [L]$ of $\fullb^{T}$  is set to $\sqrt{nP_\ell}$, and remaining entries to $0$ to obtain the decoded message $\hat{\fullb}$.
   
   The relation \eqref{eq:taut_def}, which describes how $\tau_{t+1}$ is obtained from  $\tau_{t}$,  is  called ``state evolution".  We now explain the significance of the state evolution parameters $x_t, \tau_t^2$, and the choice of  the estimation function $\eta^t$.

\subsubsection*{State evolution interpretation}
To understand the decoder, first consider the AMP update step  \eqref{eq:amp2} in which $\fullb^{t+1}$ is generated from the ``effective observation" $\mathbf{s}^t:= \fullb^t + \fullA^*\fullz^t.$  The update in \eqref{eq:amp2}  is underpinned by the following key property of the effective observation:
 \emph{$\mathbf{s}^t$ is approximately distributed as  $\fullb_0 + \tau_t \vecZ$, where $\vecZ$ is a length-$ML$ standard Gaussian random vector independent of the  message vector  $\fullb_0$.}  

In light of the above property, a natural way to generate $\fullb^{t+1}$ from $\mathbf{s}^t=\mathbf{s}$ is $\fullb^{t+1}(\mathbf{s}) =  \expec[ \fullb \, | \, \fullb + \tau_t \vecZ = \mathbf{s}]$,
i.e., $\fullb^{t+1}$ is  the Bayes optimal estimate of $\fullb$ given the observation $\mathbf{s}^t = \fullb + \tau_t \vecZ$. This conditional expectation can be computed using the independence of $\fullb$ and $\vecZ$, with  $\vecZ$ being a standard normal vector, and the location of the non-zero entry in each section of $\fullb$ uniformly distributed within the section. Then, for $i \in [ML]$, we obtain
\be
\begin{split}
 \beta^{t+1}_i(\mathbf{s})  &= \expec[\beta_i |  \fullb + \tau_t \vecZ = \mathbf{s} ]  \\
 & =  \frac{\sqrt{nP_{\sect(i)}} \, \exp\{{s_i \sqrt{n P_{\sect(i)}}}/{\tau^2_t}\}}
{\sum_{j \in \ind({\sect(i)})} \, \exp\{{s_j \sqrt{n P_{\sect(i)}}}/{\tau^2_t}\}},
\end{split}
\label{eq:cond_exp_beta0}
\ee
which is the expression in \eqref{eq:eta_def}.  Furthermore,   
$\beta^{t+1}_i(\mathbf{s})/\sqrt{nP_{\sect(i)}}$
is the posterior probability of $\beta_i$ being the non-zero entry in  its section, conditioned on
the observation $\mbf{s}= \fullb + \tau_t \vecZ$.

It is shown in \cite[Proposition 1]{RushGVIT17} that under the assumption that $\mathbf{s}^t$ has the distributional representation $\fullb + \tau_t \vecZ$, we have
\be 
x_{t+1} = \frac{1}{nP} \expec[\fullb^* \fullb^{t+1}], \qquad \frac{1}{n} \expec \norm{\fullb - \fullb^{t+1}}^2 =  P(1-x_{t+1}). 
\label{eq:norm_SErep}
\ee
The expectation in \eqref{eq:norm_SErep} is again computed over  $\fullb$ and $\vecZ$, which are independent. Using \eqref{eq:norm_SErep} in \eqref{eq:taut_def}, we see that the effective noise variance $\tau_t^2$ is the sum of the channel noise variance and the expected squared error in the estimate after step $t$.    The parameter $x_t$ can be interpreted as the power-weighted fraction of sections correctly decodable after step $t$: starting from $x_0=0$ we wish to ensure that at the termination step $T$, the parameter $x_T$ is very close to one, implying that   the expected squared error $\frac{1}{n} \expec \norm{\fullb - \fullb^{T}}^2 \approx 0$ under the distributional assumption for $\mathbf{s}^t$.  This is done in Lemma \ref{lem:lim_xt_taut}, which provides a strictly positive lower bound on the difference  $(x_{t+1} -x_t)$ for each iteration $t$ until $x_t$ reaches a value close to $1$.

The other part of the proof is establishing the validity of the key distributional assumption on the effective observation $\mbf{s}^t$. The asymptotic result in \cite{RushGVIT17} is proven by showing that the distributional assumption  holds  in the large system limit (with a suitable notion of convergence). The large system limit refers to taking $L, M, n \to \infty$, while satisfying $L \log M = nR$. In this paper, we obtain a non-asymptotic bound on the probability of decoding error by showing  that  the average squared error in each iteration $t$, given by  $\frac{1}{n}\| \fullb_0 - \fullb^t \|^2$, concentrates around the state evolution prediction $P(1-x_t)$. This concentration inequality also specifies the performance trade-offs incurred by different scaling choices for  $M$ vs.\ $L$, as  both tend to infinity. 

\subsection{Structure of the paper and main contributions}
\begin{itemize}
\item In Section \ref{sec:SE_bounds}, we derive a lower bound on the minimum increase in each step of the state evolution parameter $x_t$, for an exponentially decaying power allocation (Lemma \ref{lem:lim_xt_taut}). This in turn yields an upper bound on the number of iterations $T$. This upper bound is inversely proportional to $\Delta_R + \Delta_R^2$, where $\Delta_R := (\mc{C} -R)/\mc{C}$ is the fractional gap from capacity.

\item The main result of the paper (Theorem \ref{thm:main_amp_perf}) is a  large deviations bound on the probability of AMP decoding error.  Using this bound,  in Section  \ref{subsec:errexp}  we investigate the error exponent with AMP decoding, i.e., how fast does the error probability decay with growing block length, with $R < \mc{C}$ held fixed. We show that for an appropriate choice of code parameters, the complexity of the AMP decoder scales as a low-order polynomial in the block length $n$, while  the decoding error probability decays exponentially in  $n/(\log n)^{2T}$. Here $T$, the number of AMP iterations required for successful decoding, is bounded in terms of the gap from capacity.  In Section \ref{subsec:scaling}, we examine how fast the error probability can decay when $R$ approaches $\mc{C}$ as $n \to \infty$.

\item The proof of the main result is given in Section \ref{sec:amp_proof}, and has two main ingredients: a conditional distribution lemma (Lemma \ref{lem:hb_cond}) specifying distributional representations for the iterates (vectors) produced by the AMP decoder, and a concentration lemma (Lemma \ref{lem:main_lem}), which uses these distributional representations to obtain concentration inequalities for various inner products involving the AMP iterates. The conditional distribution lemma was already proved in \cite{RushGVIT17}, so the key technical contribution is Lemma \ref{lem:main_lem}. 

\item The proof of the concentration lemma (Lemma \ref{lem:main_lem})   is given in Section \ref{sec:lem1_proof}. In addition to strengthening the asymptotic convergence results in \cite{RushGVIT17}, Lemma \ref{lem:main_lem} also simplifies some technical aspects of the proof.  The  techniques used to derive the concentration inequalities   in Lemma \ref{lem:main_lem} are broadly similar to those used for the non-asymptotic analysis of the standard AMP recursion in \cite[Lemma 4.5]{RushV16}. However, there are a few important differences: due to the SPARC message vector $\fullb$ having a section-wise structure (with one non-zero per section), the results derived  in \cite{RushV16} for i.i.d.\ signal priors cannot be directly applied here.   
\end{itemize}

In this paper, we only consider the standard SPARC construction  described in Section \ref{sec:sparc}. Extending the analysis to the spatially-coupled SPARCs proposed in \cite{BarbKrz17,barbISIT16,barbITW16} is an interesting research direction and part of ongoing work.


\section{Bounds for state evolution parameters} \label{sec:SE_bounds}
We first derive a  lower bound for $x(\tau)$ defined in \eqref{eq:xt_tau_def}. This lower bound will be used to specify  the number of AMP iterations $T$ required for successful decoding (in terms of  $L,M, R$).  The  number of iterations $T$  determines how the  error probability bound in Theorem \ref{thm:main_amp_perf} depends on the rate $R$.
\begin{lem}
Consider any non-increasing power allocation $\{P_{\ell}\}_{\ell \in [L]}$, and let $\nu_\ell :=  {L P_\ell}/ (R \tau^2)$. Assume that there exist absolute positive constants $\bar{\mathsf a}, \underline{\sf a}$ (not depending on $L$) such that $\bar{\sf a} \geq \nu_1 \geq \nu_2 \ldots \geq \nu_{L} \geq \underline{\sf{a}}  $.   
 
\textbf{(a)}  We have $x(\tau) \geq \xtl$, where $\xtl$ is defined  as follows in terms of  constants $\alpha \in [0,1), \upsilon > 0$, that may  be arbitrarily chosen.
 \begin{align}
 \xtl & =  \sum_{\ell=1}^L \frac{P_\ell}{P}  \left[   \frac{ \mc{Q}(-  \frac{\alpha (\nu_\ell /2 -1)}{\sqrt{\nu_{\ell}}} \sqrt{\log M})  }{1 +  M^{-(1- \alpha)(\nu_\ell /2-1)}} \, \mathbf{1}\{ \nu_\ell > 2\}   \right. \nonumber \\ 
& \ \ +  \left.  \frac{\mc{Q}( 2 \upsilon/\sqrt{\nu_\ell})  }{1 +  e^{-\upsilon \sqrt{\log M}} } \, 
\mathbf{1}\Big\{ 2\Big(1- \frac{\upsilon}{\sqrt{\log M}} \Big) \leq  \nu_\ell \leq  2 \Big\} \right]\label{eq:xlb_nonasym} 
 \end{align}
 where $\kappa$ is a universal positive constant, and $\mc{Q}(\cdot)$ is the complementary distribution function of a standard Gaussian, i.e., 
 $\mc{Q}(x) = \int_{x}^{\infty} \frac{1}{\sqrt{2 \pi}} e^{-u^2/2} du$.

\textbf{(b)} For sufficiently large $M$ and  any $\delta \in (0,\frac{1}{2})$,
\begin{align}
\xtl & \geq  \Big(1- \frac{M^{-\kappa_2 \delta^2}}{\delta \sqrt{\log M}} \Big) \sum_{\ell=1}^{L} \frac{P_\ell}{P} \,
\mathbf{1}\{ \nu_\ell > 2 + \delta \}  \nonumber \\ 
& \ + \frac{1}{4}\sum_{\ell=1}^{L} \frac{P_\ell}{P} \mathbf{1}\Big\{ 2\Big(1- \frac{\kappa_3}{\sqrt{\log M}} \Big) \leq \nu_\ell \leq  2 + \delta \Big\}  \label{eq:xlb_asym}  
\end{align}
where $ \kappa_2, \kappa_3$ are universal positive constants.
\label{lem:conv_expec}
\end{lem}
\begin{IEEEproof} In Appendix \ref{app:conv_exp}. \end{IEEEproof}

Eq.\ \eqref{eq:xlb_asym}  can be interpreted as follows for large $M, L$:  if the effective noise variance at the end of step $t$ is $\tau_t^2=\tau^2$,  then any sections $\ell$ that satisfy $L P_\ell > 2 R \tau^2 $ will be decodable in step $t+1$, i.e.,  $\fullb^{t+1}_{(\ell)} \in \reals^M$ will have most of its mass on the correct non-zero entry.   

 We now evaluate the lower bound  of Lemma \ref{lem:conv_expec} for the following exponentially decaying power allocation:
 \be  
 P_\ell = P \cdot  \frac{e^{2\mc{C}/L} -1}{1- e^{-2\mc{C}}} \cdot e^{-2\mc{C}\ell/L}, \qquad  \ell \in [L].  
 \label{eq:exp_power_alloc} \ee
 For this   allocation, we have
 \be
 L P_\ell =  (P+\sigma^2) L((1 + \snr)^{1/L}-1)  \left( 1 +\snr \right)^{-\ell/L}, \ \  \ell \in [L].
\label{eq:cell}
\ee
The next lemma uses Lemma \ref{lem:conv_expec} to obtain a lower bound on how much the state evolution parameter $x_t$ increases in each iteration, for the exponentially decaying power allocation.
\begin{lem} 
Let $\delta \in (0,  \min\{ \Delta_R, \frac{1}{2} \}]$, where $\Delta_R := (\mc C - R)/\mc C$. Let $f(M) := \frac{M^{-\kappa_2 \delta^2}}{\delta \sqrt{\log M}}$, where $\kappa_2$ is the universal constant in Lemma \ref{lem:conv_expec}(b). Consider the sequence of state evolution parameters $x_0=0, x_1,  \ldots$ computed according to \eqref{eq:taut_def} and \eqref{eq:xt_tau_def} with the exponentially decaying power allocation in \eqref{eq:exp_power_alloc}. For sufficiently large $L,M$, we have:
\be
x_1 \geq  \chi_1 := (1 - f(M))  \frac{P+\sigma^2}{P} \Big( 1 - \frac{(1+ \delta/2) R}{\mc C} - \frac{5R}{L} \Big),
\label{eq:chi1}
\ee
and for $t > 1$:
\be
\begin{split}
  & x_t - x_{t-1} \\
  &  \geq  \chi := (1 - f(M)) \Big[  \frac{\sigma^2}{P} \Big( 1 - \frac{(1+ \delta/2) R}{\mc C} \Big)  \\ 
   & \hspace{0.75in} - f(M) \frac{(1+ \delta/2) R}{\mc C} \Big] - \frac{5R(1 +\sigma^2/P)}{L},
  \end{split}
    \label{eq:chi}
\ee 
until $x_t$ reaches (or exceeds) $(1-f(M))$. 
\label{lem:lim_xt_taut}
\end{lem}

\begin{IEEEproof} In Appendix \ref{app:xt_taut}. \end{IEEEproof}

\subsection{Number of iterations and the gap from capacity} \label{subsec:del_vs_CR}

We want the lower bounds $\chi_1$ and $\chi$ in \eqref{eq:chi1} and \eqref{eq:chi} to be strictly positive and depend only on the gap from capacity $\Delta_R=(\mc C - R)/\mc C$ as $M,L \to \infty$. 
For all $\delta \in (0, \Delta_R]$, we have
\be
1- \frac{(1+ \delta/2)R}{\mc{C}}  \geq 1- \Big(1+ \frac{\Delta_R}{2}\Big) (1-\Delta_R)  =\frac{\Delta_R +  \Delta_R^2}{2}.
\label{eq:key_term_lb}
\ee
Therefore, the quantities on the RHS of \eqref{eq:chi1} and \eqref{eq:chi}  can be bounded from below as
\begin{align}
\chi_1 &  \geq (1-f(M)) \frac{P+ \sigma^2}{P}\left( \frac{\Delta_R +  \Delta_R^2}{2}- \frac{5R}{L} \right),  \label{eq:chi1_lb1} \\
 \chi &  \geq (1-f(M)) \left[ \frac{\sigma^2}{P} \left(  \frac{\Delta_R +  \Delta_R^2}{2}  \right)  - f(M)\right] \nonumber \\ 
 &\qquad -  \frac{5R(1 +\sigma^2/P)}{L}.
 \label{eq:chi_lb1}
\end{align}
We take $\delta=\Delta_R$, which gives the smallest value for $f(M)$ among $\delta \in (0, \Delta_R]$.  \footnote{As Lemma \ref{lem:lim_xt_taut} assumes that $\delta \in (0, \min\{\frac{1}{2}, \Delta_R \}]$, by taking $\delta=\Delta_R$ we have assumed that $\Delta_R \leq \frac{1}{2}$, i.e., $R \geq \mc{C}/2$. This assumption can be made without loss of generality --- as the probability of error increases with rate, the large deviations bound of Theorem \ref{thm:main_amp_perf}  evaluated for $\Delta_R = \frac{1}{2}$ applies for all $R$ such that $\Delta_R < \frac{1}{2}$.
}  
 We denote this value by
 \be f_R(M) := \frac{M^{-\kappa_2 \Delta_R^2}}{\Delta_R \sqrt{\log M}}. \label{eq:f_RM} \ee  
From \eqref{eq:chi_lb1}, if  $f_R(M)/\Delta_R \to 0$ as $M \to \infty$, then  $\frac{\sigma^2}{P}\left(\frac{\Delta_R +  \Delta_R^2}{2}\right)$  will be the dominant term in $\chi$ for large enough $L,M$.  The condition $f_R(M)/\Delta_R \to 0$  will be satisfied if we choose $ \Delta_R$ such that 
\be
\Delta_R \geq \sqrt{\frac{\log \log M}{\kappa_2 \log M}},
\label{eq:DelRbnd}
\ee
where $\kappa_2$ is the universal constant from  Lemma \ref{lem:conv_expec}(b) and Lemma \ref{lem:lim_xt_taut}.  From here on, we assume that $\Delta_R$ satisfies \eqref{eq:DelRbnd}.  

Let $T$ be the number of iterations until $x_t$ exceeds $(1-f_R(M))$. We run the AMP decoder for $T$ iterations, where
\be
\begin{split}
T & := \min_t \, \{ t: \,  x_{t} \geq 1- f_R(M) \}  \\
 & \stackrel{(a)}{\leq} \frac{1-f_R(M)}{\chi}\\
 &  \stackrel{(b)}{=} \frac{P/\sigma^2}{(\Delta_R + \Delta_R^2)/2}(1 + o(1)),  \label{eq:Tdef}
\end{split}
\ee
where $o(1) \to 0$ as $M,L \to \infty$.  In \eqref{eq:Tdef}, inequality $(a)$ holds for sufficiently large $L,M$ due to Lemma \ref{lem:lim_xt_taut}, which shows for large enough $L,M$, the $x_t$ value increases by at least $\chi$ in each iteration. The equality $(b)$ follows from the lower bound on $\chi$ in \eqref{eq:chi_lb1}, and because $f_R(M)/\Delta_R = o(1)$.  

After running the decoder for $T$ iterations, the decoded message $\hat{\fullb}$ is obtained  by setting the maximum of 
$\fullb^{T}$  in each section $\ell \in [L]$ to $\sqrt{nP_{\ell}}$ and the remaining entries to $0$.    For a given $\snr$, from \eqref{eq:Tdef} we note that the number of iterations  $T$ depends only on the gap from capacity  $\Delta_R=(\mc C - R)/\mc C$, and does not grow with the problem dimensions $M, L$, or  $n$. The number of iterations increases as $R$ approaches $\mc C$.    The definition of $T$ guarantees that
  $x_T \geq (1-f_R(M))$. Therefore, using $\tau_T^2=\sigma+P(1-x_T)$ we have
    \be \sigma^2 \leq \tau^{2}_T  \leq \sigma^2 +  P f_R(M).
  \label{eq:tauT_bound} \ee 


\section{Performance of the AMP decoder} \label{sec:AMP_perf}

The \emph{section error rate}  of a decoder for a SPARC $\mc{S}$ is defined as
$
\mc{E}_{sec}(\mc{S}) := \frac{1}{L} \sum_{\ell =1}^{L}  \mathbf{1}{\{ \hat{\fullb}_{(\ell)} \neq \fullb_{0_{(\ell)}} \}}.
$
Our main result is a bound on the probability of the section error rate exceeding any fixed $\e > 0$. 

\begin{thm}
Fix any rate $R < \mc{C}$. Consider a rate $R$ SPARC $\mc{S}_n$  with block length $n$, design matrix parameters $L$ and $M$ determined according to  \eqref{eq:ml_nR}, and an exponentially decaying power allocation given by \eqref{eq:exp_power_alloc}. Furthermore, assume that $M$ is large enough that 
\[  \Delta_R \geq \sqrt{\frac{\log \log M}{\kappa_2 \log M}}, \]
where $\kappa_2$ is the universal constant used in Lemmas \ref{lem:conv_expec}(b) and \ref{lem:lim_xt_taut}. Fix any $\e > \frac{2 \snr}{\mc{C}} f_R(M)$, where $f_R(M) := \frac{M^{-\kappa_2 \Delta_R^2}}{\Delta_R \sqrt{\log M}}$.

Then, for sufficiently large $L,M$, the section error rate of the AMP decoder satisfies
\be
\begin{split}
& P \left( \mc{E}_{sec}(\mc{S}_n)  >  \e \right)  \\
& \leq K_{T} \exp\Big\{\frac{-\kappa_{T} L}{(\log M)^{2T-1}}   
\Big(\frac{\e  \sigma^2 \mc{C}}{2} - P f_R(M) \Big)^2 \Big\}
 \label{eq:pezero}
\end{split}  
\ee
where $T$ is defined in \eqref{eq:Tdef}. The constants $\kappa_T$ and $K_T$ in \eqref{eq:pezero} are given by $\kappa_T = [c^{2T} (T!)^{17}]^{-1}$ and $K_T = C^{2T} (T!)^{11}$ where $c, C > 0$ are universal constants (not depending on AMP parameters $L, M, n, $ or $\e$) but are not explicitly specified.  
\label{thm:main_amp_perf}
\end{thm}

\begin{IEEEproof} The proof   is given in Section \ref{sec:amp_proof}.
\end{IEEEproof}

 In the discussion that follows we refer to the probability $P( \mc{E}_{sec}(\mc{S}_n)  >  \e_0)$ on the left side of  \eqref{eq:pezero} as the `deviation probability' (of the section error rate), and the upper bound given by the right side of  \eqref{eq:pezero} as the `bound on the deviation probability'.

\vspace{10pt}
\textbf{Remarks}:
\begin{enumerate}
\item The probability measure in \eqref{eq:pezero} is over the Gaussian design matrix ${\fullA}$, the Gaussian channel noise $\fullw$, and   the message $\fullb$ distributed uniformly in $\mcb(P_1, \ldots, P_L)$.

\item Given $L,M$, the bound on the deviation probability given in \eqref{eq:pezero} depends on the rate $R$ only through $T$.

\item  Asymptotic convergence results of the kind given in \cite{RushGVIT17} are implied by Theorem \ref{thm:main_amp_perf}.  Indeed, for any fixed $R < \mc{C}$, consider a sequence of SPARCs $\{ \mc{S}_n\}_{n \geq 0}$ indexed by block length $n$ with $M=L^{\textsf{a}}$ for some constant $\textsf{a} > 0$. Then, from Theorem \ref{thm:main_amp_perf}  we have
 $\sum_{n=1}^{\infty} P(\mc{E}_{sec}(\mc{S}_n) \geq \e)  < \infty.$ Therefore the Borel-Cantelli lemma implies that $\lim_{n \to \infty}   \mc{E}_{sec}(\mc{S}_n) \stackrel{a.s.}{=} 0$. 
 
 We note  that for a fixed $R < \mc{C}$, there are many choices for scaling $M$ vs. $L$ that guarantee that  $\lim_{n \to \infty}   \mc{E}_{sec}(\mc{S}_n) \stackrel{a.s.}{=} 0$. Some examples, along with the tradeoffs they imply, are discussed in the  following subsection. 
 
 \item The dependence of the constants $K_T, \kappa_T$ on $T!$ arises due to the induction-based proof of the concentration lemma. These constants have not been optimized, but we believe that the dependence of  these constants on $T!$ is inevitable in any induction-based proof of the result.

 \item  As described in \cite{AntonyML}, one can obtain a small probability of codeword error, i.e.,  $P(\hat{\fullb} \neq \fullb)$,    by using a concatenated code with the SPARC as the inner code and  an outer Reed-Solomon  code.    A suitably chosen Reed-Solomon code of rate $(1-2\e)$ ensures that $\hat{\fullb} =\fullb$ whenever the section error rate $ \mc{E}_{sec} < \e$, for any $\e >0$. For such a concatenated code, the overall rate is $(1-2\e)R$ and the probability $P(\hat{\fullb} \neq \fullb)$ is bounded by the RHS of \eqref{eq:pezero}.

 \item \label{rem:twodec} The deviation probability of the section error rate for the Cho-Barron adaptive successive soft-decision decoder has also been shown to decay exponentially in $L/(\log M)^{2T -1}$ \cite[Lemma 7]{ChoB13}.   When implemented with Gaussian design matrices, both the AMP decoder and the adaptive successive soft-decision decoder have running time and memory of $O(nML)$. However, the  latter  requires an orthonormalization step in each iteration, hence the AMP decoder is faster in practice.  Moreover, the complexity and memory requirement  of the AMP decoder can be greatly improved by replacing the Gaussian design matrix with a Hadamard-based one  \cite{RushGVIT17,BarbKrz17}.  However, there is currently no theoretical analysis of the section error rate with a Hadamard-based AMP decoding scheme. 
 
 \item Though Theorem \ref{thm:main_amp_perf} is stated and proved for the exponentially decaying allocation with $P_\ell \propto e^{-2\mc{C} \ell/L}$,  a   result similar to \eqref{eq:pezero} holds for any power allocation for which a state evolution lower  bound analogous to  Lemma \ref{lem:lim_xt_taut} can be established. More precisely, consider a fixed $R< \mc{C}$  and an allocation $\{ P_\ell \}_{\ell \in [L]}$ such that the state evolution parameter $x_t$ monotonically increases until it reaches $(1- f(M))$ in a finite number of iterations $T'$. Then the deviation probability bound \eqref{eq:pezero} holds for that allocation, with $T$ replaced by $T'$.

 \end{enumerate}

\subsection{Effects of  $L,M$ on decoding performance} \label{subsec:code_params}

We now examine how varying the parameters $L,M$ affects the performance of the AMP decoder. Recall from \eqref{eq:ml_nR} that $L, M$ determine the block length via  $n = (L \log M)/R$.  For a fixed $M$ and a target section error rate $\e_0 > \frac{2 \snr}{\mc{C}} f_R(M)$, the bound on the deviation probability given in  \eqref{eq:pezero} shows that the probability of  decoding greater than a fraction $\e_0$ of sections in error  decreases exponentially in $L$.  

Next, consider fixing $L$ and increasing $M$. This has several effects.  First,  $M$ controls how small the  target section error rate in Theorem \ref{thm:main_amp_perf}  can be,  via the requirement $\e > \frac{2\snr}{\mc{C}} f_R(M)$.  Hence, in order to bound the deviation probability $P( \mc{E}_{sec}(\mc{S}_n)  >  \e_0)$ for a fixed 
$R$, there is a corresponding requirement that $M $ is large enough that $\e_0 > \frac{2\snr}{\mc{C}} f_R(M)$.  Thus, the larger the $M$, the smaller we can make the target section error rate.  Also notice from Lemmas \ref{lem:conv_expec} and \ref{lem:lim_xt_taut} that increasing $M$ tightens the lower bound on the state evolution parameter $x_t$ in each step.

\begin{figure}[t]
    \centering
    \includegraphics[width=3.5in]{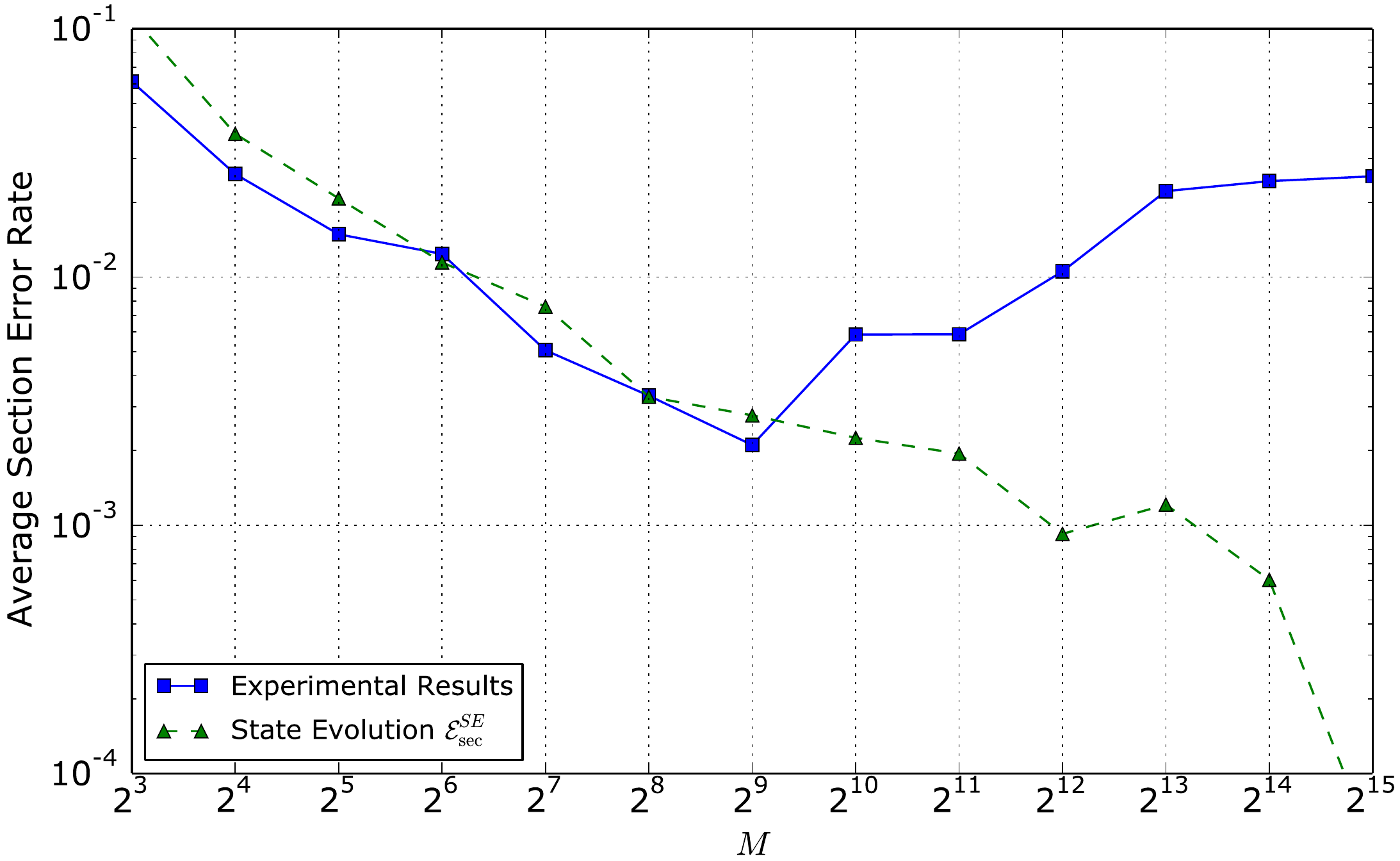}
    \caption{\small AMP error performance with increasing $M$, for
      fixed  $L=1024$, $R=1.5$ bits, and  $\snr =11.1$ (2 dB from Shannon limit). The solid line is the empirical section error rate, obtained by averaging over $200$ trials for each point. The dashed line is the section error rate predicted by state evolution. See \cite[Sec. III.A]{GreigV17} for details.  }%
        \vspace{-7pt}
\label{fig:lvsm_ser}
\end{figure}

On the other hand, the bound on the deviation probability in Theorem \ref{thm:main_amp_perf} Eq.\ \eqref{eq:pezero} worsens with increasing $M$ (with $L$ fixed). Moreover, as $R$ gets closer to $\mc{C}$,  the weakening of the bound in \eqref{eq:pezero}  due to increasing $M$ is more acute since $T$ increases with $\Delta_R^{-1}= \mc{C}/(\mc{C} - R)$ (see \eqref{eq:Tdef}).   Therefore, increasing $M$ allows the AMP decoder to have a smaller target section error rate, but  increases the probability of the observed section error rate deviating from the target by a large amount. This prediction, based on Theorem \ref{thm:main_amp_perf}, has been empirically verified recently in \cite[Sec.\ III.A]{GreigV17}. Fig.\ \ref{fig:lvsm_ser} shows a  plot from \cite{GreigV17}, with the solid curve representing the empirical section error rate for different values of $M$ with $L=1024$. We observe that the dashed curve, representing the state evolution prediction for the section error rate, approaches zero with increasing $M$. However, beyond a certain value of $M$, the empirical performance begins to diverge sharply  from the state evolution prediction. This is due to the benefit of a lower predicted section error rate being outweighed by the loss of concentration around the prediction.    

As $M$ increases with $L$ held fixed, numerical simulations for a range of rates  suggest that the empirical section error rate starts diverging from the state evolution prediction close to $M=L$ (as in Fig.\ \ref{fig:lvsm_ser}).  However, a theoretical analysis of how  $M$ should be chosen for a fixed $L$ in order to ensure the smallest section error rate (while maintaining concentration around the state evolution prediction) remains open. The challenge here is that the constants $\kappa_T, K_T$ in Theorem \ref{thm:main_amp_perf} are not optimal, and the exact dependence of the deviation probability on $T$ is not known. 

In the next two subsections, we consider the behavior of the  deviation probability bound of Theorem \ref{thm:main_amp_perf} in two different regimes. The first is where $R < \mc{C}$ is held constant as $L, M \to \infty$ (with $n= L \log M /R$) --- the so-called ``error exponent" regime. In this case, $\Delta_R$ is of constant order, so $f_R(M)$ in \eqref{eq:f_RM} decays polynomially with growing $M$.  The other regime is where $R$ approaches $\mc{C}$ as $L,M \to \infty$  (equivalently, $\Delta_R$ shrinks to $0$), while ensuring that the error probability remains small or goes to $0$. Here, \eqref{eq:DelRbnd} specifies that $\Delta_R$ should be of order at least $\sqrt{\frac{\log \log M}{ \log M}}$.

\subsection{Error exponent of AMP decoding} \label{subsec:errexp}
 For any ensemble of codes, the error exponent specifies how the codeword error probability decays with growing code length $n$ for a fixed $R < \mc{C}$ \cite{Gal68Book}. In the SPARC setting,  we wish to understand how the bound on the deviation probability in Theorem \ref{thm:main_amp_perf}  decays with $n$ for fixed values of $\e >0$ and  $R < \mc{C}$.   (As explained in Remark 5 following Theorem \ref{thm:main_amp_perf}, concatenation using an outer code can be used to extend the result to  the codeword error probability.)  With optimal encoding, it was shown in \cite{AntonyML} that deviation probability decays exponentially in $n  \min\{ \e \Delta, \Delta^2 \}$, where $\Delta=(\mc{C} -R)$.   For the AMP decoder, we consider two choices for $(M,L)$ in terms of $n$ to illustrate the trade-offs involved: 
 \begin{enumerate}
\item  $M=L^{\textsf{a}}$, for some constant $\textsf{a} >0$. Then, \eqref{eq:ml_nR} implies that $L= \Theta(\frac{n}{\log n})$ and $M=  \Theta((\frac{n}{\log n})^\mathsf{a})$. Therefore, the bound on the deviation probability in Theorem \ref{thm:main_amp_perf}  decays exponentially in $n/(\log n)^{2T}$.  

\item   $L =  {\kappa n}/{\log \log n}$, for some constant $\kappa$, which implies $M = \frac{R}{\kappa} \log n$. With this choice the bound on the deviation probability in Theorem \ref{thm:main_amp_perf}  decays exponentially in $n/(\log \log n)^{2T}$.
 \end{enumerate}
Note from \eqref{eq:Tdef}  that  for a fixed $R < \mc{C}$, the number of AMP iterations $T$ is an $\Theta(1)$ quantity that does not grow with $L, M,$ or $n$. The deviation probability decays more rapidly with $n$ for the second choice above, but this comes at the expense of much smaller $M$ (for a given $n$).  Therefore, the first choice allows for a much smaller target section error rate (due to smaller $f_R(M)$), but has a larger probability of deviation from the target. One can also compare the two cases in terms of decoding complexity, which is $O(nML T
)$ with Gaussian design matrices. The complexity in the first case is $O({n^{2+\mathsf{a}}}/{(\log n)^{1 + \mathsf{a}}})$, while in the second case it is $O({n^2 \log n}/{\log \log n})$.

\subsection{Gap from capacity with AMP decoding} \label{subsec:scaling}

We now consider how fast  $R$ can approach the capacity $\mc{C}$ with growing $n$, so that the deviation probability still decays to zero. Recall that lower bound on the gap from capacity is already specified by \eqref{eq:DelRbnd}: for the state evolution parameter $x_T$ to converge to $1$ with growing $M$ (predicting reliable decoding), we need $\Delta_R \geq \sqrt{\frac{\log \log M}{\kappa_2 \log M}}$.  When $\Delta_R$ takes this minimum value, the minimum target section error rate $f_{R}(M)$ in Theorem \ref{thm:main_amp_perf} is  
\be \underline{f_{R}}(M) =  \frac{\sqrt{\kappa_2}}{\log M  \sqrt{\log \log M} }. \label{eq:frmin0} \ee
(We note that in the lower bound for $\Delta_R$ in \eqref{eq:DelRbnd}, we can replace $\kappa_2$ by $ \kappa_2/\kappa_0$, for any $\kappa_0 > \frac{1}{2}$. This would change the factor of $\log M$ in the denominator of \eqref{eq:frmin0} to $(\log M)^{\kappa_0}$. We do not pursue this generalization in the interest of keeping the exposition simple. We just note that increasing $\kappa_0$ allows a smaller target section error rate $f_R(M)$ at the expense of a larger  gap $\Delta_R$.)

 We evaluate the bound on the deviation probability of Theorem \ref{thm:main_amp_perf} with $\Delta_R$ at the minimum value of $\sqrt{\frac{\log \log M}{\kappa_2 \log M}}$, for $\e > \frac{2 \snr}{\mc C} \underline{f_R}(M)$, with $\underline{f_R}(M)$ given in \eqref{eq:frmin0}. From \eqref{eq:Tdef}, we have the  bound 
 \be 
 T \leq \frac{2\snr}{\Delta_R}  \leq \kappa_4 \sqrt{\frac{\log M}{\log \log M}}  \label{eq:Tbnd0} 
 \ee for large enough $L,M$. Then, using Stirling's approximation to write $\log(T!) =    T\log T - T + O(\log T)$, Theorem \ref{thm:main_amp_perf} yields
 \begin{align}
&  -  \log P  ( \mc{E}_{sec}(\mc{S}_n) >  \e ) \nonumber \\
 & \geq   \frac{\kappa_5 L \e^2}{ c^{2T} (T!)^{17} (\log M)^{2T-1}}  - O(T\log T) \nonumber \\
& =   {\kappa_5 L \e^2}  \exp \big\{- 2T \log c - 17(T \log T - T) \nonumber \\
& \qquad  \qquad  - (2T-1) \log \log  M  -  O(\log T) \big \}   - O(T\log T) \nonumber \\
& \geq \frac{L \e^2}{\exp\Big\{  \kappa_6 \sqrt{(\log M) (\log \log M)} \Big(1 + O(\frac{1}{\log \log M}) \Big)\Big\}}  \nonumber \\ 
& \quad - O\Big(\sqrt{(\log M)(\log \log M)} \Big)
\label{eq:logPe_bnd}
 \end{align}
where the last inequality above follows from \eqref{eq:Tbnd0}.

We now evaluate the bound in \eqref{eq:logPe_bnd} for the case $M=L^{\textsf{a}}$  considered in Sec  \ref{subsec:errexp}. We then have $L = \Theta(\frac{n}{\log n})$ and $M=  \Theta((\frac{n}{\log n})^\mathsf{a})$. Substituting these in \eqref{eq:logPe_bnd}, we obtain  
 \begin{align}
 & -  \log P  ( \mc{E}_{sec}(\mc{S}_n) >  \e )  \nonumber \\ 
 &  \geq \frac{\kappa_7 n \e^2}{(\log n) \exp\{ \kappa_8 \sqrt{(\log n) (\log \log n)}\}}  \nonumber \\
 & = \kappa_7 \exp \{ \log n -  \kappa_8\sqrt{(\log n) (\log \log n)} - \log \log n \} \e^2 \nonumber \\
 & = \kappa n^{1 - O(\sqrt{\frac{\log \log n}{\log n}} + \frac{\log \log n}{\log n})} \e^2.
 \end{align}
 Therefore, for the case $M= L^{\textsf{a}}$, with a gap from capacity ($\Delta_R$) that is of order $\sqrt{\frac{\log \log n}{\log n}}$, we can achieve a deviation probability that decays as
 $\exp\{-\kappa n^{1 - O(\sqrt{{\log \log n}/{\log n}})} \e^2\}$. Furthermore, from  \eqref{eq:frmin0} we see that $\e$ must be of  order at least $\frac{1}{\log n  \sqrt{\log \log n}}$.
 
 We note that this gap from capacity is of a larger order than polar codes, which have a polynomial gap to capacity \cite{GuruXia15polar}. Guruswami and Xia showed in\cite{GuruXia15polar} that for binary input, symmetric memoryless channels, polar codes  of block length $n$ with gap from capacity of  order $\frac{1}{n^{\mu}}$ can achieve a block error probability decaying as $2^{-n^{0.49}}$ with a decoding algorithm whose complexity scales as $n \cdot  \text{poly}(\log n)$. (Here $0 < \mu < \frac{1}{2}$ is a universal constant.) We remark that for AWGN channels, there is no known coding scheme that provably achieves a polynomial gap to capacity with efficient decoding.

 Recall the lower bound on the gap to capacity arises from the condition \eqref{eq:chi_lb1} which is required to ensure that the (deterministic) state evolution sequence $x_1, x_2, \ldots $ is guaranteed to increase by at least an amount proportional to $\Delta_R$ in each iteration. It was shown in \cite[Sec.\ 4.18]{AntonyThesis} that for the iterative hard-decision decoder that the gap to capacity can be improved to $O(\frac{\log \log M}{\log M})$ by modifying the exponential power allocation: the idea is to  flatten the power allocation for a certain number of sections at the end. We expect such a modification to yield a similar improvement in the capacity gap for the AMP decoder, but we do not detail this analysis as it is involves additional technical details. 
 
 To summarize, the AMP decoder (as well as the adaptive hard-decision/soft-decision decoders) are efficient and achieve near-exponential decay of  error probability in the regime where $R < \mc {C}$ remains fixed. When $\Delta_R$ shrinks to $0$ with growing $M$, these decoders are no longer efficient as they require $M$ to increase exponentially in $1/\Delta_R$ (cf.\ \eqref{eq:DelRbnd}). An interesting open question is whether spatially coupled SPARCs with AMP decoding have a smaller gap from capacity. The analysis of the state evolution equations for spatially coupled SPARCs (via potential functions) in \cite{barbISIT16,barbITW16} indicates that they achieve capacity with AMP decoding, but a rigorous analysis of the error probability of these  spatially coupled SPARCs is still an open question.
 

\section{Proof of Theorem \ref{thm:main_amp_perf}} \label{sec:amp_proof}

The main ingredients in the proof of Theorem \ref{thm:main_amp_perf} are two technical lemmas (Lemma \ref{lem:hb_cond} and Lemma \ref{lem:main_lem}).  After laying down some definitions and notation that will be used in the proof, we state the two lemmas and use them to prove Theorem \ref{thm:main_amp_perf}.

\subsection{Definitions and Notation for the Proof}
For consistency with earlier analyses of AMP, we use notation similar to  \cite{BayMont11, RushGVIT17}.  Define the following column vectors recursively for $t\geq 0$, starting with $\fullb^0=\textbf{0}$ and $\fullz^0=\fully$.
\begin{equation}
\begin{split}
\fullh^{t+1}  := \fullb_0 - (\fullA^*\fullz^t + \fullb^t), \qquad &  \fullq^t  :=\fullb^t - \fullb_0, \\
\fullbamp^t := \fullw-\fullz^t,\qquad & \fullm^t :=-\fullz^t.
\end{split}
\label{eq:hqbm_def}
\end{equation}
Recall that $\fullb_0$ is the message vector chosen by the transmitter. The vector $\fullh^{t+1}$ is the noise in the effective observation $\fullA^*\fullz^t + \fullb^t$, while $\fullq^t$ is the error in the estimate $\fullb^t$.  The proof will show that $\fullh^{t+1}$ is approximately i.i.d.\ $\mc{N}(0, \tau_t^2)$, while $\fullbamp^t$ is approximately i.i.d.\ $\mc{N}(0, \tau_t^2 - \sigma^2)$.

Define  $\mathscr{S}_{t_1, t_2}$ to be the sigma-algebra generated by
\[ \fullbamp^0, ..., \fullbamp^{t_1 -1}, \fullm^0, ..., \fullm^{t_1 - 1}, \fullh^1, ..., \fullh^{t_2}, \fullq^0, ..., \fullq^{t_2}, \,  \fullb_0, \fullw. \]
Lemma \ref{lem:hb_cond} iteratively computes the conditional distributions $\fullbamp^t |_{ \mscrs_{t, t}}$ and $\fullh^{t+1} |_{ \mscrs_{t+1, t}}$. Lemma \ref{lem:main_lem} then uses this conditional distributions to show the concentration of various inner products involving $\fullh^{t+1}, \fullq^t, \fullbamp^t$, and $\fullm^t$ to deterministic constants.  

For $t \geq 1$, let
\be
\lambda_t := \frac{-1}{\tau^2_{t-1}}\Big( P - \frac{\norm{\fullb^t}^2}{n} \Big).
\label{eq:lambda_t_def}
\ee
We then have
\begin{equation}
\fullbamp^{t} + \lambda_t \fullm^{t-1} = \fullA \fullq^t, \quad \text{ and } \quad \fullh^{t+1} + \fullq^t = \fullA^* \fullm^t,
\label{eq:bmq}
\end{equation}
which follows from \eqref{eq:amp1} and \eqref{eq:hqbm_def}. From \eqref{eq:bmq}, we have the matrix equations 
 \be  
 \fullB_t +  [\mathbf{0} | \fullM_{t-1}] \fullL_t = \fullA \fullQ_t  \quad \text{ and } \quad  \fullH_t + \fullQ_{t} = \fullA^* \fullM_t, \label{eq:XtYt_rel} 
 \ee
where for $t \geq 1$
\be
\begin{split}
&\fullM_t  := [\fullm^0 \mid \ldots \mid \fullm^{t-1} ],  \qquad \fullQ_t  :=  [\fullq^0 \mid \ldots \mid \fullq^{t-1} ] \\ 
& \fullB_t := [\fullbamp^0 | \ldots | \fullbamp^{t-1}], \quad  \fullH_t = [\fullh^1 | \ldots | \fullh^{t}], \\
&  \fullL_t := \text{diag}(\lambda_0, \ldots, \lambda_{t-1}). 
\end{split}
\label{eq:XYMQt}
\ee
The notation $[\textbf{c}_1 \mid \textbf{c}_2 \mid \ldots \mid \textbf{c}_k]$ is used to denote the matrix with columns $\textbf{c}_1, \ldots, \textbf{c}_k$.  We define $\fullM_0, \fullQ_0, \fullB_0$, $\fullH_0$, and $\fullL_0$ to be all-zero vectors.  

 We use  $\fullm^t_{\|}$ and $\fullq^t_{\|}$ to denote the projection of $\fullm^t$ and $\fullq^t$ onto the column space of $\fullM_t$ and $\fullQ_t$, respectively. Let
 $\fulla_t := (\alpha^t_0, \ldots, \alpha^t_{t-1})^*$ and $\fullg_t :=  (\gamma^t_0, \ldots, \gamma^t_{t-1})^*$ be the coefficient vectors of these projections, i.e.,
 \be
 \fullm^t_{\| } = \sum_{i=0}^{t-1} \alpha^t_i \fullm^i, \quad  \fullq^t_{\|} = \sum_{i=0}^{t-1} \gamma^t_i \fullq^i.
 \label{eq:mtqt_par}
 \ee
 The projections of $\fullm^t$ and $\fullq^t$ onto the orthogonal complements of $\fullM^t$ and $\fullQ^t$, respectively,  are denoted by
 \be
 \fullm^t_{\perp} := \fullm^t - \fullm^t_{\|}, \quad  \fullq^t_{\perp} := \fullq^t - \fullq^t_{\|}
  \label{eq:mtqt_perp}
 \ee
 
 Lemma \ref{lem:main_lem} shows that for large $n$, the entries of $\fulla_t$ and $\fullg_t$ concentrate around constants. We now specify these constants. With $\tau^2_t$  and $x_t$ as defined in \eqref{eq:taut_def} and \eqref{eq:xt_tau_def}, for $t \geq 0$ define
\be
\sigma^2_t : = \tau_t^2 - \sigma^2 = P(1-x_t).
\label{eq:sigt_def}
 \ee
Define matrices $\tilde{\fullC}^t, \breve{\fullC}^t \in \mathbb{R}^{t \times t}$ for $t \geq 1$ such that for $0\leq i,j \leq t-1$,
\be
\tilde{C}^{t}_{i+1,j+1} = \sigma_{\max(i,j)}^2, \quad \text{ and } \quad \breve{C}^{t}_{i+1,j+1} = \tau_{\max(i, j)}^2. 
\label{eq:Ct_def}
\ee
The concentrating values for $\fullg^t$ and $\fulla^t$ are
\be
\begin{split}
\hat{\fullg}^{t} &:= \sigma_{t}^2 (\tilde{\fullC}^{t})^{-1}  (1, \ldots, 1)^* \stackrel{(a)}{=} (0,\ldots, 0, \sigma_t^2/\sigma_{t-1}^2)^* \in \mathbb{R}^t, \\
\hat{\fulla}^{t} &:= \tau_{t}^2 (\breve{\fullC}^{t})^{-1}  (1, \ldots, 1)^* \stackrel{(b)}{=} (0,\ldots, 0, \tau_t^2/\tau_{t-1}^2)^*  \in \mathbb{R}^t.
\label{eq:hatalph_hatgam_def}
\end{split}
\ee
To see that $(a)$ holds,  we observe that $(\tilde{\fullC}^{t})^{-1} \tilde{\fullC}^{t} = \mathsf{I}_{t}$ implies that $(\tilde{\fullC}^{t})^{-1}  (\sigma_{t-1}^2, \ldots, \sigma_{t-1}^2)^*  = (0, \ldots, 0, 1) \in \mathbb{R}^t$. The equality $(b)$ is obtained similarly.
Let $(\sigma^{\perp}_0)^2 := \sigma_0^2$ and $(\tau^{\perp}_0)^2 := \tau_0^2$, and for $t > 0$ define 
\be
\begin{split}
& (\sigma_{t}^{\perp})^2 := \sigma_{t}^2 \Big(1 - \frac{ \sigma_{t}^2 }{\sigma_{t-1}^2 }\Big), \quad \text{and} \quad (\tau^{\perp}_{t})^2 := \tau_{t}^2 \Big(1 - \frac{\tau_{t}^2}{\tau_{t-1}^2}\Big).
\label{eq:sigperp_defs}
\end{split}
\ee

\begin{lem} \label{lem:sigmatperp}
For sufficiently large $L,M$, the constants $(\sigma_k^{\perp})^2$ and $(\tau_k^{\perp})^2$ are bounded below by a  positive constant $c_* > 0$ for $0 \leq k < T$.  The value of $c_*$ depends on the ratio $R/\mathcal{C}$, with $c_*$ approaching $0$ as the rate approaches the capacity.  \end{lem}

\begin{IEEEproof}
For $k=0$, the lower bounds are immediate since $(\tau^{\perp}_{0})^2= \sigma^2 + P$ and $(\sigma^{\perp}_{0})^2 = \sigma^2$. For $1 \leq k \leq T$, we write
\be 
\begin{split}
(\tau^{\perp}_{k})^2  & = \frac{\tau_k^2}{\tau_{k-1}^2}(P(x_k -x_{k-1})) \\
&  \stackrel{(a)}{\geq}  \frac{\sigma^2 P}{\sigma^2+P} \, \chi 
 \stackrel{(b)}{\geq} \frac{\sigma^2 P}{\sigma^2+P} \, \kappa_0 (\Delta_R + \Delta_R^2/2),  
 \label{eq:taupk_bnd} 
 \end{split}
 \ee
where $\kappa_0 >0$ is a universal constant. Here, $(a)$ is due to \eqref{eq:chi}, and $(b)$ follows from \eqref{eq:chi_lb1} for large enough $L,M$ because $f(M)$ is of smaller order than $\Delta_R$ when \eqref{eq:DelRbnd} is satisfied.

  The lower bound on $(\sigma^{\perp}_{k})^2$ follows in a similar manner by writing
\begin{align}
& (\sigma^{\perp}_{k})^2 = \frac{\sigma^2_k}{\sigma^2_{k-1}} (\sigma^2_k - \sigma^2_{k-1})=\frac{ 1- x_k}{1- x_{k-1}}  P(x_k - x_{k-1})  \nonumber \\
&  \stackrel{(a)}{\geq} \chi \cdot P \chi \stackrel{(b)}{\geq} P \kappa^2_0 (\Delta_R + \Delta_R^2/2)^2, \label{eq:sigpk_bnd}
\end{align}
where $(a)$ is obtained as follows. Since $T$ is the first iteration in which the $x$ value exceeds $(1-f(M))$, and the increase in each iteration is at least $\chi$, for $1 \leq k <T$ we have $1 > x_k \geq  x_{k-1} + \chi$. The inequality $(b)$ follows from the same argument as  the last inequality in \eqref{eq:taupk_bnd}.
From \eqref{eq:taupk_bnd} and \eqref{eq:sigpk_bnd}, it is clear that the bounds  tend to $0$ as $R$ approaches $\mc C$. 
\end{IEEEproof}

\begin{lem} \label{lem:Ct_invert}
If the $(\sigma_k^{\perp})^2$ and $(\tau_k^{\perp})^2$ are bounded below by some positive constants for $0 \leq k < T$, then the matrices $\tilde{\fullC}^k$ and $\breve{\fullC}^k$ defined in \eqref{eq:Ct_def} are invertible for $1 \leq k \leq T$. \end{lem}

\begin{IEEEproof}
The proof can be found in \cite[Lemma 2]{RushV16}.
\end{IEEEproof}

We use the following notation.  Given two random vectors $\textbf{X}, \textbf{Y}$ and a sigma-algebra $\mscrs$, we write $\textbf{X} |_\mscrs \stackrel{d}{=} \textbf{Y}$ when the conditional distribution of $\textbf{X}$  given $\mscrs$ equals the distribution of $\textbf{Y}$. For a matrix $\fullA$ with full column rank, $\proj_{\fullA} := \fullA(\fullA^* \fullA)^{-1} \fullA^*$ denotes the orthogonal projection matrix onto the column space of $\fullA$, and $\proj^{\perp}_\fullA := \mathsf{I} - \proj_\fullA$. We also recall that $N=ML$.

The following lemma, which was also used for the asymptotic result in \cite{RushGVIT17}, characterizes   the conditional distribution of the vectors $\fullh^{t+1}$ and $\fullbamp^t$ given the matrices in \eqref{eq:XYMQt} as well as $\fullb_0$ and $\fullw$.

\begin{lem} [Conditional Distribution Lemma {\cite[Lemma 4]{RushGVIT17}}] 
For the vectors $\fullh^{t+1}$ and $\fullbamp^t$ defined in \eqref{eq:hqbm_def}, the following hold for $1 \leq t \leq T$, provided $n >T$, and $\fullM_t$ and $\fullQ_t$ have full column rank.
\begin{align}
& \fullh^{1} \lvert_{\mscrs_{1, 0}} \stackrel{d}{=} \tau_0 \vecZ_0 + \fullD_{1,0}, \\
& \fullh^{t+1} \lvert_{\mscrs_{t+1, t}} \stackrel{d}{=} \frac{\tau_t^2}{\tau_{t-1}^2} \fullh^{t} + \tau_{t}^{\perp} \, \vecZ_t + \fullD_{t+1,t}, \label{eq:Ha_dist} 
\end{align}
and 
\begin{align}
& \fullbamp^{0} \lvert_{\mscrs_{0, 0}} \stackrel{d}{=} \sigma_0 \vecZ'_0, \\
&  \fullbamp^{t} \lvert_{\mscrs_{t, t}}\stackrel{d}{=} \frac{\sigma_t^2}{\sigma_{t-1}^2} \fullbamp^{t-1} +  \sigma_{t}^{\perp} \, \vecZ'_t + \fullD_{t,t}. \label{eq:Ba_dist}
\end{align}
where $\vecZ_0, \vecZ_t \in \mathbb{R}^N$ and $\vecZ'_0, \vecZ'_t \in \mathbb{R}^n$ are i.i.d.\ standard Gaussian random vectors that are independent of the corresponding conditioning sigma algebras. The deviation terms are $\fullD_{0,0}=0$,
\begin{align}
\fullD_{1,0} &= \Big[ \Big(\frac{\norm{\fullm^0}}{\sqrt{n}}  - \tau_0\Big)\mathsf{I} -\frac{\norm{\fullm^0}}{\sqrt{n}} \proj_{\fullq^0}\Big]  \vecZ_0  \nonumber \\ 
& \ + \fullq^0 \Big(\frac{\norm{\fullq^0}^2}{n}\Big)^{-1} \Big(\frac{(\fullb^0)^*\fullm_0}{n} - \frac{\norm{\fullq^0}^2}{n}\Big), \label{eq:D10}
\end{align}
and for $t >0$,
\begin{align}
& \fullD_{t,t} =  \, \sum_{r=0}^{t-2} \gamma^t_r \fullbamp^r  + \Big( \gamma^t_{t-1} - \frac{\sigma^2_t}{\sigma^2_{t-1}} \Big)\fullbamp^{t-1} \nonumber  \\
& \quad + \Big[  \Big(\frac{\norm{\fullq^t_{\perp}}}{\sqrt{n}} - \sigma_{t}^{\perp}\Big) \mathsf{I}  - \frac{\norm{\fullq^t_{\perp}} }{\sqrt{n}} \proj_{\fullM_t}\Big]\vecZ'_t  \nonumber \\
& \quad + \fullM_t\Big(\frac{\fullM_{t}^* \fullM_{t}}{n}\Big)^{-1} \Bigg(\frac{\fullH_t \fullq^t_{\perp}}{n} \nonumber  \\
& \qquad \quad - \frac{\fullM_t}{n}^* \Big[\lambda_t \fullm^{t-1} - \sum_{r=1}^{t-1} \lambda_{r} \gamma^t_{r} \fullm^{r-1} \Big]\Bigg),\label{eq:Dtt} 
\end{align}
\begin{align}
& \fullD_{t+1,t} = \,  \sum_{r=0}^{t-2} \alpha^t_r \fullh^{r+1} + \Big( \alpha^t_{t-1} - \frac{\tau^2_t}{\tau^2_{t-1}} \Big) \fullh^{t} \nonumber \\
& \quad   +  \Big[\Big(\frac{\norm{\fullm^t_{\perp}}}{\sqrt{n}} - \tau_{t}^{\perp}\Big)  \mathsf{I} -\frac{\norm{\fullm^t_{\perp}}}{\sqrt{n}} \proj_{\fullQ_{t+1}} \Big]\vecZ_t \nonumber \\
&\quad + \fullQ_{t+1} \Big(\frac{\fullQ_{t+1}^* \fullQ_{t+1}}{n}\Big)^{-1} \Bigg(\frac{\fullB^*_{t+1} \fullm^t_{\perp}}{n} \nonumber \\
& \qquad \quad  - \frac{\fullQ_{t+1}^*}{n} \Big[\fullq^t - \sum_{i=0}^{t-1} \alpha^t_i \fullq^i \Big]\Bigg).\label{eq:Dt1t}  
\end{align} 
\label{lem:hb_cond}
\end{lem}

The next lemma uses the representation in Lemma \ref{lem:hb_cond} to show that for each $t \geq 0$, $\fullh^{t+1}$ is the sum of an i.i.d.\ $\mc{N}(0, \tau_t^2)$ random vector plus a deviation term.  Similarly $\fullbamp^t$ is the sum of an i.i.d.\ $\mc{N}(0, \sigma_t^2)$ random vector and a deviation term.  
\begin{lem}
For $t \geq 0$, the conditional distributions in Lemma \ref{lem:hb_cond} can be expressed as 
\be
\fullh^{t+1} \lvert_{\mscrs_{t+1, t}} \stackrel{d}{=} \tilde{\fullh}^{t+1} + \tilde{\fullD}_{t+1} , \qquad  \fullbamp^{t} \lvert_{\mscrs_{t, t}}\stackrel{d}{=} \breve{\fullbamp}^{t} + \breve{\fullD}_{t},
\label{eq:htil_rep}
\ee
where
\begin{align}
& \tilde{\fullh}^{t+1} := \tau_{t}^2 \sum_{r=0}^{t} \Big(\frac{\tau^{\perp}_r}{\tau_{r}^2}\Big) \vecZ_r, 
\quad \ \tilde{\fullD}_{t+1} := \tau_{t}^2 \sum_{r=0}^{t} \Big(\frac{1}{\tau_{r}^2}\Big) \fullD_{r+1, r},  \label{eq:htilde_def} \\
& \breve{\fullbamp}^{t}:= \sigma_{t}^2 \sum_{r=0}^{t} \Big(\frac{\sigma^{\perp}_r}{\sigma_{r}^2}\Big) \vecZ'_r,
\quad \  \breve{\fullD}_{t} := \sigma_{t}^2 \sum_{r=0}^{t} \Big(\frac{1}{\sigma_{r}^2}\Big) \fullD_{r, r}.
\label{eq:btilde_def}
\end{align}
\label{lem:ideal_cond_dist}
Here $\vecZ_r \in \reals^N$, $\vecZ'_r \in \reals^n$ are the independent standard Gaussian vectors defined in Lemma \ref{lem:hb_cond}.

Consequently, $\tilde{\fullh}^{t+1}  \stackrel{d}{=} \tau_t \tilde{\vecZ}_t$, and $\breve{\fullbamp}^{t}  \stackrel{d}{=} \sigma_t \breve{\vecZ}_t$, where $\tilde{\vecZ}_t \in \reals^N$ and $ \breve{\vecZ}_t \in \reals^n$ are standard  Gaussian random vectors such that for any $j  \in [N]$ and  $i \in [n]$,  the vectors $(\tilde{Z}_{0,j}, \ldots, \tilde{Z}_{t,j})$ and  $(\breve{Z}_{0,i}, \ldots, \breve{Z}_{t,i})$ are each jointly Gaussian with 
\be \expec[\tilde{Z}_{r,j} \tilde{Z}_{s,j}] = \frac{\tau_s}{\tau_r}, \qquad  \expec[ \breve{Z}_{r,i} \breve{Z}_{s,i}] = \frac{\sigma_s}{\sigma_r} \qquad \text{ for }0 \leq r \leq s \leq t. \ee
\end{lem}
\begin{IEEEproof}
We give the proof for the distributional representation of $\fullh^{t+1}$, with the proof for $\fullbamp^t$ being similar. The representation in  \eqref{eq:htil_rep} can be directly obtained by using Lemma \ref{lem:hb_cond} Eq.\ \eqref{eq:Ha_dist} to recursively write $\fullh^{t}$ in terms of $(\fullh^{t-1}, \vecZ_{t-1}, \fullD_{t, t-1})$, then $\fullh^{t-1}$ in terms of $(\fullh^{t-2}, \vecZ_{t-2}, \fullD_{t-1, t-2})$, and so on. 

Using \eqref{eq:htilde_def}, we write $\tilde{\fullh}^{t+1}= \tau_t  \tilde{\vecZ}_t$, where $\tilde{\vecZ}_t = \tau_{t} \sum_{r=0}^{t}({\tau^{\perp}_r}/{\tau_{r}^2}) \vecZ_r$ is a Gaussian random vector with i.i.d.\ entries, with zero mean and variance equal to
\be
\begin{split}
 \tau_{t}^2\sum_{r=0}^{t} \frac{ (\tau^{\perp}_r)^2}{\tau_{r}^4} &  = \frac{ \tau_{t}^2}{\tau_0^2} + \sum_{r=1}^{t} \Big(\frac{ \tau_{t}^2}{\tau_{r}^2} \Big) \Big(1 - \frac{\tau_r^2}{\tau_{r-1}^2}\Big)  \\
 &=  \frac{\tau_{t}^2}{\tau_0^2} + \sum_{r=1}^{t} \Big(\frac{\tau_{t}^2}{\tau_{r}^2} - \frac{\tau_{t}^2}{\tau_{r-1}^2}\Big) = 1.
\label{eq:barhvar}
\end{split}
\ee
For $j \in [N]$ the covariance between the $j$th entries of $\tilde{\vecZ}_r$ and $\tilde{\vecZ}_s$, for $0 \leq r \leq s \leq t$, is 
\be
\begin{split}
 \expec[\tilde{Z}_{r,j} \tilde{Z}_{s,j}] 
 & = \tau_{r} \tau_{s} \sum_{u=0}^{r} \sum_{v=0}^{s} \Big(\frac{\tau^{\perp}_u}{\tau_{u}^2} \Big) \Big(\frac{\tau^{\perp}_v}{\tau_{v}^2} \Big) 
\mathbb{E}\{Z_{u_j} Z_{v_j}\} \\
& \stackrel{(a)}{=} \tau_{r} \tau_{s} \sum_{u=0}^{r} \frac{(\tau^{\perp}_u)^2}{\tau_{u}^4}
\stackrel{(b)}{=} \frac{\tau_s}{\tau_r},
\end{split}
\ee
where $(a)$ follows from the independence of $Z_{u_j}$ and $Z_{v_j}$ and $(b)$ from the calculation in \eqref{eq:barhvar}.
\end{IEEEproof}

The next lemma shows that the deviation terms in Lemma \ref{lem:hb_cond} are small, in the sense that their section-wise maximum absolute value and norm concentrate around $0$.  Lemma \ref{lem:main_lem} also provides concentration results for various inner products and functions involving $\{\fullh^{t+1}, \fullq^t, \fullbamp^t, \fullm^t\}$.

Let  $c, C >0$ be universal constants not depending on $n$, $\e$, or $t$.  For $t \geq 0$, let
\be 
\begin{split}
& K_t = C^{2t} (t!)^{11} , \quad \kappa_t = \frac{1}{c^{2t} (t!)^{17}},  \\
&  K_t' = C(t+1)^5 K_t , \quad \kappa_t' = \frac{\kappa_t}{c(t+1)^7}. 
\end{split}
\label{eq:Ktkapt_defs} \ee
To keep the notation compact, we use $K, K',  \kappa,$ and $\kappa'$ to denote generic positive universal  constants throughout the lemma statement and proof.

 \begin{lem}
The following  statements hold for $1 \leq t < T$, where $T$ is defined in \eqref{eq:Tdef} and $\e \in (0,1)$.  Let $X_n \doteq c$ be shorthand for $$P(\abs{X_n - c} \geq \e) \leq t^3 K K_{t-1} \exp\Big\{ \frac{-\kappa \kappa_{t-1} L \e^2}{t^3 (\log M)^{2t-1}}\Big\},$$  and let $X_n \circeq c$ be shorthand for $$P(\abs{X_n - c} \geq \e) \leq t^4 K K'_{t-1} \exp\Big\{\frac{-\kappa \kappa'_{t-1} L \e^2}{t^6 (\log M)^{2t+1}}\Big\}.$$

\begin{enumerate}[(a)]
\item 
\begin{align}
& P \Big( \Big[ \frac{1}{L} \sum_{\ell = 1}^L \max_{j \in \ind(\ell)} \abs{[\Delta_{t+1,t}]_{j}} \Big]^2 \geq \epsilon \Big) \nonumber \\
& \quad \leq  P \Big(\frac{1}{L} \sum_{\ell = 1}^L \max_{j \in \ind(\ell)} ([\Delta_{t+1,t}]_{j})^2 \geq \epsilon \Big)   \label{eq:Ha} \\
& \quad \leq t^3 K K_{t-1}' \exp\Big\{\frac{- \kappa \kappa_{t-1}'  L \e}{t^4 (\log M)^{2t}}\Big\},  \label{eq:Ha1} 
\\
& \frac{1}{n}\norm{\fullD_{t,t}}^2 \doteq 0,  \label{eq:Ba}
\end{align}

\item
\begin{align}
& P\Big(\frac{1}{n}\Big \lvert (\fullh^{t+1})^* \fullq^0 \Big \lvert \geq \epsilon \Big) \nonumber \\
&\quad \leq t^3 K K'_{t-1} \exp\Big\{\frac{ -\kappa \kappa'_{t-1} L \e^2}{t^4 (\log M)^{2t-1}}\Big\}, \quad \label{eq:Hb} \\
& \frac{1}{n}(\fullbamp^t)^* \fullw \doteq 0 , \quad \frac{1}{n}(\fullm^t)^* \fullw \doteq - \sigma^2 .  \label{eq:Bb}
\end{align}


\item For all $0 \leq r \leq t$,
\begin{align}
\frac{1}{n} (\fullq^{r})^* \fullq^{t+1}  & \circeq \sigma_{t+1}^2, \quad \frac{1}{n}\norm{\fullq^{t+1}}^2 \circeq \sigma_{t+1}^2,
\label{eq:Hc} \\
\frac{1}{n}(\fullbamp^r)^*\fullbamp^t &\doteq \sigma_t^2.
\label{eq:Bc}
\end{align}


\item For all $0 \leq r,s \leq t$, 
\begin{align}
\frac{1}{n} (\fullh^{s+1})^*\fullq^{r+1} &\circeq - \frac{\sigma_{r+1}^2 \tau_{\max(r,s)}^2}{\tau_r^2}, \label{eq:Hd}\\
\frac{1}{n} (\fullbamp^r)^*\fullm^s &\doteq \sigma_{\max(r,s)}^2 , \label{eq:Bd}
\end{align}


\item For all $0 \leq r \leq t$, 
\begin{align}
 \lambda_{t+1} &\circeq \frac{\sigma_{t+1}^2}{\tau_t^2},  \label{eq:He} \\
 \frac{1}{n}(\fullm^r)^* \fullm^t &\doteq \tau_{t}^2. \label{eq:Be}
\end{align}

\item  Let $\Qmat_{t+1} :=  \frac{1}{n}\fullQ_{t+1}^* \fullQ_{t+1}$ and $\Mmat_t := \frac{1}{n}\fullM_{t}^* \fullM_{t}$.  Then
\begin{align}
&P(\Qmat_{t+1} \text{ is singular}) \leq t K'_{t-1} \exp\Big\{ \frac{-\kappa \kappa'_{t-1} L}{(\log M)^{2t+1}}\Big\}, \label{eq:Qsing}\\
&P(\Mmat_{t} \text{ is singular}) \leq t K_{t-1} \exp\Big\{ \frac{-\kappa \kappa_{t-1} L}{(\log M)^{2t-1}} \Big\}. \label{eq:Msing}
\end{align} 
For all $1 \leq i,j \leq t+1$ and $1 \leq i', j' \leq t$:
\begin{align}
& P\Big(\Big \lvert [\Qmat_{t+1} ^{-1} - (\tilde{\fullC}^{t+1})^{-1}]_{i, j} \Big \lvert \geq \epsilon \, \, \Big \lvert \,\, \Qmat_{t+1} \text{ invertible} \Big) \nonumber \\
 & \quad \leq K K'_{t-1} \exp\Big\{ \frac{- \kappa \kappa'_{t-1}  L \epsilon^2}{(\log M)^{2t-1}}\Big\}, \label{eq:Hf1}\\
& P\Big(\Big \lvert \gamma^{t+1}_{i-1} - \hat{\gamma}^{t+1}_{i-1}  \Big \lvert \geq \epsilon  \, \, \Big \lvert \,\, \Qmat_{t+1} \text{ invertible} \Big) \nonumber  \\
 & \quad \leq t^5 K K'_{t-1} \exp\Big\{\frac{-\kappa \kappa'_{t-1}L \e^2}{t^8(\log M)^{2t+1}}\Big\}, \label{eq:Hf} \\
& P\Big(\Big \lvert [\Mmat_{t} ^{-1} - (\breve{\fullC}^{t})^{-1}]_{i', j'} \Big \lvert \geq \epsilon \, \, \Big \lvert \,\, \Mmat_{t} \text{ invertible} \Big) \nonumber  \\
& \quad \leq K K_{t-1} \exp\Big\{ \frac{- \kappa \kappa_{t-1}  L \epsilon^2}{(\log M)^{2t-3}}\Big\},  \label{eq:Bf1}\\ 
& P\Big(\Big \lvert \alpha^{t}_{i'-1} - \hat{\alpha}^{t}_{i'-1}  \Big \lvert \geq \epsilon \, \, \Big \lvert \,\, \Mmat_{t} \text{ invertible} \Big)  \nonumber  \\ 
& \quad  \leq t^4 K K_{t-1} \exp\Big\{ \frac{-\kappa \kappa_{t-1} L \epsilon^2}{t^5 (\log M)^{2t-1}}\Big\}, \quad t \geq 1. \label{eq:Bf}
\end{align}
Terms $\tilde{\fullC}_{t+1}$ and $\breve{\fullC}_{t}$ are defined in \eqref{eq:Ct_def} and $\hat{\fullg}^{t+1}$ and $\hat{\fulla}^{t}$ are defined in \eqref{eq:hatalph_hatgam_def}.


\item For terms $(\sigma^{\perp}_{t+1})^2$ and $(\tau^{\perp}_t)^2$ defined in \eqref{eq:sigperp_defs} and shown to be positive in Lemma \ref{lem:sigmatperp},
\begin{align}
& P\Big(\Big \lvert \frac{1}{n}\norm{\fullq^{t+1}_{\perp}}^2 - (\sigma^{\perp}_{t+1})^2 \Big \lvert \geq \epsilon \Big)  \nonumber \\ 
& \quad \leq  t^6 K K'_{t-1}\exp\Big\{\frac{ -\kappa \kappa'_{t-1}  L \epsilon^2}{t^{10}(\log M)^{2t+1}}\Big\}, \label{eq:Hg} \\
& P\Big(\Big \lvert \frac{1}{n}\norm{\fullm^t_{\perp}}^2 - (\tau^{\perp}_t)^2 \Big \lvert \geq \epsilon \Big) \nonumber \\
 & \quad \leq   t^5 K K_{t-1} \exp\Big\{ \frac{-\kappa \kappa_{t-1} L \epsilon^2}{t^7 (\log M)^{2t-1}}\Big\}. \label{eq:Bg}
\end{align}


\item
\begin{align}
& P\Big(\frac{1}{L}\sum_{\ell = 1}^L \max_{j \in \ind(\ell)} \abs{h^{t+1}_{j}} \geq \tau_0 \sqrt{3\log M} + \e \Big) \nonumber \\
&\quad \leq t^4 K K'_{t-1}\exp\Big\{\frac{-\kappa \kappa'_{t-1}  L \epsilon^2}{ t^6 (\log M)^{2t}}\Big\}, \label{eq:Hh0} \\
& P\Big(\frac{1}{L}\sum_{\ell = 1}^L \max_{j \in \ind(\ell)} (h^{t+1}_{j})^2 \geq 6\tau_0^2 \log M + \e \Big) \nonumber \\
&\quad \leq t^4 K K'_{t-1}\exp\Big\{\frac{-\kappa \kappa'_{t-1}  L \epsilon}{t^6 (\log M)^{2t}}\Big\}.    \label{eq:Hh}
\end{align}

\end{enumerate}
\label{lem:main_lem}
\end{lem}

The lemma is proved in Section \ref{sec:lem1_proof}. 

\subsection{Comments on Lemma \ref{lem:main_lem}} \label{subsec:lem_comments}

The proof of Lemma \ref{lem:main_lem} is inductive: the concentration results for time step $t$ depend on the results at times $t= 0, 1, \ldots, t-1$. To prove Theorem \ref{thm:main_amp_perf}, the main result we need from Lemma \ref{lem:main_lem} is that for each $t \leq T$,  the squared error $\frac{1}{n} \| \fullq^t \|^2 = 
\frac{1}{n}\| \fullb^t - \fullb_0 \|^2$ concentrates on $\sigma_t^2$. This result is used in Section \ref{subsec:proof_thm1} below to prove Theorem \ref{thm:main_amp_perf}.  The concentration of $\frac{1}{n}\| \fullq^t \|^2$ is shown in part (c) of Lemma \ref{lem:main_lem}, but the proof of part (c) requires the other concentration results in the lemma to hold. For example, we prove part (c) by noting that $\fullq^t = \eta^{t-1}(\fullb_0 - \fullh^{t}) - \fullb_0$, appealing to Lemma \ref{lem:hb_cond} to find the conditional distribution of $\fullh^{t}$, and then using other parts of the induction to show that the terms in the conditional distribution of $\fullh^t$ concentrate.

While the concentration inequalities   in Lemma \ref{lem:main_lem} are broadly similar to those in \cite[Lemma 4.5]{RushV16}, there are a few important differences.  The first is that the denoising functions $\{\eta^t\}_{t \geq 0}$ (defined in \eqref{eq:eta_def}) act section-wise on their vector input due to the fact that the message vector $\fullb_0$ has a section-wise structure.  In contrast, the analysis in \cite{RushV16} considers only separable denoising functions that act component-wise on vector inputs. This is because it is assumed in \cite{RushV16} that the signal has i.i.d.\ entries.  However, the ``signal" (the message vector) considered here is only section-wise independent, with the section size $M$ approaching infinity in the large system limit. (In the large system limit $L, M, n$ all tend to infinity, with the constraint $L \log M = n R$.)  A related consequence is that the sampling ratio ($n/N$) of the measurement matrix \cite{RushV16} is assumed to be of constant order, while in the SPARC setting, $\frac{n}{N} = \frac{n}{ML} \rightarrow 0$ in the large system limit.

\emph{The concentration constants}: The dependence on $t$ of the constants $\kappa_t, \kappa'_t, K_t, K'_t$ in \eqref{eq:Ktkapt_defs} is determined by the induction used in the proof:  the concentration results for step $t$ depend on those corresponding to all the previous steps. The $t!$ terms in the constants arise due to quantities that can be expressed as a sum of $t$ terms with step indices $1,\ldots, t$, e.g., $\Delta_{t,t}$ and $\Delta_{t+1,t}$ in \eqref{eq:Dtt} and \eqref{eq:Dt1t}.  The concentration results for such quantities have $1/t$ and $t$ multiplying the exponent and pre-factor, respectively, in each step $t$ (see Lemma \ref{sums}), which results in the $t!$ terms in $K_t$ and $\kappa_t$. Similarly, the $(C_2)^t$ and $(c_2)^t$ terms in  $\kappa_t, K_t$  arise due to quantities  that are the \emph{product} of two terms, for each of which we have a concentration result available from the induction hypothesis. (see Lemma \ref{products}). 

Finally, the $\log M$ factor in the denominator of the exponent in each of the concentration inequalities is due to a dependence on the magnitude of the largest entry in each section of $\fullD_{t+1,t}$. (Each section has $M$ entries, and the maximum of $M$ i.i.d.\ standard Gaussians is close to $\sqrt{2 \log M}$.)

\subsection{Proof of  Theorem \ref{thm:main_amp_perf} } \label{subsec:proof_thm1}
The event that the section error rate exceeds $\e$ is
$ \{ \mc{E}_{sec}(\mc{S}_n)  > \e \} =  \Big\{ \sum_{\ell }  \mathbf{1} \{ \hat{\fullb}_{(\ell)} \neq \fullb_{0_{(\ell)}} \} > L \e \Big\}$. 
It is shown in \cite[Sec.\ V.E]{RushGVIT17} that
\be
\{ \mc{E}_{sec}(\mc{S}_n)  > \e \} \ \Rightarrow \  \Big\{ \frac{ \| \fullb^{T} - \fullb_0 \|^2 }{n}  \geq \frac{\e  \sigma^2 \mc{C}}{2} \Big\},
\label{eq:sec_error_exp}
\ee
where $\fullb^T$ is the AMP estimate at the termination step $T$. (Recall that the largest entry within each section of $\fullb^T$ is chosen to produce $\hat{\fullb}$.)

Now, from \eqref{eq:Hc} of Lemma \ref{lem:main_lem}(c), we know that for any $\tilde{\e} \in (0,1)$:
\begin{align}
& P\Big( \frac{\| \fullb^{T} - \fullb_0 \|^2}{n} \geq \sigma_{T}^2 + \tilde{\e} \Big) = P\Big( \frac{\| \fullq^{T} \|^2}{n} \geq \sigma_{T}^2 + \tilde{\e} \Big)  \nonumber \\
& \quad \leq  K_{T}\exp\Big\{ \frac{-\kappa_{T}L \tilde{\e}^2}{(\log M)^{2T-1}}\Big\}.
\label{eq:betaTst_conc}
\end{align}
From  the definition of $T$ and \eqref{eq:tauT_bound}, we have  $\sigma_{T}^2 = \tau_{T}^2 - \sigma^2 \leq 
P f_R(M)$. 
Hence, \eqref{eq:betaTst_conc} implies
\be
\begin{split}
& P\Big( \frac{\| \fullb^{T} - \fullb_0 \|^2}{n}  \geq P f_R(M) + \tilde{\e} \Big) \\
&  \leq P\Big(\frac{\| \fullb^{T} - \fullb_0 \|^2}{n} \geq \sigma_{T}^2 + \tilde{\e} \Big)  \\
&  \leq  K_{T} \exp\Big\{\frac{-\kappa_{T} L \tilde{\e}^2}{(\log M)^{2T-1}}\Big\}.
\end{split}
\label{eq:norm_bound}
\ee
Now take  $\tilde{\e} = \frac{\e  \sigma^2 \mc{C}}{2} - Pf_R(M)$, noting that this $\tilde{\e}$ is strictly positive whenever $\e > 2 \snr f_R(M)/\mc{C}$, the condition specified in the theorem statement.   Finally, combining \eqref{eq:sec_error_exp} and \eqref{eq:norm_bound} we obtain
\ben
\begin{split}
& P( \mc{E}_{sec}(\mc{S}_n)  >  \e) \\
 & \leq  K_{T} \exp\Big\{ \frac{-\kappa_{T} L}{(\log M)^{2T-1}}  \Big(\frac{\e  \sigma^2 \mc{C}}{2} - P f_R(M) \Big)^2\Big\}. 
\end{split}
\een
\qed 

\section{Proof of Lemma  \ref{lem:main_lem}} \label{sec:lem1_proof}

The proof proceeds by induction on $t$.  We label as $\mathcal{H}^{t+1}$ the results \eqref{eq:Ha}, \eqref{eq:Ha1}, \eqref{eq:Hb}, \eqref{eq:Hc}, \eqref{eq:Hd}, \eqref{eq:He}, \eqref{eq:Qsing}, \eqref{eq:Hf1}, \eqref{eq:Hf}, \eqref{eq:Hg}, \eqref{eq:Hh0}, and  \eqref{eq:Hh}. We similarly label as $\mathcal{B}^t$ the results \eqref{eq:Ba}, \eqref{eq:Bb}, \eqref{eq:Bc}, \eqref{eq:Bd}, \eqref{eq:Be}, \eqref{eq:Msing}, \eqref{eq:Bf1}, \eqref{eq:Bf}, \eqref{eq:Bg}.  The proof consists of four steps: (1) $\mathcal{B}_0$ holds, (2) $\mathcal{H}_1$ holds, (3) if $\mathcal{B}_r, \mathcal{H}_s$ hold for all $r < t $ and $s \leq t $, then $\mathcal{B}_t$ holds, and (4) if $\mathcal{B}_r, \mathcal{H}_s$ hold for all $r \leq t $ and $s \leq t$, then $\mathcal{H}_{t+1}$ holds.

Appendix \ref{app:conc_lemmas} lists a few basic concentration inequalities, and Appendix \ref{app:useful_lemmas} contains other lemmas that are used in the proof. 
To keep the notation compact, we use $K, K',\kappa$, and $\kappa'$ to denote generic positive universal
constants throughout the proof, with the values changing as the proof progresses.

\subsection{Step 1: Showing $\mathcal{B}_0$ holds} \label{subsub:step1}

\textbf{(a) [Eq.\ \eqref{eq:Ba} for $t=0$].} $\fullD_{0,0} = 0$ so there is nothing to prove.

\noindent \textbf{(b)  [Eq.\ \eqref{eq:Bb} for $t=0$].} We first show concentration of $(\fullbamp^0)^*\fullw/n$. From Lemma \ref{lem:hb_cond} and the distribution of the channel noise, we note that $\fullbamp^0 \stackrel{d}{=} \sigma_0 \vecZ'_0$ and $\fullw \overset{d}{=} \sigma \vecZ$, where $\vecZ'_0, \vecZ \in \mathbb{R}^n$ are independent standard Gaussian random vectors.  Then applying Lemma \ref{lem:max_abs_normals}, we obtain
\begin{align}
P\Big(\frac{1}{n} \Big \lvert  (\fullbamp^0)^*\fullw\Big \lvert \geq \epsilon \Big) &= P\Big(\frac{1}{n}\Big \lvert (\vecZ'_0)^*\vecZ \Big \lvert \geq \frac{\epsilon}{ \sigma \sigma_0} \Big) \nonumber \\
&  \leq 2\exp\Big\{\frac{-n \e^2}{3\sigma^2 \sigma_0^2}\Big\}.
\label{eq:Bf101}
\end{align}

To show concentration for $(\fullm^0)^*\fullw/n$ recall $\fullm^0 = \fullbamp^0 - \fullw$, and therefore $(\fullm^0)^*\fullw = (\fullbamp^0)^*\fullw- \norm{\fullw}^2 \overset{d}{=} (\fullbamp^0)^*\fullw - \sigma^2 \norm{\vecZ}^2$.  The result then follows by applying Lemma \ref{sums}, \eqref{eq:Bf101}, and Lemma \ref{lem:max_abs_normals}.

\noindent  \textbf{(c)  [Eq.\ \eqref{eq:Bc} for $t=0$].}  From Lemma \ref{lem:hb_cond}, it follows $\| \fullbamp^0 \|^2 \overset{d}{=} \sigma_0^2 \|{\vecZ'_0} \|^2$ and therefore by Lemma \ref{lem:max_abs_normals},
\begin{align*}
P\Big(\Big \lvert \frac{1}{n} \| \fullbamp^0 \|^2 - \sigma_0^2 \Big \lvert \geq \epsilon \Big) &= P\Big(\Big \lvert \frac{1}{n}\| \vecZ'_0 \|^2 - 1 \Big \lvert \geq \frac{\epsilon}{\sigma_0^2} \Big) \nonumber \\
& \leq 2\exp\Big\{-\frac{n \e^2}{8 \sigma_0^4}\Big\}.
\end{align*}

\noindent  \textbf{(d)  [Eq.\ \eqref{eq:Bd} for $t=0$].} Recall that $\fullm^0 = \fullbamp^0 - \fullw$.
The result follows from Lemma \ref{sums}, $\mc{B}_0(b)$ and $\mc{B}_0(c)$. 

\noindent  \textbf{(e)  [Eq.\ \eqref{eq:Be} for $t=0$].} Since  $\fullw \overset{d}{=} \sigma \vecZ$, where $\vecZ \in \mathbb{R}^n$ is standard Gaussian, we have
\ben
\begin{split}
\|{\fullm^0} \|^2 = \norm{\fullbamp^0 - \fullw}^2 & = \| \fullbamp^0 \|^2 + \|\fullw \|^2 - 2 (\fullbamp^0)^*\fullw \\
&  \overset{d}{=} \| \fullbamp^0 \|^2 + \sigma^2 \| \vecZ \|^2 - 2(\fullbamp^0)^*\fullw.
\end{split}
\een
Using the above,
\begin{align*}
&P\Big(\Big \lvert \frac{\| \fullm^0 \|^2}{n} - \tau_0^2 \Big \lvert \geq \epsilon \Big) \\
&  = P\Big(\Big \lvert \frac{\| \fullbamp^0 \|^2}{n} + \frac{\sigma^2 \| \vecZ \|^2}{n} - \frac{2(\fullbamp^0)^*\fullw}{n} - (\sigma_0^2 + \sigma^2) \Big \lvert \geq \epsilon \Big) \\
&\overset{(a)}{\leq} P\Big(\Big \lvert \frac{\| \fullbamp^0 \|^2}{n} - \sigma_0^2 \Big \lvert \geq \frac{\epsilon}{3} \Big) + P\Big(\Big \lvert \frac{\norm{\vecZ}^2}{n} - 1 \Big \lvert \geq \frac{\epsilon}{3\sigma^2} \Big) \\
& \quad + P\Big(\Big \lvert \frac{(\fullbamp^0)^*\fullw}{n} \Big \lvert \geq \frac{\epsilon}{6} \Big) \\
&\overset{(b)}{\leq} 2\exp\Big\{\frac{-n \e^2}{72 \sigma_0^4}\Big\} + 2\exp\Big\{\frac{-n \e^2}{72 \sigma^4}\Big\} + 2\exp\Big\{\frac{-n \e^2}{108 \sigma^2 \sigma_0^2}\Big\}.
\end{align*}
Step (a) follows from Lemma \ref{sums}, and step (b) from $\mc{B}(b)$, $\mc{B}(c)$, and Lemma \ref{lem:max_abs_normals}.

\noindent  \textbf{(f)  [Eqs.\ \eqref{eq:Msing}, \eqref{eq:Bf1}, and \eqref{eq:Bf} for $t=0$].} There is nothing to prove here.

\noindent  \textbf{(g) [Eq.\ \eqref{eq:Hg} for $t=0$].} Since $\fullm^0_{\perp} = \fullm^0$ this result is the same as $\mc{B}(e)$.


\subsection{Step 2: Showing $\mathcal{H}_1$ holds}   \label{subsub:step2}

\textbf{(a)  [Eqs.\ \eqref{eq:Ha} - \eqref{eq:Ha1} for $t=0$].} Eq.\ \eqref{eq:Ha} follows from the Cauchy-Schwarz inequality.  We now prove \eqref{eq:Ha1}. From the definition of $\fullD_{1,0}$ given in Lemma \ref{lem:hb_cond} \eqref{eq:D10}, and noting 
that $\| \fullq^0 \|^2 = nP$,   we can write
\begin{align*}
\fullD_{1,0} &= \vecZ_0 \Big(\frac{\norm{\fullm^0}}{\sqrt{n}}  - \tau_0\Big) -  \frac{\norm{\fullm^0}}{\sqrt{n}}   \frac{\fullq^0}{\sqrt{nP}}  Z  \\
& \   + \frac{\fullq^0}{P} \Big(\frac{(\fullbamp^0)^*\fullm^0}{n} - P\Big), \label{eq:D10}
\end{align*}
where we have used the fact that $ \proj_{\fullq^0} \vecZ_0 \overset{d}{=}   \frac{\fullq^0}{\sqrt{nP}} Z$ where $Z \sim \mc{N}(0,1)$ by Lemma \ref{lem:gauss_p0}.  
 Consider a single element $j \in [N]$ of $\fullD_{1,0}$.  Using Lemma \ref{lem:squaredsums} and the bound $q^0_{j} \leq \sqrt{nP_{\ind(\sect(j))}}$, it follows
\be
\begin{split}
([\Delta_{1,0}]_{j})^2 & \leq 3\Big \lvert [Z_{0}]_j \Big \lvert^2 \cdot \Big \lvert \frac{\norm{\fullm^0}}{\sqrt{n}} - \tau_0 \Big \lvert^2  \\
& + 3\frac{nP_{\ind(\sect(j))}}{P^2} \cdot \Big \lvert \frac{(\fullbamp^0)^* \fullm^0}{n} - P\Big \lvert^2 \\
&  + 3\frac{\norm{\fullm^0}^2}{n} \cdot  \frac{nP_{\ind(\sect(j))}}{P} \cdot \frac{Z^2}{n}.
\end{split}
\label{eq:Del10j_bnd}
\ee
Using \eqref{eq:Del10j_bnd}, we  have the following bound:
\begin{align}
& P \Big(\frac{1}{L} \sum_{\ell=1}^L \max_{j \in \ind(\ell)} ([\Delta_{1,0}]_{j})^2 \geq \epsilon \Big)  \nonumber \\
&   \overset{(a)}{\leq} P \Big( \Big \lvert \frac{\norm{\fullm^0}}{\sqrt{n}} - \tau_0 \Big \lvert^2 \frac{1}{L} \sum_{\ell=1}^L \max_{j \in \ind(\ell)} ([Z_{0}]_j)^2 \geq \frac{\epsilon}{9} \Big) \nonumber  \\
& \quad + P \Big(c \log M\Big \lvert \frac{(\fullbamp^0)^* \fullm^0}{n} - P\Big \lvert^2 \geq \frac{\epsilon}{9} \Big) \nonumber  \\
& \quad + P \Big(c \log M \cdot \frac{\norm{\fullm^0}^2}{n} \cdot \frac{Z^2}{n} \geq \frac{\epsilon}{9} \Big),
\label{eq:sum_del2_bnd1}
\end{align}
where the inequality follows from Lemma \ref{sums}, and the fact that $nP_{\ell}/P \leq c \log M$ for all $\ell \in [L]$, where  $c>0$ is an absolute constant.  Label the terms on the RHS of \eqref{eq:sum_del2_bnd1} as $T_1, T_2, T_3$.  To complete the proof, we show that each term is upper bounded by  $K'_0\exp\{- \kappa'_0 L \e\}$.  First,
\begin{align*}
T_1 &\leq P \Big( \Big \lvert \frac{\norm{\fullm^0}}{\sqrt{n}} - \tau_0 \Big \lvert \geq \frac{\sqrt{\epsilon}}{3\sqrt{3 \log M}} \Big) \\
& \quad + P \Big(\frac{1}{L} \sum_{\ell=1}^L \max_{j \in \ind(\ell)} \abs{[Z_{0}]_j}^2 \geq 3 \log M \Big) \\
& \  \leq K'_0e^{- \kappa'_0 L \e} + e^{-\kappa L \log M}.
\end{align*}
The above follows from Lemma \ref{sqroots}, result $\mc{B}_0(e)$, and Lemma \ref{lem:max_abs_normals}.
Next, the second term $T_2$ in \eqref{eq:sum_del2_bnd1} has the desired bound from $\mathcal{B}_0 (d)$.
Finally, for the last term $T_3$  we have
\begin{align*}
&T_3 \leq  P \Big(\Big(\Big \lvert \frac{\norm{\fullm^0}^2}{n} - \tau_0^2 \Big \lvert + \tau_0^2\Big) \frac{Z^2}{n} \geq \frac{\epsilon}{9c \log M} \Big) \\
&\leq P \Big(\Big \lvert \frac{\norm{\fullm^0}^2}{n} - \tau^2_0 \Big \lvert \geq \sqrt{\epsilon} \Big) \\ 
& \quad + P \Big(\frac{\abs{Z}}{\sqrt{n}} \geq \frac{\sqrt{\epsilon}}{3\sqrt{c \log M}} \min \Big\{1, \frac{1}{\abs{\tau_0}}\Big\} \Big) \\
& \overset{(b)}{\leq} K'_0 e^{- \kappa'_0 L \e} + 2e^{-\kappa L \e }. \label{eq:Hb15}
\end{align*}
Step $(b)$ follows from result $\mathcal{B}_0 (e)$, and Lemma \ref{lem:normalconc}.  This completes the proof of \eqref{eq:Ha1}.

\noindent  \textbf{(b)  [Eq.\ \eqref{eq:Hb} for $t=0$].} Using the conditional distribution of $\fullh^1$ stated in Lemma \ref{lem:hb_cond} and Lemma \ref{sums}, we have
\begin{align*}
& P\Big(\Big \lvert \frac{(\fullh^1)^*\fullq^0}{n} \Big \lvert \geq \epsilon \Big) = P\Big(\Big \lvert \frac{\tau_0 \vecZ_0^*\fullq^0}{n} + \frac{\fullD_{1,0}^*\fullq^0}{n} \Big \lvert \geq \epsilon \Big) \\
&  \leq P\Big(\Big \lvert \frac{\vecZ_0^*\fullq^0}{n} \Big \lvert \geq \frac{\epsilon}{2\tau_0} \Big) + P\Big(\Big \lvert\frac{\fullD_{1,0}^*\fullq^0}{n} \Big \lvert \geq \frac{\epsilon}{2} \Big).
\end{align*}
Label the two terms on the right side of the above as $T_1$ and $T_2$.  
To complete the proof we will show each term is upper bounded by $ K'_0 \exp\{- \kappa'_0 L \e^2\}.$  
Since $\fullq^0$ is independent of $\vecZ_0$, we have $\vecZ_0^*\fullq^0 \stackrel{d}{=} \| \fullq^0 \| Z$, where $Z$ is a $\mc{N}(0,1)$ random variable. Therefore, 
\begin{align*}
T_1 = P\Big(\frac{\norm{\fullq^0}}{\sqrt{n}}\frac{\abs{Z}}{\sqrt{n}} \geq \frac{\epsilon}{2 \tau_0} \Big) &  = P\Big( \abs{Z} \geq \frac{\epsilon \sqrt{n}}{2 \tau_0 \sqrt{P}} \Big) \\
& \leq 2\exp\Big\{-\frac{n \e^2}{8 \tau_0^2 P}\Big\}.
\end{align*}
where the last inequality follows from Lemma \ref{lem:normalconc}.  Finally,
\begin{align*}
T_2 &= P\Big(\Big \lvert\frac{1}{n} \sum_{\ell = 1}^{L} [\fullD_{1,0}]_{(\ell)} \fullq^0_{(\ell)} \Big \lvert \geq \frac{\epsilon}{2} \Big) \\
&  \leq P\Big(\frac{1}{n}\sum_{\ell = 1}^L \sqrt{nP_{\ell}} \max_{j \in \ind(\ell)} \abs{[\Delta_{1,0}]_{j}} \geq \frac{\epsilon}{2} \Big) \\
& \overset{(a)}{\leq} P\Big(\frac{1}{L}\sum_{\ell = 1}^L \max_{j \in \ind(\ell)} \abs{[\Delta_{1,0}]_{j}} \geq \frac{\epsilon \sqrt{\log M}}{2Rc} \Big) %
\\
& \overset{(b)}{\leq} K'_0\exp\{- \kappa'_0 L \min\{ \e^2 \log M, 1\}\}.
\end{align*}
Step $(a)$ follows from the fact that for all $\ell \in [L]$, $\sqrt{nP_{\ell}} \leq c\sqrt{\log M}$ for some constant $c>0$ and the fact that $nR = L \log M$ and step $(b)$ from $\mathcal{H}_1 (a)$.

\noindent  \textbf{(c)  [Eq.\ \eqref{eq:Hc} for $t=0$].} We begin by showing the result for $\norm{\fullq^1}^2/n$ .  Recalling that  $\fullq^1 = \eta^0(\fullb_0 - \fullh^1) - \fullb_0$, and using the conditional distribution of $\fullh^1$ stated in Lemma \ref{lem:hb_cond}, we write
\begin{align*}
& P\Big( \Big \lvert \frac{1}{n}\norm{\fullq^1}^2 - \sigma_1^2\Big \lvert \geq \epsilon \Big) \\
 &= P\Big( \Big \lvert \frac{1}{n}\norm{\eta^0(\fullb_0 - \tau_0 \vecZ_0 - \fullD_{1,0}) - \fullb_0}^2 - \sigma_1^2\Big \lvert \geq \epsilon \Big)\\
& \leq P\Big(\Big \lvert \frac{1}{n}\norm{\eta^0(\fullb_0 - \tau_0 \vecZ_0) - \fullb_0}^2 - \sigma_1^2\Big \lvert \geq \frac{\epsilon}{2} \Big) \nonumber \\
&\ + P\Big(\frac{1}{n} \Big \lvert \norm{\eta^0(\fullb_0 - \tau_0 \vecZ_0 - \fullD_{1,0}) - \fullb_0}^2  \\
& \hspace{0.6in}  - \norm{\eta^0(\fullb_0 - \tau_0 \vecZ_0) - \fullb_0}^2 \Big \lvert \geq \frac{\epsilon}{2} \Big).
\end{align*}
Label the two terms on the RHS as $T_1$ and $T_2$.  We will show that each of these is upper bounded by $K'_0 \exp\{\frac{- \kappa'_0 L \e^2}{\log M}\}$.   The bound for $T_1$ is obtained using Hoeffding's inequality (Lemma \ref{lem:hoeff_lem}). To verify the conditions to apply Hoeffding's inequality,  first write  
\begin{align*} 
& \frac{1}{n} \norm{\eta^0(\fullb_0 - \tau_0 \vecZ_0) - \fullb_0}^2  \\
&= \frac{1}{L} \sum_{\ell = 1}^L  \frac{ 1}{\log M}\Big(R \, \norm{\eta^0_{(\ell)}(\fullb_0 - \tau_0 \vecZ_0) - \fullb_{0_{(\ell)}} }^2\Big),
\end{align*}
and note that for each $\ell \in [L]$:
\ben
\begin{split}
0  & \leq \frac{R}{\log M}  \norm{\eta_{(\ell)}^0(\fullb_0 - \tau_0 \vecZ_0) - \fullb_{0_{(\ell)}} }^2 \\
& \leq \frac{2R }{\log M} \Big(\norm{\eta_{(\ell)}^0(\fullb_0 - \tau_0 \vecZ_0)}^2 + \norm{\fullb_{0_{(\ell)}}}^2\Big) \\
& \leq 4R \Big(\frac{nP_{\ell}}{\log M}\Big) \leq c,
\end{split}
\een
for some absolute constant $c > 0$.  Next, the expectation given by Hoeffding's inequality can be written as 
\begin{align*} 
\mathbb{E}_{\vecZ_0}\norm{\eta^0(\fullb_0 - \tau_0 \vecZ_0) - \fullb_0}^2 & = \mathbb{E}_{\fullb, \vecZ_0}\norm{\eta^0(\fullb - \tau_0 \vecZ_0) - \fullb}^2 \\
& = n\sigma_1^2
\end{align*}
where the first equality is true for each $\fullb_0 \in \mathcal{B}_{M,L}$ because of the uniform distribution of the non-zero entry in each section of $\fullb$ over the $M$ possible locations and the i.i.d.\ distribution of $\vecZ_0$. The second equality follows from Lemma \ref{lem:expect_etar_etas}.  

Next, we bound term $T_2$.  First, write
\begin{align*}
&\norm{\eta^0(\fullb_0 - \tau_0 \vecZ_0 - \fullD_{1,0}) - \fullb_0}^2  \hspace{-1pt} -  \hspace{-1pt} \norm{\eta^0(\fullb_0 - \tau_0 \vecZ_0) - \fullb_0}^2 \\
&= [\eta^0(\fullb_0 - \tau_0 \vecZ_0 - \fullD_{1,0}) - \fullb_0]^* \\
& \qquad \  [\eta^0(\fullb_0 - \tau_0 \vecZ_0 - \fullD_{1,0}) - \eta^0(\fullb_0 - \tau_0 \vecZ_0)] \\
& \  +  [\eta^0(\fullb_0 - \tau_0 \vecZ_0) - \fullb_0]^* \\ 
& \qquad \ [\eta^0(\fullb_0 - \tau_0 \vecZ_0 - \fullD_{1,0}) - \eta^0(\fullb_0 - \tau_0 \vecZ_0)].
\end{align*}
Now using the above and Lemma \ref{sums}, 
\[ T_2 \leq T_{2,a} + T_{2,b} \]
where 
\begin{align*} 
& T_{2,a} \\
&  = P\Big(\frac{1}{n} \Big \lvert [\eta^0(\fullb_0 - \tau_0 \vecZ_0 - \fullD_{1,0}) - \fullb_0]^* \\ 
& \qquad \qquad [\eta^0(\fullb_0 - \tau_0 \vecZ_0 - \fullD_{1,0}) - \eta^0(\fullb_0 - \tau_0 \vecZ_0)] \Big \lvert \geq \frac{\epsilon}{4} \Big) 
\end{align*}
and 
\begin{align*}
&T_{2,b}  \\
& = P\Big( \frac{1}{n}\Big \lvert[\eta^0(\fullb_0 - \tau_0 \vecZ_0) - \fullb_0]^* \\ 
& \qquad \qquad  [\eta^0(\fullb_0 - \tau_0 \vecZ_0 - \fullD_{1,0}) - \eta^0(\fullb_0 - \tau_0 \vecZ_0)] \Big \lvert \geq \frac{\epsilon}{4} \Big).
\end{align*}
 Then,
\ben \begin{split}
& T_{2,a} = P\Big(\frac{1}{n} \sum_{\ell = 1}^L \Big \lvert [\eta^0_{(\ell)}(\fullb_0 - \tau_0 \vecZ_0 - \fullD_{1,0}) - \fullb_{0_{(\ell)}}]^* \\
& \qquad \quad  [\eta^0_{(\ell)}(\fullb_0 - \tau_0 \vecZ_0 - \fullD_{1,0}) - \eta^0_{(\ell)}(\fullb_0 - \tau_0 \vecZ_0)]  \Big \lvert \geq \frac{\epsilon}{4} \Big) \\ 
&\leq P\Big(\frac{1}{n}  \sum_{\ell = 1}^L 2\sqrt{nP_{\ell}} \sum_{i \in \ind(\ell)} \Big \lvert \eta^0_{i}(\fullb_0 - \tau_0 \vecZ_0 - \fullD_{1,0})   \\
& \hspace{1.5in} -\eta^0_{i}(\fullb_0 - \tau_0 \vecZ_0) \Big \lvert \geq \frac{\epsilon}{4} \Big) \\
&\overset{(a)}{\leq} P\Big(\frac{1}{L}\sum_{\ell = 1}^L \max_{j \in \ind(\ell)} \abs{[\Delta_{1,0}]_{j}} \geq \frac{\epsilon \tau_0^2}{16Rc\sqrt{\log M}} \Big) \\
&  \overset{(b)}{\leq} K'_0\exp\Big\{- \frac{ \kappa'_0 L \e^2}{\log M}\Big\}.
\end{split} \een
Step $(a)$ follows from Lemma \ref{lem:BC9} applied to each section using $(nP_{\ell})^{3/2} \leq c (\log M)^{3/2}$ for $\ell \in [L]$,  for some constant $c>0$.  Step $(b)$ follows from $\mathcal{H}_1 (a)$.  Term $T_{2,b}$ has the same upper bound which can be shown as above using Lemma \ref{lem:BC9}. This proves the result for $\norm{\fullq^1}^2/n$.

Proving the  concentration result for $(\fullq^0)^*\fullq^1/n$ is similar: we use Lemma \ref{sums} followed by Hoeffding's inequality and Lemma \ref{lem:BC9}.

\noindent  \textbf{(d)  [Eq.\ \eqref{eq:Hd} for $t=0$].}  Recalling that $\fullq^1 = \eta^0(\fullb_0 - \fullh^1) - \fullb_0$, we write
\ben
\begin{split}
(\fullh^1)^* \fullq^1 &  = (\fullh^1)^* [\eta^0(\fullb_0 - \fullh^1) - \fullb_0]  \\
& = (\fullh^1)^* \eta^0(\fullb_0 - \fullh^1) + (\fullh^1)^*\fullq^0.
\end{split}
\een
Using the above,
\begin{align}
& P\Big(\Big \lvert \frac{(\fullh^1)^*\fullq^1}{n} + \sigma_1^2 \Big \lvert \geq \epsilon \Big) \nonumber  \\
 &= P\Big(\Big \lvert \frac{(\fullh^1)^* \eta^0(\fullb_0 - \fullh^1)}{n} + \frac{(\fullh^1)^*\fullq^0}{n} + \sigma_1^2 \Big \lvert \geq \epsilon \Big)  \nonumber \\
 &\leq   P\Big(\Big \lvert \frac{(\fullh^1)^* \eta^0(\fullb_0 - \fullh^1)}{n} + \sigma_1^2 \Big \lvert \geq \frac{\e}{2} \Big)  \nonumber \\
 & \qquad + P\Big(\Big \lvert \frac{(\fullh^1)^*\fullq^0}{n} \Big \lvert \geq \frac{\e}{2} \Big). \label{eq:h1q1_split}
\end{align}
By $\mathcal{H}_1 (b)$, the second term in \eqref{eq:h1q1_split} is  bounded by $K'_0 \exp\{-\kappa'_0 n \e^2\}$.  To bound the first term,  using the conditional distribution of $\fullh^1$ stated in Lemma \ref{lem:hb_cond} we write
\begin{align}
&P\Big( \Big \lvert  \frac{1}{n} (\fullh^1)^* \eta^0(\fullb_0 - \fullh^1) + \sigma_1^2 \Big \lvert \geq \frac{\epsilon}{2} \Big)  \nonumber \\
& = P\Big( \Big \lvert  \frac{1}{n} (\tau_0 \vecZ_0 + \fullD_{1,0})^* \eta^0(\fullb_0 - \tau_0 \vecZ_0 - \fullD_{1,0}) + \sigma_1^2 \Big \lvert \geq \frac{\epsilon}{2} \Big) \nonumber  \\ 
& \leq P\Big( \frac{1}{n}\Big \lvert \fullD_{1,0}^* \eta^0(\fullb_0 - \tau_0 \vecZ_0 - \fullD_{1,0}) \Big \lvert \geq \frac{\epsilon}{6} \Big) \nonumber \\
& \ + P\Big( \Big \lvert  \frac{1}{n}  \tau_0 \vecZ_0^* \eta^0(\fullb_0 - \tau_0 \vecZ_0) + \sigma_1^2 \Big \lvert \geq \frac{\epsilon}{6} \Big) \nonumber \\
&\ + P\Big(  \frac{1}{n} \Big \lvert \tau_0 \vecZ_0^* [\eta^0(\fullb_0 - \tau_0 \vecZ_0 - \fullD_{1,0}) \nonumber \\ 
& \qquad \qquad - \eta^0(\fullb_0 - \tau_0 \vecZ_0)] \Big \lvert \geq \frac{\epsilon}{6} \Big). \label{eq:h1_eta0}
\end{align}
 Label the terms of the above as $T_1 , T_2, T_3$.  To complete the proof, we will show that  each term is bounded by $K'_0 \exp\{- \kappa'_0 L \e^2/\log M\}$.   We begin by  bounding $T_1$.  
\begin{align*}
T_{1} &= P\Big(\frac{1}{n} \sum_{\ell=1}^L \Big \lvert ([\fullD_{1,0}]_{(\ell)})^*\eta^0_{(\ell)}(\fullb_0 - \tau_0 \vecZ_0 - \fullD_{1,0})\Big \lvert \geq \frac{\epsilon}{6} \Big) \\
&  \leq P\Big(\frac{1}{n}\sum_{\ell=1}^L\sqrt{nP_{\ell}} \max_{j \in \ind(\ell)} \abs{[\Delta_{1,0}]_{j}} \geq \frac{\epsilon}{6} \Big) \\
&\overset{(i)}{\leq} P\Big(\frac{1}{L}\sum_{\ell = 1}^L \max_{j \in \ind(\ell)} \abs{[\Delta_{1,0}]_{j}} \geq \frac{\epsilon \sqrt{\log M}}{6Rc} \Big) \\
& \overset{(j)}{\leq}K'_0\exp\{- \kappa'_0 L \min\{ \e^2 \log M, 1\}\}.
\end{align*}
Step $(i)$ follows from the fact that $\sqrt{nP_{\ell}} \leq c \sqrt{\log M}$ for some constant $c>0$ and all $\ell \in [L]$, and step $(j)$ from $\mathcal{H}_1 (a)$. 

Next consider $T_2$.  Because of the uniform distribution of the non-zero entry in each section of $\fullb$ over the $M$ possible locations and the i.i.d.\ distribution of $\vecZ_0$,  for any $\fullb_0 \in \mathcal{B}_{M,L}$, we have $\mathbb{E}_{\vecZ_0}\{\tau_0 \vecZ_0^*\eta^0(\fullb_0 -  \tau_0 \vecZ_0)\}  = \mathbb{E}_{\vecZ_0, \fullb}\{\tau_0 \vecZ_0^*\eta^0(\fullb -  \tau_0 \vecZ_0)\} $.  This expectation can then computed as 
\be
\begin{split}
\frac{1}{n}\mathbb{E}\{\tau_0 \vecZ_0^*\eta^0(\fullb -  \tau_0 \vecZ_0)\} & \stackrel{(a)}{=} 
\frac{1}{n}\mathbb{E}\norm{\eta^0(\fullb -  \tau_0 \vecZ_0)}^2 - P \\
&  \stackrel{(b)}{=}  - P(1 - x_1) = -\sigma_1^2,
\end{split}
\label{eq:Hd12a}
\ee
where equality $(a)$ is obtained using Stein's lemma, Lemma \ref{lem:stein} (see  \cite[p.1491, Eqs.\ (102) -- (104)]{RushGVIT17} for details). The equality $(b)$ follows from Lemma \ref{lem:expect_etar_etas}. Now, from Lemma \ref{lem:Hd_convergence} and \eqref{eq:Hd12a}, we have
\ben
\begin{split}
& T_2  = \\
& P\Big( \frac{1}{n}\Big \lvert \tau_0 \vecZ_0^* \eta^0(\fullb_0 - \tau_0 \vecZ_0) - \mathbb{E}\{\tau_0 \vecZ_0^*\eta^0(\fullb -  \tau_0 \vecZ_0)\}  \Big \lvert \geq \frac{\epsilon}{6} \Big) \\
&  \leq \exp\{-\kappa L \e^2\}.
\end{split}
\een

Finally consider the term $T_3$.
\begin{align*}
T_{3} &= P\Big(\frac{1}{n}\sum_{\ell = 1}^L \Big \lvert ([\vecZ_{0}]_{(\ell)})^*\Big[\eta_{(\ell)}^0(\fullb_0 - \tau_0 \vecZ_0 - \fullD_{1,0})  \\
& \hspace{1in} - \eta_{(\ell)}^0(\fullb_0 - \tau_0 \vecZ_0)\Big] \Big \lvert \geq \frac{\epsilon}{6\tau_0} \Big) \\
&\leq P\Big(\frac{1}{n} \sum_{\ell = 1}^L \max_{k \in \ind(\ell)} \abs{[Z_{0}]_k} \sum_{i \in \ind(\ell)} \abs{\eta_{i}^0(\fullb_0 - \tau_0 \vecZ_0 - \fullD_{1,0}) \\
&\hspace{1in} - \eta_{i}^0(\fullb_0 - \tau_0 \vecZ_0)} \geq \frac{\epsilon}{6\tau_0} \Big) \\
&\overset{(e)}{\leq} P\Big(\frac{1}{L}\sum_{\ell = 1}^L \max_{k \in \ind(\ell)} \abs{[Z_{0}]_k} \max_{j \in \ind(\ell)} \abs{[\Delta_{1,0}]_{j}} \geq \frac{\epsilon \tau_0}{12Rc} \Big) \\
&  \overset{(f)}{\leq}  K'_0 \exp\Big\{\frac{-\kappa'_0 L\e^2}{\log M}\Big\}.
\end{align*}
Step $(e)$ follows from Lemma \ref{lem:BC9} applied to each section and the fact that $nP_{\ell} \leq c \log M$ for some constant $c>0$ and all $\ell \in [L]$. Step $(f)$  is obtained as follows.
\begin{align*}
&P \Big(\frac{1}{L} \sum_{\ell = 1}^L \max_{j \in \ind(\ell)} \abs{[\Delta_{1,0}]_{j}} \max_{k \in \ind(\ell)} \abs{[Z_{0}]_k} \geq \e\Big) \\
&  \overset{(g)}{\leq} P \Big(\frac{1}{L} \sum_{\ell = 1}^L \max_{k \in \ind(\ell)} ([Z_{0}]_j)^2 \,  \frac{1}{L} \sum_{\ell = 1}^L \max_{j \in \ind(\ell)} ([\Delta_{1,0}]_{j})^2  \geq \e^2 \Big) \\
&\leq P \Big(\frac{1}{L} \sum_{\ell = 1}^L \max_{j \in \ind(\ell)} ([Z_{0}]_j)^2  \geq 3 \log M \Big)  \\
& \quad + P \Big( \frac{1}{L} \sum_{\ell = 1}^L \max_{j \in \ind(\ell)} ([\Delta_{1,0}]_{j})^2  \geq \frac{\e^2}{3 \log M} \Big)  \\
&\overset{(h)}{\leq} e^{-\kappa L \log M} + K'_0 e^{-\frac{\kappa'_0 L \e^2}{\log M}}. 
\end{align*} 
Step $(g)$ follows from Cauchy-Schwarz and step $(h)$  from Lemma \ref{lem:max_abs_normals} and $\mathcal{H}_1 (a)$.  Using the bounds for $T_1, T_2, T_3$, the three terms in \eqref{eq:h1_eta0}, completes the proof.

\noindent  \textbf{(e)  [Eq.\ \eqref{eq:He} for $t=0$].} By definition, $$\lambda_1 = -\frac{1}{\tau_0^2}\Big(P - \frac{\norm{\fullb^1}^2}{n}\Big),$$
and so it follows that:
\be
P\Big(\Big \lvert \lambda_1 + \frac{\sigma_1^2}{\tau_0^2} \Big \lvert \geq \epsilon \Big) = P \Big(\Big \lvert \Big(\frac{1}{n}\norm{\fullb^1}^2 - P\Big) + \sigma_1^2 \Big \lvert \geq \e \tau_0^2 \Big).
\label{eq:Hj11}
\ee
Note that, $\norm{\fullq^1}^2 - 2 (\fullq^0)^* \fullq^1 =  \norm{\fullb^1 - \fullb_0}^2 + 2\fullb_0^*(\fullb^1 - \fullb_0) = \norm{\fullb^1}^2 - nP.$  Using this in \eqref{eq:Hj11},
\begin{align*}
&P\Big(\Big \lvert \lambda_1 + \frac{\sigma_1^2}{\tau_0^2} \Big \lvert \geq \epsilon \Big) \\
&  = P \Big(\Big \lvert \Big(\frac{\norm{\fullq^1}^2}{n} - 2 \frac{(\fullq^0)^* \fullq^1}{n} \Big) - (\sigma_1^2 - 2 \sigma_1^2) \Big \lvert \geq \epsilon \tau_0^2 \Big) \\
&  \leq  P \Big(\Big \lvert \frac{\norm{\fullq^1}^2}{n} - \sigma_1^2 \Big \lvert \geq \frac{\epsilon \tau_0^2}{2} \Big) + P \Big(\Big \lvert \frac{(\fullq^0)^* \fullq^1}{n} - \sigma_1^2 \Big \lvert \geq \frac{\epsilon \tau_0^2}{4} \Big) \\
&  \leq   K'_0 \exp\Big\{ \frac{- \kappa'_0 L \epsilon^2}{\log M}\Big\},
\end{align*}
where the last inequality follows from $\mathcal{H}_1 (c)$.

\noindent  \textbf{(f)  [Eqs.\ \eqref{eq:Qsing}, \eqref{eq:Hf1}, and \eqref{eq:Hf} for $t=0$].} Note that $\textbf{Q}_1 = \frac{1}{n} \norm{\fullq^0}^2 = P = \sigma_0^2 = \tilde{\fullC}^1$ so the inverse concentration in \eqref{eq:Hf1} is trivially true.  Recall that $\gamma^1_0 = \frac{(\fullq^0)^*\fullq^1}{nP}$. Then using $\mathcal{H}_1 (c)$ we have
\ben \begin{split}
& P\Big(\Big \lvert \gamma^1_0 - \frac{\sigma_1^2}{\sigma_0^2}\Big \lvert \geq \epsilon \Big) = P\Big(\Big \lvert  \frac{(\fullq^0)^*\fullq^1}{nP} - \frac{\sigma_1^2}{\sigma_0^2}\Big \lvert \geq \epsilon \Big) \\
& = P\Big(\Big \lvert \frac{(\fullq^0)^*\fullq^1}{n} - \sigma_1^2 \Big \lvert \geq \epsilon P \Big) 
  \leq  K'_0 \exp\Big\{\frac{- \kappa'_0 L \e^2}{\log M}\Big\}.
\end{split} \een

\noindent  \textbf{(g)  [Eq.\ \eqref{eq:Hg} for $t=0$].}  By definition, $\norm{\fullq^1_{\perp}}^2 = \norm{\fullq^1}^2 - (\gamma^1_0)^2 \norm{\fullq^0}^2 = \norm{\fullq^1}^2 - nP (\gamma^1_0)^2.$  Using this and the fact that $(\sigma_1^{\perp})^2 = \sigma_1^2(1 - (\sigma_1^2/\sigma_0^2))$, we find the following upper bound.
\begin{align*}
P&\Big(\Big \lvert \frac{\norm{\fullq^1_{\perp}}^2}{n} - (\sigma_1^{\perp})^2 \Big \lvert \geq \epsilon\Big) \\
&  = P\Big(\Big \lvert \frac{\norm{\fullq^1}^2}{n} - P (\gamma^1_0)^2 -  \sigma_1^2\Big(1 - \frac{\sigma_1^2}{\sigma_0^2}\Big) \Big \lvert \geq \epsilon\Big)\\
&\overset{(a)}{\leq} P\Big(\Big \lvert \frac{\norm{\fullq^1}^2}{n} - \sigma_1^2 \Big \lvert \geq \frac{\epsilon}{2} \Big) + P\Big(\Big \lvert P(\gamma_0^1)^2 - \frac{ \sigma_1^4}{\sigma_0^2}\Big \lvert \geq \frac{\epsilon}{2}\Big) \\
&  \overset{(b)}{\leq}  K'_0 \exp\Big\{ \frac{- \kappa'_0 L \epsilon^2}{\log M}\Big\}.
\end{align*}
Step $(a)$ follows from Lemma \ref{sums} and step $(b)$ from $\mathcal{H}_1 (c)$ and $\mathcal{H}_1 (f)$ along with Lemma \ref{powers}.

\noindent  \textbf{(h)  [Eqs.\ \eqref{eq:Hh0}- \eqref{eq:Hh} for $t=0$].}  For any element $i \in [N]$, using the conditional distribution of $\fullh^1$ stated in Lemma \ref{lem:hb_cond} and the Triangle Inequality, it follows $\abs{h^1_{i}} \leq \tau_0 \abs{[Z_{0}]_i} + \abs{[\Delta_{1,0}]_{i}}.$  We then have the following upper bound:
\begin{align*}
& P\Big(\frac{1}{L} \sum_{\ell = 1}^L \max_{j \in \ind(\ell)} \abs{h^{1}_{j}} \geq  \tau_0 \sqrt{3 \log M} + \e \Big) \\
&\leq P\Big( \frac{1}{L} \sum_{\ell = 1}^L \Big[\tau_0 \max_{j \in \ind(\ell)} \abs{ [Z_{0}]_j} +  \max_{j \in \ind(\ell)} \abs{[\Delta_{1,0}]_{j}} \Big]   \\
& \hspace{1.5in}  \geq  \tau_0 \sqrt{3 \log M} + \e \Big) \\
& \leq P\Big(\frac{1}{L} \sum_{\ell = 1}^L \max_{j \in \ind(\ell)} \abs{ [Z_{0}]_j} \geq \sqrt{3 \log M} \Big)  \\
& \qquad + P\Big(\frac{1}{L} \sum_{\ell = 1}^L \max_{j \in \ind(\ell)} \abs{[\Delta_{1,0}]_j} \geq \e \Big)  \\
&   \leq e^{-\kappa L \log M} + K'_0 e^{-\kappa'_0 L \e^2}
\end{align*}
where the last inequality is obtained using Lemma \ref{lem:max_abs_normals} and $\mc{H}_1 (a)$.  The result for the squared terms can be shown similarly.


\subsection{Step 3: Showing $\mathcal{B}_t$ holds} \label{subsub:step3}

 We prove the statements in $\mc{B}_t$ assuming that $\mc{B}_{0}, \ldots, \mc{B}_{t-1}$, and $\mc{H}_1, \ldots, \mc{H}_t$ hold due to the induction hypothesis.  We begin with a lemma that is used to prove $\mc{B}_t (a)$.  Parts $(a) - (g)$ of step $\mc{B}_{t}$ assume the invertibility of $\Mmat_1, \ldots, \Mmat_t$, but for the  sake of brevity, we do not explicitly specify the conditioning.

\begin{lem}
\label{lem:Mv_conc}
Let $$\mathbf{v} := \frac{1}{n}\fullH_t^* \fullq^t_{\perp} - \frac{1}{n}\fullM_t^*\Big[\lambda_t \fullm^{t-1} - \sum_{i=1}^{t-1} \lambda_{i} \gamma^t_{i} \fullm^{i-1}\Big]$$ and $\Mmat_t := \frac{1}{n}\fullM_t^* \fullM_t$.  For $j \in [t]$,
\ben
\begin{split}
& P\Big(\Big \lvert [\Mmat_t^{-1} \mathbf{v}]_{j} \Big \lvert \geq \e \,\, \Big \lvert \,\, \Mmat_1 , \ldots, \Mmat_t \text{ invertible}\Big) \\ 
& \leq  t^2 K K_{t-1} \exp\Big\{\frac{- \kappa \kappa_{t-1} L \epsilon^2}{t^2(\log M)^{2t-1}}\Big\}.
\end{split}
\een
\end{lem}

\begin{IEEEproof}
The proof is similar to that of \cite[Lemma 5.1]{RushV16} and is therefore omitted. \end{IEEEproof}

\noindent \textbf{(a)  [Eq.\ \eqref{eq:Ba}].} The proof of $\mc{B}_t (a)$ follows closely to that of \cite[$\mc{B}_t (a)$]{RushV16} and is therefore omitted.

\noindent  \textbf{(b)  [Eq.\ \eqref{eq:Bb}].} We begin by showing concentration for $(\fullbamp^t)^*\fullw/n$. From the conditional representation of $\fullbamp^t$ given in Lemma \ref{lem:hb_cond} Eq. \eqref{eq:Ba_dist}, we have
\begin{align*}
&P\Big(\Big \lvert \frac{(\fullbamp^t)^*\fullw}{n} \Big \lvert \geq \epsilon \Big) \\
&  = P\Big(\Big \lvert \frac{\sigma_t^2}{\sigma_{t-1}^2} \frac{(\fullbamp^{t-1})^*\fullw}{n} + \frac{\sigma_t^{\perp} (\vecZ'_t)^*\fullw}{n} + \frac{\fullD_{t,t}^*\fullw}{n} \Big \lvert \geq \epsilon \Big) \\
&\leq P\Big(\Big \lvert \frac{(\fullbamp^{t-1})^* \fullw}{n} \Big \lvert \geq \frac{\epsilon \sigma_{t-1}^2}{3\sigma_t^2}\Big) + P\Big(\Big \lvert \frac{(\vecZ'_t)^*\fullw}{n} \Big \lvert \geq \frac{\epsilon}{3\sigma^{\perp}_t} \Big)  \\
& \qquad + P\Big(\Big \lvert \frac{\fullD_{t,t}^*\fullw}{n} \Big \lvert \geq \frac{\epsilon}{3} \Big).
\end{align*}
Label the three terms on the right side of the above as $T_1, T_2, T_3$. The proof is completed by showing each is upper bounded by $t^3 K K_{t-1}\exp\Big\{ \frac{-\kappa \kappa_{t-1} L \e^2}{t^3 (\log M)^{2t-1}}\Big\}$.  First, $T_1$ has the desired upper bound by inductive hypothesis $\mathcal{B}_{t-1} (b)$. For $T_2$, we recall that $\vecZ'_t$ is independent of $\fullw$,  and $\sigma_t^{\perp}$ is bounded from below (Lemma \ref{lem:sigmatperp}). Hence,  using Lemma \ref{lem:max_abs_normals},  $T_2$ is upper bounded by $2\exp\{- \kappa n \e^2\}$ . Finally, using $\abs{\fullD_{t,t}^*\fullw} \leq \norm{\fullD_{t,t}} \norm{\fullw}$, $T_3$ can be bounded as
\begin{align*}
&T_3 \leq P\Big(\frac{\norm{\fullD_{t,t}}\norm{\fullw}}{n} \geq \frac{\epsilon}{3} \Big)  \\
&  \leq P\Big(\frac{\norm{\fullD_{t,t}}}{\sqrt{n}}\Big(\Big \lvert \frac{\norm{\fullw}}{\sqrt{n}} - \sigma \Big \lvert + \sigma \Big)\geq \frac{\epsilon}{3} \Big) \\
&\overset{(a)}{\leq} P\Big( \frac{\norm{\fullD_{t,t}}}{\sqrt{n}} \geq \frac{\epsilon}{6} \min \{1, \frac{1}{\sigma}\} \Big) + P\Big( \Big \lvert \frac{\norm{\fullw}}{\sqrt{n}} - \sigma \Big \lvert \geq \sqrt{\e} \Big) \\
&  \overset{(b)}{\leq} t^3 K K_{t-1} \exp\Big\{ \frac{-\kappa \kappa_{t-1} L \epsilon^2}{t^3(\log M)^{2t-1}} \Big\} + 2e^{-\kappa n \e}.
\end{align*}
Step $(a)$ follows from Lemma \ref{products},  and step $(b)$ from $\mathcal{B}_{t} (a)$ and Lemma \ref{lem:max_abs_normals}. 

Now consider concentration for $(\fullm^t)^*\fullw/n$. By definition, $\fullm^t = \fullbamp^t - \fullw$ and so the result follows from Lemma \ref{sums}, the concentration result  for ${(\fullbamp^t)^*\fullw}/{n}$, and Lemma \ref{lem:max_abs_normals}.


\noindent  \textbf{(c)  [Eq.\ \eqref{eq:Bc}].}  We prove the concentration result for $\norm{\fullbamp^t}^2/n$, with the proof for $(\fullbamp^t)^* \fullbamp^r/n$ with $0 \leq r \leq t-1$ following similarly. Using the conditional distribution of $\fullbamp^t$ from Lemma \ref{lem:hb_cond} Eq.\ \eqref{eq:Ba_dist}, we have
$\norm{\fullbamp^t}^2 = \| ({\sigma_t^2}/{\sigma_{t-1}^2}) \fullb^{t-1} + \sigma^{\perp}_t \vecZ'_t + \fullD_{t,t} \|^2$.
Expanding this expression for $\| \fullbamp^t \|^2$, and recalling that $(\sigma^{\perp}_t)^2 = \sigma_t^2(1- \sigma_t^2/\sigma^2_{t-1})$,  we use Lemma \ref{sums} to write
\begin{align}
&P\Big(\Big \lvert \frac{\norm{\fullbamp^t}^2}{n} - \sigma_t^2 \Big \lvert \geq \epsilon \Big) \nonumber \\
&\leq P\Big(\Big(\frac{\sigma_t^2}{\sigma_{t-1}^2}\Big)^2 \abs{ \frac{\norm{\fullbamp^{t-1}}^2}{n} - \sigma_{t-1}^2} \geq \frac{\epsilon}{4} \Big) \nonumber \\
& \quad +  P\Big( (\sigma^{\perp}_t)^2 \abs{\frac{ \norm{\vecZ'_t}^2}{n} - 1 } \geq \frac{\epsilon}{4} \Big) + P\Big(\frac{\norm{\fullD_{t,t}}^2}{n} \geq \frac{\epsilon}{4} \Big) \nonumber \\
& \quad + P\Big( \frac{2 \sigma_t^2 \sigma^{\perp}_t}{\sigma_{t-1}^2} \frac{\abs{(\fullbamp^{t-1})^*\vecZ'_t}}{n}   
+ 2 \sigma^{\perp}_t  \frac{\abs{ \fullD_{t,t}^* \, \vecZ'_t}}{n} \nonumber\\
& \hspace{1.2in}   +  \frac{2 \sigma_t^2}{\sigma_{t-1}^2} \frac{\abs{ \fullD_{t,t}^* \, \fullb^{t-1}}}{n}  \geq \frac{\epsilon}{4} \Big).
\label{eq:Bct7}
\end{align}
Label the probabilities in \eqref{eq:Bct7} as $T_1, T_2, T_3, T_4$. To prove the result we show each term is upper bounded by 
$t^3 K K_{t-1} \exp\Big\{ \frac{-\kappa \kappa_{t-1} L \e^2}{t^3 (\log M)^{2t-1}}\Big\}$. This is true for term $T_1$ and $T_3$ using inductive hypothesis $\mathcal{B}_{t-1} (c)$ and $\mc{B}_t (a)$, respectively.  Next, $T_2 \leq 2\exp\{-\kappa n \epsilon^2\}$ using Lemma \ref{lem:max_abs_normals}. 

For $T_4$, we note that $( \fullbamp^{t-1} )^*\vecZ'_t \stackrel{d}{=}\| \fullbamp^{t-1} \| Z$, and $ \fullD_{t,t}^* \vecZ'_t \stackrel{d}{=} \| \fullD_{t,t} \| \tilde{Z}$, where $Z,\tilde{Z}$ are standard normal random variables. We therefore have
\be
\begin{split}
T_4 & \leq P\Big(  \frac{2 \sigma_t^2 \sigma^{\perp}_t}{ \sigma_{t-1}^2} \frac{\| \fullbamp^{t-1} \| \abs{Z}}{n}  \geq \frac{\e}{12} \Big)  \\
& \ +  P\Big(  2 \sigma^{\perp}_t  \frac{\| \fullD_{t,t} \| |\tilde{Z}| }{n} \geq \frac{\e}{12}  \Big)  \\
& \  +   P\Big( \frac{2 \sigma_t^2}{\sigma_{t-1}^2} \frac{\| \fullD_{t,t} \| \| \fullbamp^{t-1} \| }{n} \geq \frac{\e}{12}  \Big).
\end{split}
\label{eq:T4_bnd1}
\ee

Letting $c_1= \frac{\sigma_{t-1}^2}{ 24 \sigma_t^2 \sigma^{\perp}_t}$, the first term on the RHS of \eqref{eq:T4_bnd1} can be written as 
\begin{align*} 
&  P\Big( \frac{\| \fullbamp^{t-1} \|}{\sqrt{n}} \cdot \frac{\abs{Z}}{\sqrt{n}}  \geq c_1 \e \Big)  \nonumber \\
& =    P\Big(  \Big( \frac{\norm{\fullbamp^{t-1}}}{\sqrt{n}} - \sigma_{t-1}  + \sigma_{t-1} \Big)  \frac{\abs{Z}}{ \sqrt{n} } \geq c_1 \e \Big) \nonumber \\
&  \leq  P\Big(  \frac{\norm{\fullbamp^{t-1} }}{\sqrt{n}} - \sigma_{t-1}  \geq \sigma_{t-1} \Big) +  
P \Big(  \frac{\abs{Z}}{\sqrt{n}} \geq \frac{c_1 \e}{2 \sigma_{t-1} } \Big) \nonumber \\
 & \stackrel{(a)}{ \leq } t^3 K K_{t-1} \exp\Big\{ \frac{- \kappa \kappa_{t-1} L }{t^3 (\log M)^{2t-1}}\Big\}  + 2 \exp \Big\{ \frac{- n \e^2  c_1^2}{8 \sigma_{t-1}^2 } \Big\},
\end{align*}  
where step $(a)$ follows from  Lemma \ref{sqroots}, inductive hypothesis $\mathcal{B}_{t-1} (c)$, and  Lemma \ref{lem:normalconc}.  

Letting $c_2= \frac{1}{24\sigma^{\perp}_t}$, the second term on the RHS of \eqref{eq:T4_bnd1} can be bounded as 
\ben
\begin{split} 
&  P\Big(   \frac{\| \fullD_{t,t} \|}{\sqrt{n}} \frac{|\tilde{Z}| }{\sqrt{n}} \geq c_2 \e  \Big)\nonumber  \\ 
  & \leq   P\Big(   \frac{\| \fullD_{t,t} \|}{\sqrt{n}} \geq \sqrt{c_2 \e}  \Big)  + P\Big( |\tilde{Z}|  \geq  \sqrt{n c_2 \e} \Big) \nonumber  \\
& \leq  t^3 K K_{t-1} \exp\Big\{\frac{-\kappa \kappa_{t-1} L \e^2}{t^3 (\log M)^{2t-1}}\Big\} + 2 \exp \Big\{ \frac{-n \e c_2}{2} \Big\},
\end{split}  
\een
where the last inequality  follows from $\mc{B}_t (a)$ and Lemma \ref{lem:normalconc}.

The concentration inequality for the last term in \eqref{eq:T4_bnd1} follows in a similar manner using the concentration results for
 $\| \fullD_{t,t} \| / \sqrt{n}$ and $\norm{\fullbamp^{t-1}}/\sqrt{n}$. We have therefore shown that $T_4$ is bounded by $  t^3 K K_{t-1} \exp\Big\{ \frac{-\kappa \kappa_{t-1} L \e^2}{t^3 (\log M)^{2t-1}}\Big\} $, which completes the proof of the concentration result for $\norm{\fullbamp^t}^2/{n}$.

\noindent  \textbf{(d)  [Eq.\ \eqref{eq:Bd}].} We show concentration of $(\fullbamp^r)^* \fullm^s/n$ for $0 \leq r, s \leq t$ when either $r = t$, $s = t$, or both $r = s = t$. By definition, $\fullm^s = \fullbamp^s - \fullw$ so $(\fullbamp^r)^* \fullm^s = (\fullbamp^r)^*\fullbamp^s - (\fullbamp^r)^*\fullw$. Then it follows:
\begin{align*}
&P\Big(\Big \lvert \frac{(\fullbamp^r)^* \fullm^s}{n} - \sigma_{\max(r,s)}^2 \Big \lvert \geq \epsilon \Big) \\
& = P\Big(\Big \lvert \frac{(\fullbamp^r)^* \fullbamp^s}{n} -  \frac{(\fullbamp^r)^*\fullw}{n}  - \sigma_{\max(r,s)}^2 \Big \lvert \geq \epsilon \Big) \\
&\overset{(a)}{\leq} P\Big(\Big \lvert \frac{(\fullbamp^r)^* \fullbamp^s}{n} - \sigma_{\max(r,s)}^2 \Big \lvert \geq \frac{\epsilon}{2} \Big) + P\Big(\Big \lvert \frac{(\fullbamp^r)^*\fullw}{n} \Big \lvert \geq \frac{\epsilon}{2} \Big) \\
& \overset{(b)}{\leq} t^3 K K_{t-1} \exp\Big\{ \frac{-\kappa \kappa_{t-1} L \epsilon^2}{t^3 (\log M)^{2t-1}}\Big\}.
\end{align*}
Step $(a)$ follows from Lemma \ref{sums} and step $(b)$ using $\mathcal{B}_t (c)$ and $\mathcal{B}_{0} (b)$ - $\mathcal{B}_{t} (b)$.

\noindent  \textbf{(e)  [Eq.\ \eqref{eq:Be}].} By definition, $\fullm^t = \fullbamp^t - \fullw$, and so it follows:
\begin{align*}
& P\Big(\Big \lvert \frac{(\fullm^{r})^* \fullm^t}{n} - \tau_t^2 \Big \lvert \geq \epsilon \Big) \\
& = P\Big(\Big \lvert \frac{(\fullm^{r})^*\fullbamp^t}{n} - \frac{(\fullm^{r})^*\fullw}{n} - \tau_t^2 \Big \lvert \geq \epsilon \Big) \\
&\overset{(a)}{\leq}P\Big(\Big \lvert \frac{(\fullm^{r})^* \fullbamp^t}{n} - \sigma_t^2 \Big \lvert \geq \frac{\epsilon}{2} \Big) + P\Big(\Big \lvert \frac{(\fullm^{r})^*\fullw}{n} + \sigma^2\Big \lvert \geq \frac{\epsilon}{2} \Big) \\
&   \overset{(b)}{\leq} t^3 K K_{t-1} \exp\Big\{ \frac{-\kappa \kappa_{t-1} L \epsilon^2}{t^3 (\log M)^{2t-1}}\Big\}.
\end{align*}
Step $(a)$ follows from Lemma \ref{sums} and step $(b)$ from $\mc{B}_t (d)$ and $\mathcal{B}_{0} (b)$ - $\mathcal{B}_{t} (b)$.

\noindent  \textbf{(f), (g)  [Eqs.\ \eqref{eq:Msing}, \eqref{eq:Bf1}, \eqref{eq:Bf}, and \eqref{eq:Bg}].} The proofs of $\mc{B}_t (f), (g)$ follow closely to that of \cite[$\mc{B}_t (g), (h)$]{RushV16} and are not included here.  For result \eqref{eq:Msing} we note that $(\tau_t^{\perp}) > 0$ for $0 \leq t < T$ by Lemma \ref{lem:sigmatperp}, while \cite{RushV16} used the stopping criterion to show this fact.

\subsection{Step 4: Showing $\mathcal{H}_{t+1}$ holds} \label{subsub:step4}
We begin with a lemma that is used in the proof of $\mc{H}_{t+1} (a)$.  Parts $(a) - (h)$ of step $\mc{H}_{t+1}$ assume the invertibility of $\Qmat_1, \ldots, \Qmat_{t+1}$, but for the  sake of brevity, we do not explicitly specify the conditioning.

\begin{lem}
\label{lem:Qv_conc}
Let $\mathbf{v} := \frac{1}{n}\fullB^*_{t+1} \fullM_t^{\perp} - \frac{1}{n}\fullQ_{t+1}^*[\fullq^t - \sum_{i=0}^{t-1} \alpha^t_i \fullq^i]$ and $\Qmat_{t+1} := \frac{1}{n}\fullQ_{t+1}^* \fullQ_{t+1}$.   For $r \in [t+1]$,
\ben
\begin{split}
& P\Big(\Big \lvert [\Qmat_{t+1}^{-1} \mathbf{v}]_{r}\Big \lvert \geq \e \,\, \Big \lvert \,\, \Qmat_1, \ldots, \Qmat_{t+1} \text{ invertible} \Big) \\
&  \leq  t^2 K K'_{t-1} \exp\Big\{ \frac{- \kappa \kappa'_{t-1} L\e^2}{t^2(\log M)^{2t-1}} \Big\}.
\end{split}
\een
\end{lem}
\begin{IEEEproof}
The proof follows similarly to that of Lemma \ref{lem:Mv_conc} and \cite[Lemma 5.2]{RushV16}.
\end{IEEEproof}

\noindent  \textbf{(a) [Eqs.\ \eqref{eq:Ha} - \eqref{eq:Ha1}].} Eq.\ \eqref{eq:Ha} follows from the Cauchy-Schwarz inequality.   To show \eqref{eq:Ha1}, we recall the definition of $\fullD_{t+1,t}$ from  Lemma \ref{lem:hb_cond} Eq.\ \eqref{eq:Dt1t}:
\begin{align}
\fullD_{t+1,t} &  =   \sum_{r=0}^{t-1} (\alpha^t_r - \hat{\alpha}^t_r)\fullh^{r+1} + \Big(\frac{\norm{\fullm^t_{\perp}}}{\sqrt{n}} - \tau_{t}^{\perp}\Big) \vecZ_t  \nonumber  \\
& \quad -  \frac{\norm{\fullm^t_{\perp}} \tilde{\fullQ}_{t+1} \tilde{\vecZ}}{n} + \sum_{r=0}^{t} \fullq^r \left[\Qmat^{-1} \mathbf{v} \right]_{r+1},
\label{eq:delta_rew}
\end{align} 
where we have used that $\fullQ_{t+1}\Qmat^{-1} \mathbf{v} = \sum_{r=0}^{t} \fullq^r [\Qmat^{-1} \mathbf{v}]_{r+1}$ and $(\|\fullm^t_{\perp}\|/\sqrt{n}) \proj_{\fullQ_{t+1}} \vecZ_t \overset{d}{=}   (\|\fullm^t_{\perp}\| \tilde{\fullQ}_{t+1} \tilde{\vecZ})/n$ which follows by Lemma \ref{lem:gauss_p0} where $\tilde{\vecZ} \in \mathbb{R}^{t+1}$ is a vector of i.i.d.\ standard Gaussians and 
\be
\tilde{\fullQ}_{t+1} = 
[ \tilde{\fullq}^0 \mid \ldots \mid \tilde{\fullq}^t ] := \sqrt{n}\left[\frac{\fullq^0_{\perp}}{ \|\fullq^0_{\perp} \|} \mid \ldots \mid \frac{\fullq^t_{\perp}}{ \| \fullq^t_{\perp} \|} \right].
\label{eq:Hat2a}
\ee
Consider a single element $j \in [N]$ of $\fullD_{t+1,t}$.   Using the triangle inequality to bound the expression in \eqref{eq:delta_rew}, we obtain
\begin{align}
& \abs{[\Delta_{t+1,t}]_j}   \leq  \sum_{r=0}^{t-1} | \alpha^t_r - \hat{\alpha}^t_r | | \fullh^{r+1}_j | + |[Z_{t}]_j| \Big \lvert \frac{ \| \fullm^t_{\perp} \| }{\sqrt{n}}  \nonumber  \\ 
& \quad  - \tau_{t}^{\perp}\Big \lvert + \frac{ \| \fullm^t_{\perp}\|  }{\sqrt{n}} \sum_{s=0}^t 
\frac{ |\tilde{Z}_{s}| | \tilde{q}^s_j |}{\sqrt{n}} + c \sqrt{\log M} \sum_{u=1}^{t+1} \Big \lvert [\Qmat^{-1} \mathbf{v} ]_u \Big \lvert.
\label{eq:del_j_bnd}
\end{align}
In the last term  above, we have used the fact that for each for $i \in [N]$ we have the upper bound $|q^r_{i}| \leq 2 \sqrt{nP_{\ind(\sect(i))}} \leq c \sqrt{\log M}$ for constant $c >0$. Squaring \eqref{eq:del_j_bnd} and applying Lemma \ref{lem:squaredsums} to bound the RHS, we obtain
\begin{align*}
([\Delta_{t+1,t}]_j)^2 &\leq 3(t+1)\sum_{r=0}^{t-1} \abs{\alpha^t_r - \hat{\alpha}^t_r}^2 (\fullh^{r+1}_j)^2  \\
&  + 3(t+1)([Z_{t}]_j)^2 \Big \lvert \frac{ \| \fullm^t_{\perp} \| }{\sqrt{n}} - \tau_{t}^{\perp}\Big \lvert^2 \\
&+ 3(t+1)\frac{ \| \fullm^t_{\perp} \|^2}{n} \sum_{s=0}^t \frac{\tilde{Z}_{s}^2}{n} (\tilde{q}^s_j)^2  \\
& + 3(t+1)c \log M \sum_{u=1}^{t+1} [\Qmat^{-1} \mathbf{v} ]_u^2.
\end{align*}
Therefore, setting $\e' = \frac{\e}{9(t+1)^2}$ and using Lemma \ref{sums},  we have the following upper bound:
\be
\begin{split}
& P \Big(\frac{1}{L} \sum_{\ell=1}^L \max_{j \in \ind(\ell)} ([\Delta_{t+1,t}]_{j})^2 \geq \epsilon \Big) \\
&\leq \sum_{r=0}^{t-1} P\Big(\Big \lvert \alpha^t_{r} - \hat{\alpha}^t_r \Big \lvert^2 \frac{1}{L} \sum_{\ell=1}^L \max_{j \in \ind(\ell)} (h^{r+1}_{j})^2 \geq \e'  \Big)    \\
& \ + P\Big(\Big \lvert \frac{\norm{\fullm^t_{\perp}}}{\sqrt{n}} - \tau_{t}^{\perp}\Big \lvert^2 \frac{1}{L} \sum_{\ell=1}^L \max_{j \in \ind(\ell)} ([Z_{t}]_j)^2 \geq \e'  \Big) \\
& \ + \sum_{k=0}^t P\Big(\frac{\norm{\fullm^t_{\perp}}^2}{n} \cdot \frac{\tilde{Z}_{k}^2}{n} \cdot \frac{1}{L} \sum_{\ell=1}^L \max_{j \in \ind(\ell)} (\tilde{q}^k_j)^2 \geq \e'  \Big)   \\
& \ + \sum_{u=1}^{t+1} P\Big(  \Big \lvert [\Qmat^{-1} \mathbf{v} ]_{u}\Big \lvert \geq \frac{\sqrt{\e'}}{2\sqrt{c\log M}} \Big).
\label{eq:Hbt2_2}
\end{split}
\ee
Label the terms on the RHS of \eqref{eq:Hbt2_2} as $T_{1}, T_2, T_3, T_{4}$. To complete the proof, we will show that each term is upper bounded by $t^3 K K'_{t-1} \exp\Big\{\frac{ -\kappa \kappa'_{t-1} L \e}{t^4 (\log M)^{2t}}\Big\}$. 

First, for $0 \leq r \leq t-1$,
\begin{align*}
 T_{1}  &  \leq \sum_{r=0}^{t-1}  P\Big(\Big \lvert \alpha^t_r - \hat{\alpha}^t_r\Big \lvert \geq \sqrt{\epsilon'/{7 \tau_0^2 \log M}} \Big) \nonumber  \\
&  \quad + \sum_{r=0}^{t-1}  P\Big(\frac{1}{L} \sum_{\ell=1}^L \max_{j \in \ind(\ell)} (h^{r+1}_j)^2 \geq 7 \tau_0^2 \log M  \Big) \nonumber \\
&\overset{(b)}{\leq} t \cdot t^4 K K_{t-1} \exp\Big\{ \frac{ - \kappa \kappa_{t-1} L \e}{(t+1)^2(t-1)^5(\log M)^{2t}}\Big\}  \nonumber \\
& \quad  + t K K_{t-1}  \exp\Big\{\frac{- \kappa \kappa_{t-1} L }{ (\log M)^{2t-2}}\Big\}.
\end{align*}
Step $(b)$ follows from $\mathcal{B}_t (f)$ and inductive hypotheses $\mathcal{H}_1 (h)$ - $\mathcal{H}_{t} (h)$. Next consider $T_2$.
\begin{align*}
& T_2 \leq P\Big(\Big \lvert \frac{\norm{\fullm^t_{\perp}}}{\sqrt{n}} - \tau_{t}^{\perp}\Big \lvert \geq \sqrt{\frac{\epsilon'}{3\log M} }\Big)  \nonumber 
\\ 
& \qquad + P\Big(\frac{1}{L} \sum_{\ell=1}^L \max_{j \in \ind(\ell)} ([Z_{t}]_j)^2 \geq 3 \log M \Big) \nonumber \\
&\overset{(c)}{\leq} K K_{t-1} \exp\Big\{\frac{ -\kappa \kappa_{t-1}  L \e}{27(t+1)^2 t^7(\log M)^{2t}}\Big\}  +e^{-\kappa L \log M}.
\end{align*}
Step $(c)$ follows from $\mathcal{B}_t (g)$, Lemma \ref{sqroots}, and Lemma \ref{lem:max_abs_normals}. From Lemma \ref{lem:Qv_conc}, term $T_{4}$ is upper bounded by $t^3 K K'_{t} \exp\Big\{ \frac{- \kappa \kappa'_{t} L\e^2}{t^4(\log M)^{2t}} \Big\}$.

Finally, consider $T_{3}$ for $0 \leq k \leq t$.   Recall that $\tilde{\fullq}^k = \sqrt{n} \, \fullq^k_\perp \| \fullq^k_\perp \|$, and for each section $\ell \in [L]$ we have
\ben
\max_{j \in \ind(\ell)}  ([\tilde{q}^k_\perp]_j)^2 \leq \| [\fullq^k_\perp]_{(\ell)} \|^2 \leq \| \fullq^k_{(\ell)} \|^2 \leq 2nP_\ell \leq c \log M,
\een
for a universal constant $c >0$. Recalling $\e' = \frac{\e}{9(t+1)^2}$, we therefore have
\begin{align*}
& T_3   \leq  \sum_{k=0}^t P\Big(\frac{\norm{\fullm^t_{\perp}}^2}{n} \cdot \frac{\tilde{Z}_{k}^2}{n}  \geq \frac{\e'}{c \log M}  \Big)  \nonumber \\
& \leq \sum_{k=0}^t P\Big(  \frac{\tilde{Z}_{k}^2}{n} \geq \frac{\e'}{ 2 (\tau_t^{\perp})^2 c \log M} \Big) + 
 P\Big( \frac{\norm{\fullm^t_{\perp}}^2}{n} \geq  2 (\tau_t^{\perp})^2 \Big) \nonumber \\
 & \stackrel{(d)}{\leq}  2(t+1) \exp\Big\{ \frac{-\kappa L\e }{t^2} \Big\} + t^5 K K_{t-1} \exp\Big\{ \frac{-\kappa \kappa_{t-1} L}{t^7 (\log M)^{2t-1}}\Big\}
\end{align*}
where step $(d)$ is obtained using $\mc{B}_t (g)$ Lemma \ref{lem:normalconc} (and recalling that $nR/\log M = L$).  Thus, using the definitions of $\kappa_t, \kappa'_t$ in \eqref{eq:Ktkapt_defs},  we have the required bound for each of the terms in \eqref{eq:Hbt2_2}.

\noindent  \textbf{(b) [Eq.\ \eqref{eq:Hb}].} Using the conditional distribution of $\fullh^{t+1}$ in Lemma \ref{lem:hb_cond} Eq. \eqref{eq:Ha_dist}, and Lemma \ref{sums}, we have
\begin{align*}
&P\Big(\frac{1}{n}\Big \lvert (\fullh^{t+1})^*\fullq^{0} \Big \lvert \geq \epsilon \Big) \\
& = P\Big(\frac{1}{n} \Big \lvert \frac{\tau_t^2}{\tau_{t-1} ^2} (\fullh^{t})^*\fullq^{0} +  \tau^{\perp}_t \vecZ_t ^*\fullq^{0} +\fullD_{t+1,t}^*\fullq^{0}\Big \lvert \geq \epsilon \Big) \\
&\leq P\Big(\frac{1}{n} \Big \lvert (\fullh^{t})^*\fullq^{0} \Big \lvert \geq \frac{\epsilon \tau_{t-1}^2}{3 \tau_{t}^2} \Big) + P\Big(\frac{1}{n} \Big \lvert \vecZ_t^* \fullq^0 \Big \lvert \geq \frac{\epsilon}{3\tau^{\perp}_t} \Big) \\
 &\quad + P\Big(\frac{1}{n} \Big \lvert \fullD_{t+1,t}^*\fullq^0 \Big \lvert \geq \frac{\epsilon}{3} \Big).
\end{align*}
Label the terms on the right side of the above as $T_1, T_2, T_3$.  
First, by the induction hypothesis $\mathcal{H}_{t} (b)$,  $T_{1}$ is bounded by $t^3 K K'_{t-1} \exp\Big\{\frac{-\kappa  \kappa'_{t-1} L \e^2}{t^4 (\log M)^{2t-3}}\Big\}$.  Next consider term $T_2$.  Since $\fullq^0$ and $\vecZ_t$ are independent, we have $\vecZ_t^*\fullq^0 \stackrel{d}{=} \|\fullq^0\| Z$, where $Z \in \mathbb{R}$ is standard Gaussian.  Therefore, using Lemma \ref{lem:normalconc} and recalling that $\| \fullq^0\|^2 = nP$, we obtain
\begin{align*}
T_2  =   P\Big(\frac{\norm{\fullq^0}}{\sqrt{n}}\frac{\abs{Z}}{\sqrt{n}} \geq \frac{\epsilon}{3\tau^{\perp}_t} \Big) & = P\Big(\frac{\abs{Z}}{\sqrt{n}} \geq \frac{\epsilon}{3 \tau^{\perp}_t\sqrt{P}} \Big)  \\
& \leq 2\exp\Big\{\frac{-n \epsilon^2}{18 P(\tau_t^{\perp})^2}\Big\}.
\end{align*}
Finally,
\begin{align*}
T_3 & = P\Big(\Big \lvert\sum_{\ell = 1}^L \frac{([\fullD_{t+1,t}]_{(\ell)})^*\fullq^0_{(\ell)}}{n} \Big \lvert \geq \frac{\epsilon}{3} \Big) \\
& \overset{(a)}{\leq} P\Big(\frac{1}{L} \sum_{\ell = 1}^L \max_{j \in \ind(\ell)} \abs{[\Delta_{t+1,t}]_{j}} \geq \frac{\epsilon\sqrt{\log M}}{3Rc} \Big) \\
& \overset{(b)}{\leq} t^3 K K'_{t-1} \exp\Big\{ \frac{-\kappa \kappa'_{t-1} L \e^2}{t^4 (\log M)^{2t-1}}\Big\} .
\end{align*}
Step $(a)$ follows since $nR = L \log M$ with $\sum_{i \in \ind(\ell)} \abs{q^0_i} = \sqrt{nP_{\ell}}$ and $\sqrt{nP_{\ell}} \leq c \sqrt{\log M}$ for some constant $c > 0$ and each $\ell \in [L]$ and step $(b)$ from $\mathcal{H}_{t+1} (a)$.


\noindent  \textbf{(c) [Eq.\ \eqref{eq:Hc}].} We will  show the  concentration result for $(\fullq^r)^*\fullq^{t+1}/n$ when $0 \leq r \leq t+1$. 
Recalling that $\fullq^{r} = \eta^{r-1}(\fullb_0 - \fullh^{r}) - \fullb_0$, it follows that
\ben
\begin{split}
& P\Big( \Big \lvert \frac{1}{n}(\fullq^r)^*\fullq^{t+1} - \sigma_{t+1}^2\Big \lvert \geq \epsilon \Big) \\
& = P\Big( \Big \lvert \frac{1}{n}(\eta^{r-1}(\fullb_0 - \fullh^{r}) - \fullb_0)^*(\eta^t(\fullb_0 - \fullh^{t+1}) - \fullb_0)  \\
& \hspace{2in}- \sigma_{t+1}^2 \Big \lvert \geq \epsilon \Big).
\end{split}
\een
Using the representation in Lemma \ref{lem:ideal_cond_dist} Eq. \eqref{eq:htil_rep}, and Lemma \ref{sums} we write
\be
\begin{split}
&P\Big( \Big \lvert \frac{1}{n}(\eta^{r-1}(\fullb_0 - \fullh^{r}) - \fullb_0)^*(\eta^t(\fullb_0 - \fullh^{t+1}) - \fullb_0) \\
& \hspace{2in} - \sigma_{t+1}^2 \Big \lvert \geq \epsilon \Big) \\
&= P\Big(\Big \lvert \frac{1}{n}(\eta^{r-1}(\fullb_0 - \tilde{\fullh}^{r} - \tilde{\fullD}_r) - \fullb_0)^* \\
& \hspace{0.5in} (\eta^t(\fullb_0 - \tilde{\fullh}^{t+1} - \tilde{\fullD}_{t+1}) - \fullb_0) - \sigma_{t+1}^2 \Big \lvert \geq \e \Big) \\
&\leq P\Big( \frac{1}{n} \Big \lvert[\eta^{r-1}(\fullb_0 - \tilde{\fullh}^{r} - \tilde{\fullD}_r) - \fullb_0]^* \\
& \hspace{0.5in} \left[\eta^t(\fullb_0 - \tilde{\fullh}^{t+1} - \tilde{\fullD}_{t+1}) -\eta^t(\fullb_0 - \tilde{\fullh}^{t+1})\right] \Big \lvert \geq \frac{\e}{3} \Big) \\
& \ +  P\Big( \frac{1}{n} \Big \lvert \left[\eta^{r-1}(\fullb_0 - \tilde{\fullh}^{r} - \tilde{\fullD}_r)- \eta^{r-1}(\fullb_0 - \tilde{\fullh}^{r})\right]^* \\ 
& \hspace{0.6in} [\eta^t(\fullb_0 - \tilde{\fullh}^{t+1}) - \fullb_0] \Big \lvert \geq \frac{\e}{3} \Big) \\
& \ + P\Big(\Big \lvert  \frac{1}{n} [\eta^{r-1}(\fullb_0 - \tilde{\fullh}^{r}) - \fullb_0]^*[\eta^t(\fullb_0 - \tilde{\fullh}^{t+1}) - \fullb_0]  \\
& \hspace{2in} - \sigma_{t+1}^2 \Big \lvert \geq \frac{\e}{3} \Big).
\label{eq:replaced_probs_2}
\end{split}
\ee
We label the three terms on the RHS of \eqref{eq:replaced_probs_2} as $T_1, T_2, T_3$ and bound each term separately.  We first note the following bound about $\tilde{\fullD}_{r}$  that will be used repeatedly. For 
$1 \leq r \leq t+1$,
\be
\begin{split}
&P\Big( \frac{1}{L}  \sum_{\ell = 1}^L\max_{j \in \ind(\ell)} \Big \lvert [\tilde{\Delta}_{r}]_j \Big \lvert \geq\e\Big) \\
& = P\Big( \frac{1}{L}  \sum_{\ell = 1}^L \max_{j \in \ind(\ell)} \Big \lvert   \sum_{k=0}^{r-1} \Big(\frac{\tau_{r-1}^2}{\tau_{k}^2}\Big) [\Delta_{k+1,k}]_j \Big \lvert \geq \e \Big) \\
&\overset{(a)}{\leq} \sum_{k=0}^{r-1}  P\Big(\frac{1}{L}  \sum_{\ell = 1}^L   \max_{j \in \ind(\ell)}  \Big \lvert  [\Delta_{k+1,k}]_j \Big \lvert \geq \frac{\e}{r} \Big) \\
&  \overset{(b)}{\leq}  t^4 K K'_{t-1} \exp\Big\{ \frac{- \kappa \kappa'_{t-1} L \e^2 }{t^4(t+1)^2 (\log M)^{2t}}\Big\}.
\label{eq:maxtildedelta}
\end{split}
\ee
In the above, step $(a)$ follows by the triangle inequality, Lemma \ref{sums}, and the fact that $\tau_{r-1}^2/\tau_k^2 \leq 1$ for $k \leq r-1 $, and step $(b)$ from $\mc{H}_{t+1}(a)$ and the fact that $r \leq (t+1)$.

First consider $T_1$, the first term of \eqref{eq:replaced_probs_2}.
\be
\begin{split}
& T_1 = P\Big( \frac{1}{n} \Big \lvert \sum_{\ell = 1}^L [\eta^{r-1}_{(\ell)}(\fullb_0 - \tilde{\fullh}^{r} - \tilde{\fullD}_r) - \fullb_{0_{(\ell)}}]^* \\
& \qquad   \left[\eta^t_{(\ell)}(\fullb_0 - \tilde{\fullh}^{t+1} - \tilde{\fullD}_{t+1}) -\eta^t_{(\ell)}(\fullb_0 - \tilde{\fullh}^{t+1})\right] \Big \lvert \geq \frac{\e}{3} \Big)  \\
&\leq  P\Big( \frac{1}{n}  \sum_{\ell = 1}^L 2\sqrt{nP_{\ell}} \sum_{j \in \ind(\ell)} \Big \lvert \eta^t_{j}(\fullb_0 - \tilde{\fullh}^{t+1} - \tilde{\fullD}_{t+1}) \\ 
& \hspace{1in} -\eta^t_{j}(\fullb_0 - \tilde{\fullh}^{t+1}) \Big \lvert \geq \frac{\e}{3} \Big)  \\
&\overset{(a)}{\leq}  P\Big( \frac{1}{n}  \sum_{\ell = 1}^L \frac{4 (nP_{\ell})^{3/2}}{\tau_t^2}\max_{j \in \ind(\ell)} \Big \lvert [\tilde{\Delta}_{t+1}]_j \Big \lvert \geq \frac{\e}{3} \Big) \\
&\overset{(b)}{\leq}  P\Big( \frac{1}{L}  \sum_{\ell = 1}^L\max_{j \in \ind(\ell)} \Big \lvert [\tilde{\Delta}_{t+1}]_j \Big \lvert \geq \frac{\e \tau_t^2}{12cR\sqrt{\log M}} \Big) \\
&  \overset{(c)}{\leq}  t^4 K K'_{t-1} \exp\Big\{ \frac{ - \kappa \kappa'_{t-1} L \e^2}{t^4(t+1)^2 (\log M)^{2t+1}}\Big\}.
\label{eq:T1_new}
\end{split}
\ee
In the above, step $(a)$ follows from Lemma \ref{lem:BC9} applied to each section, step $(b)$ from the fact that $(nP_{\ell})^{3/2} \leq c (\log M)^{3/2}$ for some constant $c > 0$ and all $\ell \in [L]$, and step $(c)$ from \eqref{eq:maxtildedelta}.  The same upper bound as that shown in \eqref{eq:T1_new} for $T_1$ also holds (shown similarly) for term $T_2$ of \eqref{eq:replaced_probs_2}.

Finally, consider the last  term $T_3$ of \eqref{eq:replaced_probs_2}.  Using the definition of $\tilde{\fullh}^r$ in \eqref{eq:htilde_def}, we have
\begin{align}
& T_3  =  P\Big(\Big \lvert \frac{1}{n}[\eta^{r-1}(\fullb_0 -  \tau_{r-1} \tilde{\vecZ}_{r-1}) - \fullb_0]^* \nonumber \\
& \qquad \qquad  [\eta^t(\fullb_0 - \tau_{t} \tilde{\vecZ}_t) - \fullb_0] - \sigma_{t+1}^2 \Big \lvert \geq \frac{\e}{3} \Big)  \nonumber \\ 
& = P\Big( \Big \lvert \frac{1}{L}\sum_{\ell = 1}^L \frac{R}{\log M} [\eta^{r-1}_{(\ell)}(\fullb_0 + \tau_{r-1} \tilde{\vecZ}_{r-1}) - \fullb_{0_{(\ell)}}]^* \nonumber \\
& \qquad [\eta^t_{(\ell)}(\fullb_0 + \tau_t \tilde{\vecZ}_t) - \fullb_{0_{(\ell)}}] - \sigma_{t+1}^2\Big \lvert \geq \frac{\epsilon}{3} \Big),  \nonumber \\
& \leq 2\exp\{-{L \e^2}/(9c^2)\},
\label{eq:Hct2}
\end{align}
where the last inequality is obtained using Hoeffding's inequality (Lemma \ref{lem:hoeff_lem}), which can be applied after verifying two conditions. 
 First,
\begin{align*}
&\mathbb{E}_{\tilde{\vecZ}_{r-1}, \tilde{\vecZ}_{t}}  [\eta^{r-1}(\fullb_0 + \tau_{r-1} \tilde{\vecZ}_{r-1}) - \fullb_0]^*[\eta^t(\fullb_0 + \tau_t \tilde{\vecZ}_t) - \fullb_0]  \\
&=\mathbb{E}_{\tilde{\vecZ}_{r-1}, \tilde{\vecZ}_{t}, \fullb}  [\eta^{r-1}(\fullb+ \tau_{r-1} \tilde{\vecZ}_{r-1}) \hspace{-1pt} -  \hspace{-1pt} \fullb]^*[\eta^t(\fullb + \tau_t \tilde{\vecZ}_t)  \hspace{-1pt} -  \hspace{-1pt} \fullb]  \\
& = n\sigma_{t+1}^2,
\end{align*}
where the first equality is true for each $\fullb_0 \in \mathcal{B}_{M,L}$ because of the uniform distribution of the non-zero entry in each section of $\fullb$ over the $M$ possible locations and the entrywise i.i.d.\ distributions of $\tilde{\vecZ}_{r-1}$ and $\tilde{\vecZ}_{t}$ and the second equality by Lemma \ref{lem:expect_etar_etas}.  Second, for each section $\ell \in [L]$, there exists a constant $c>0$ such that,
\ben
\begin{split}
& 0 \leq  \\
& \frac{R  \,  [\eta^{r-1}_{(\ell)}(\fullb_0 + \tau_{r-1} \tilde{\vecZ}_{r-1}) - \fullb_{0_{(\ell)}}]^*[\eta^t_{(\ell)}(\fullb_0 + \tau_t \tilde{Z}_t) - \fullb_{0_{(\ell)}}] }{\log M} \\
&  \leq c.
\end{split}
\een

We have thus shown that each term of \eqref{eq:replaced_probs_2} is upper bounded by $t^4 K K'_{t-1} \exp\Big\{\frac{ -\kappa \kappa'_{t-1} L \e^2}{t^6 (\log M)^{2t+1}}\Big\}$, which gives the desired result.


\noindent  \textbf{(d) [Eq.\ \eqref{eq:Hd}].} Recalling that $\fullq^{s+1} = \eta^s(\fullb_0 - \fullh^{s+1}) - \fullb_0$, we have
\ben
\begin{split}
& (\fullh^{r+1})^* \fullq^{s+1} = (\fullh^{r+1})^* [\eta^s(\fullb_0 - \fullh^{s+1}) - \fullb_0] \\
 &= (\fullh^{r+1})^* \eta^s(\fullb_0 - \fullh^{s+1}) + (\fullh^{r+1})^*\fullq^0.
\label{eq:Hdt0}
\end{split}
\een
Using the above and Lemma \ref{sums}, we obtain
\be
\begin{split} 
& P\Big(\Big \lvert \frac{(\fullh^{r+1})^*\fullq^{s+1}}{n} + \frac{\sigma_{s+1}^2 \tau_{\max(r,s)}^2}{\tau_s^2} \Big \lvert \geq \epsilon \Big)   \\
&\leq P\Big(\Big \lvert \frac{(\fullh^{r+1})^* \eta^s(\fullb_0 - \fullh^{s+1})}{n} + \frac{\sigma_{s+1}^2 \tau_{\max(r,s)}^2}{\tau_s^2} \Big \lvert \geq \frac{\e}{2} \Big)  \\
& \quad + P\Big(\Big \lvert \frac{(\fullh^{r+1})^*\fullq^0}{n} \Big \lvert \geq \frac{\e}{2} \Big) \\ 
& = P\Big(\Big \lvert \frac{(\tilde{\fullh}^{r+1} + \tilde{\fullD}_{r+1})^* \eta^s(\fullb_0 - \tilde{\fullh}^{s+1} - \tilde{\fullD}_{s+1})}{n}    \\ 
& \qquad + \frac{\sigma_{s+1}^2 \tau_{\max(r,s)}^2}{\tau_s^2} \Big \lvert \geq \frac{\e}{2} \Big) + P\Big(\Big \lvert \frac{(\fullh^{r+1})^*\fullq^0}{n} \Big \lvert \geq \frac{\e}{2} \Big)  
\label{eq:Hdt1a},
\end{split}
\ee
where the last equality is obtained using the representation in Lemma \ref{lem:ideal_cond_dist}, Eq.\ \eqref{eq:htil_rep}.
The second term on the RHS of \eqref{eq:Hdt1a} is upper bounded by  $t^3 K K'_{t-1} \exp\Big\{\frac{- \kappa \kappa'_{t-1} L \e^2}{t^4 (\log M)^{2t-1}}\Big\}$ by $\mathcal{H}_{t+1} (b)$. In what follows we upper bound the first term of \eqref{eq:Hdt1a}, denoted by $T_1$.    Using Lemma \ref{sums}, we have 
\begin{align}
& T_1 \leq  P\Big(\Big \lvert \frac{(\tilde{\fullh}^{r+1})^* \eta^s(\fullb_0 - \tilde{\fullh}^{s+1})}{n}  \frac{\sigma_{s+1}^2 \tau_{\max(r,s)}^2}{\tau_s^2} \Big \lvert \geq \frac{\e}{6} \Big) \nonumber  \\
&  + P\Big(\Big \lvert \frac{\tilde{\fullD}_{r+1}^* \eta^s(\fullb_0 - \tilde{\fullh}^{s+1} - \tilde{\fullD}_{s+1})}{n} \Big \lvert \geq \frac{\e}{6} \Big) \nonumber \\
&  +  P\Big(\Big \lvert \frac{(\tilde{\fullh}^{r+1})^*[\eta^s(\fullb_0 - \tilde{\fullh}^{s+1} - \tilde{\fullD}_{s+1})- \eta^s(\fullb_0 - \tilde{\fullh}^{s+1})]}{n} \Big \lvert \nonumber \\
& \hspace{2in} \geq \frac{\e}{6} \Big).
\label{eq:Hdt13_b}
\end{align}
We label the terms on the RHS of \eqref{eq:Hdt13_b} as $T_{1a}, T_{1b}, T_{1c}$, and bound each separately. Using Lemma \ref{lem:ideal_cond_dist}, the first term can be written as 
\be
T_{1a} = P\Big( \Big \lvert \frac{\tau_r \tilde{\vecZ}_r^* \eta^s(\fullb_0 - \tau_s \tilde{\vecZ}_s)}{n} + \frac{\sigma_{s+1}^2 \tau_{\max(r,s)}^2}{\tau_s^2} \Big \lvert \geq \frac{\epsilon}{6} \Big),
\label{eq:Hdt9}
\ee
where $\tilde{\vecZ}_r, \tilde{\vecZ}_s \in \mathbb{R}^N$ are standard Gaussian vectors, with $\expec[ \tilde{Z}_{r_j}, \tilde{Z}_{s_j}]= \frac{\tau_{\max(r,s)}}{\tau_{\min(r,s)}}$, for $j \in [N]$. We then observe that
\begin{align}
&\frac{\tau_r \mathbb{E}_{\tilde{\vecZ}_r, \tilde{\vecZ}_s}\{\tilde{\vecZ}_r^*\eta^s(\fullb_0 - \tau_s \tilde{\vecZ}_s)\}}{n} \nonumber  \\
&  \overset{(a)}{=} \frac{\tau_r \mathbb{E}_{\tilde{\vecZ}_r, \tilde{\vecZ}_s, \fullb }\{\tilde{\vecZ}_r^*\eta^s(\fullb - \tau_s \tilde{\vecZ}_s)\}}{n}  \nonumber \\
&  \overset{(b)}{=} \frac{\tau_{\max(r,s)}^2}{\tau_s^2}  \Big[\frac{\mathbb{E}_{\tilde{\vecZ}_s, \fullb}  \| \eta^s(\fullb + \tau_s \tilde{\vecZ}_{s}) \|^2}{n} - P \Big] \nonumber   \\
&  \overset{(c)}{=}  \frac{- \sigma_{s+1}^2 \tau_{\max(r,s)}^2}{\tau_s^2}.
\label{eq:Hdt9a}
\end{align}
In the above, step $(a)$ follows for each $\fullb_0 \in \mathcal{B}_{M,L}$ because of the uniform distribution of the non-zero entry in each section of $\fullb$ over the $M$ possible locations and the entry-wise i.i.d.\ distributions of $\tilde{\vecZ}_{r}$ and $\tilde{\vecZ}_{s}$, step $(b)$ by Stein's Lemma (see  \cite[p.1491, Eqs.\ (102) -- (104)]{RushGVIT17} for details), and step $(c)$ from Lemma \ref{lem:expect_etar_etas}.  Using \eqref{eq:Hdt9a} in \eqref{eq:Hdt9},  it is shown in Lemma \ref{lem:Hd_convergence}  that $T_{1a} \leq \exp\{-\kappa L \e^2\}$.

Next consider $T_{1b}$:
\ben
\begin{split}
&T_{1b} \\
&  = P\Big(\frac{1}{n}\sum_{\ell = 1}^L \Big \lvert  ([\tilde{\fullD}_{r+1}]_{(\ell)})^* \, \eta_{(\ell)}^s(\fullb_0 - \tilde{\fullh}^{s+1} - \tilde{\fullD}_{s+1}) \Big \lvert \geq \frac{\epsilon}{6} \Big)  \\
& \leq P\Big(\frac{1}{n}\sum_{\ell = 1}^L \sqrt{nP_{\ell}}\max_{j \in \ind(\ell)} \abs{[\tilde{\Delta}_{r+1}]_{j}} \geq \frac{\epsilon}{6} \Big) 
\\
&  \overset{(a)}\leq P\Big(\frac{1}{L }\sum_{\ell = 1}^L \max_{j \in \ind(\ell)} \abs{[\tilde{\Delta}_{r+1}]_{j}} \geq \frac{\epsilon\sqrt{\log M}}{6Rc} \Big) \\
 & \overset{(b)}\leq t^4 K K'_{t-1} \exp\Big\{\frac{- \kappa \kappa'_{t-1} L \e^2}{t^6 (\log M)^{2t-1}}\Big\}.
\end{split}
\een
Here, step $(a)$ follows from the fact that $\sqrt{nP_{\ell}} \leq c \sqrt{\log M}$ for some constant $c > 0$ for each section $\ell \in [L]$ and $nR = L \log M$, and step $(b)$ from \eqref{eq:maxtildedelta}.   Finally, consider the last term $T_{1c}$ of \eqref{eq:Hdt13_b}.
\be
\begin{split}
& T_{1c} = P\Big(\frac{1}{n}\Big \lvert \sum_{\ell = 1}^L (\tilde{\fullh}^{r+1}_{(\ell)})^* \left[\eta^s_{(\ell)}(\fullb_0 - \tilde{\fullh}^{s+1} \right. \\
& \hspace{1in} \left. - \tilde{\fullD}_{s+1})- \eta^s_{(\ell)}(\fullb_0 - \tilde{\fullh}^{s+1})\right] \Big \lvert \geq \frac{\e}{6} \Big) \\
&\leq P\Big(\frac{1}{n}\sum_{\ell = 1}^L \max_{i \in \ind(\ell)}  \abs{\tilde{h}^{r+1}_{i}} \sum_{j \in \ind(\ell)} \Big \lvert \eta_{j}^s(\fullb_0 - \tilde{\fullh}^{s+1} - \tilde{\fullD}_{s+1})   \\
& \hspace{1in} - \eta_{j}^s(\fullb_0 - \tilde{\fullh}^{s+1}) \Big \lvert \geq \frac{\epsilon}{6} \Big)\\
& \overset{(c)}\leq P\Big(\frac{1}{L}\sum_{\ell = 1}^L\max_{i \in \ind(\ell)} \abs{\tilde{h}^{r+1}_{i}} \max_{j \in \ind(\ell)} \abs{[\tilde{\Delta}_{s+1}]_{j}} \geq \frac{\epsilon \tau_t^2}{12Rc} \Big) \\
&\overset{(d)}{=} P\Big(\frac{\tau_{s+1}}{L}\sum_{\ell = 1}^L\max_{i \in \ind(\ell)} | [\tilde{Z}_{r}]_i | \max_{j \in \ind(\ell)} \abs{[\tilde{\Delta}_{s+1}]_{j}} \geq \frac{\epsilon \tau_t^2}{12Rc} \Big).
\label{eq:Hdt17a}
\end{split}
\ee
In the above, step $(c)$ follows from Lemma \ref{lem:BC9} applied to each section $\ell \in [L]$,  using $nP_{\ell} \leq c \log M$ for a universal constant $c>0$; step $(d)$ holds since $\tilde{\fullh}^{u} \overset{d}{=} \tau_{u-1} \tilde{\vecZ}_{u-1}$ where $\tilde{\vecZ}_{u-1} \in \mathbb{R}^N $ is standard Gaussian, as shown in Lemma \ref{lem:ideal_cond_dist}.  Now considering the probability on the RHS of \eqref{eq:Hdt17a}, we find:
\ben
\begin{split}
& T_{1c} \leq P\Big(\frac{\tau_{s+1}}{L}\sum_{\ell = 1}^L\max_{i \in \ind(\ell)} | [\tilde{Z}_{r}]_i| \max_{j \in \ind(\ell)} \abs{[\tilde{\Delta}_{s+1}]_{j}} \geq \frac{\epsilon \tau_t^2}{12Rc} \Big) \\
&\overset{(e)}{\leq} P\Big(\frac{1}{L}\sum_{\ell = 1}^L\max_{i \in \ind(\ell)} ( [\tilde{Z}_{r}]_i)^2  \\
& \qquad \qquad \cdot \frac{1}{L}\sum_{\ell = 1}^L \max_{j \in \ind(\ell)} \abs{[\tilde{\Delta}_{s+1}]_{j}}^2 \geq \Big(\frac{\epsilon \tau_t^2}{12\tau_{1}Rc}\Big)^2 \Big) \\
&\leq P\Big(\frac{1}{L}\sum_{\ell = 1}^L\max_{i \in \ind(\ell)} ([\tilde{Z}_{r}]_i)^2 \geq 3 \log M\Big)  \\
& \quad + P\Big(\frac{1}{L}\sum_{\ell = 1}^L \max_{j \in \ind(\ell)} \abs{[\tilde{\Delta}_{s+1}]_{j}}^2 \geq \Big(\frac{\epsilon \tau_t^2}{12\tau_{1}Rc\sqrt{3 \log M}}\Big)^2\Big)\\
&\overset{(f)}{\leq} e^{- \kappa L \log M} + t^4 K K'_{t-1} \exp\Big\{ \frac{- \kappa \kappa'_{t-1} L \e^2}{t^6 (\log M)^{2t+1}}\Big\}.
\end{split}
\een
Step $(e)$ follows by Cauchy-Schwarz and step $(f)$ from Lemma \ref{lem:max_abs_normals} and the fact that
for $0 \leq s \leq t$,
\be
\begin{split}
&P\Big( \frac{1}{L}  \sum_{\ell = 1}^L\max_{j \in \ind(\ell)} \Big \lvert [\tilde{\Delta}_{s+1}]_j \Big \lvert^2 \geq\e\Big) \\
&  = P\Big( \frac{1}{L}  \sum_{\ell = 1}^L \max_{j \in \ind(\ell)} \Big \lvert  \tau_{s}^2 \sum_{r=0}^{s} \frac{1}{\tau_{r}^2} [\Delta_{r+1}]_j \Big \lvert^2 \geq \e \Big) \\
&\overset{(g)}{\leq} \sum_{r=0}^{s}  P\Big(\frac{1}{L}  \sum_{\ell = 1}^L   \max_{j \in \ind(\ell)}  \Big \lvert  [\Delta_{r+1}]_j \Big \lvert^2 \geq \frac{\e}{(s+1)^2} \Big) \\
&  \overset{(h)}{\leq}  t^4 K K'_{t-1} \exp\Big\{ \frac{ - \kappa \kappa'_{t-1} L \e }{t^4(t+1)^2 (\log M)^{2t}}\Big\}.
\end{split}
\label{eq:del_til_bnd}
\ee
In the above, step $(g)$ follows by Lemma \ref{lem:squaredsums}, Lemma \ref{sums}, and the fact that $\tau_{s}^2/\tau_i^2 \leq 1$ for $0 \leq i \leq s$ and step $(h)$ from $\mc{H}_{t+1}(a)$ and the fact that $0 \leq s \leq t$. We have now shown that the three terms of \eqref{eq:Hdt13_b} are bounded by $ t^4 K K'_{t-1} \exp\Big\{\frac{- \kappa \kappa'_{t-1} L \e^2 }{t^6 (\log M)^{2t+1}}\Big\}$, which completes the proof.


\noindent  \textbf{(e) [Eq.\ \eqref{eq:He}].} This result follows similarly to $\mc{H}_1$(e) by noting $\| \fullb^{t+1} \|^2 - nP = \| \fullq^{t+1} \|^2 - 2 (\fullq^0)^* \fullq^{t+1}$ and appealing to $\mc{H}_{t+1} (c)$.


\noindent  \textbf{(f), (g) [Eqs.\ \eqref{eq:Qsing}, \eqref{eq:Hf1}, \eqref{eq:Hf}, \eqref{eq:Hg}].} These results follow along the same lines as $\mc{B}_{t} (f), (g)$.

\noindent  \textbf{(h) [Eqs.\ \eqref{eq:Hh0} - \eqref{eq:Hh}].} Using the representation of $\fullh^{t+1}$ from Lemma \ref{lem:ideal_cond_dist} Eq.\ \eqref{eq:htil_rep}, we write
\ben
\begin{split}
&P\Big(\frac{1}{L} \sum_{\ell = 1}^L \max_{j \in \ind(\ell)}  | h^{t+1}_j | \geq \tau_0 \sqrt{3\log M} + \e \Big) \\
& = P\Big(\frac{1}{L} \sum_{\ell = 1}^L \max_{j \in \ind(\ell)} \abs{\tilde{h}^{t+1}_j + [\tilde{\Delta}_{t+1}]_j} \geq \tau_0 \sqrt{3\log M} + \e \Big) \\
&\leq P\Big(\frac{1}{L} \sum_{\ell = 1}^L \max_{j \in \ind(\ell)}  | \tilde{h}^{t+1}_j |  \\
& \qquad  \qquad + \frac{1}{L} \sum_{\ell = 1}^L \max_{j \in \ind(\ell)}\abs{[\tilde{\Delta}_{t+1}]_j} \geq  \tau_0 \sqrt{3\log M} + \e \Big) \\
&\leq P\Big(\frac{1}{L} \sum_{\ell = 1}^L \max_{j \in \ind(\ell)} | \tilde{h}^{t+1}_j | \geq  \tau_0 \sqrt{3\log M} \Big)  \\
& \quad + P\Big(\frac{1}{L} \sum_{\ell = 1}^L \max_{j \in \ind(\ell)}\abs{[\tilde{\Delta}_{t+1}]_j} \geq \e \Big) \\
&\overset{(a)}{\leq} e^{-\kappa L \log M} + t^4 K K'_{t-1} \exp\Big\{\frac{- \kappa \kappa'_{t-1} L \e^2}{t^6 (\log M)^{2t}}\Big\}.
\end{split}
\een
In step $(a)$, the first term on the RHS follows from Lemma \ref{lem:max_abs_normals} since $\tilde{\fullh}^{t+1} \overset{d}{=} \tau_{t} \tilde{\vecZ}_{t}$ where $\tau_t < \tau_0$ and $\tilde{\vecZ}_{t} \in \mathbb{R}^N $ is standard Gaussian. The second term on the RHS follows from the bound in  \eqref{eq:del_til_bnd}.
where the final inequality follows since $\tau_t/\tau_0 \leq 1$.  The result for the squared terms follows similarly.

\appendices
\renewcommand{\theequation}{A.\arabic{equation}}

\section{Proof of Lemma \ref{lem:conv_expec}} \label{app:conv_exp}
\textbf{(a)} Recall the definition of $x(\tau)$ given in \eqref{eq:xt_tau_def} which can be written as
\be
x(\tau) = \sum_{\ell = 1}^L \frac{P_{\ell}}{P} \mc{E}_{\ell}(\tau), 
\label{eq:xtau_def}
\ee
where
\[
\mc{E}_\ell(\tau) := \expec \left[
\frac{ e^{U^{\ell}_1 \sqrt{n P_\ell}/{\tau}}}
{ e^{U^{\ell}_1 \sqrt{n P_\ell}/{\tau}}  +  e^{{-n P_\ell}/{\tau^2}} \sum_{j = 2}^M e^{U^{\ell}_j \sqrt{n P_\ell}/{\tau}} } \right].
\] 
Using the relation $nR =  {L \ln M}$, we write $\frac{n P_\ell}{\tau^2} =  \nu_\ell \ln M$, where $\nu_\ell = LP_\ell /(R \tau^2)$.
Then $\mc{E}_\ell(\tau)$ can be expressed as
\begin{equation}
\mc{E}_\ell(\tau)  = \expec\Big[ \Big(1 +  e^{-\sqrt{\nu_\ell \ln M} \, U^{\ell}_1} M^{-\nu_\ell} \sum_{j=2}^M e^{\sqrt{\nu_\ell \ln M} \, U^{\ell}_j } \Big)^{-1}   \Big].
\label{eq:Eell_iter}
\end{equation}

 Letting
\be
\begin{split}
& X := M^{-\nu_\ell} \sum_{j=2}^M \exp\{\sqrt{\nu_\ell \ln M} \, U^{\ell}_j\},  \\
&  V := \exp\{-\sqrt{\nu_\ell \ln M} \, U^{\ell}_1\},
\end{split}
\label{eq:Xrv_def}
\ee
we use iterated expectation and the independence of $X, V$ to write
\be \mc{E}_\ell(\tau) = \mathbb{E}_V \mathbb{E}[(1 + VX)^{-1} | V] \geq \mathbb{E}_V [(1 + V\mathbb{E}X)^{-1}], \label{eq:Eltau_bnd} \ee
where the last step follows from Jensen's inequality.  The expectation of $X$ is
\ben
\begin{split}
 \expec X & = M^{-\nu_\ell} \sum_{j=2}^M \expec \Big[ \exp\{\sqrt{\nu_\ell \ln M} \, U^{\ell}_j \} \Big]\\
& \stackrel{(a)}{=} M^{-\nu_\ell} (M-1)  M^{\nu_\ell /2}  \leq M^{1 - \nu_\ell /2}
\end{split}
\een
where $(a)$ is obtained using the moment generating function of a Gaussian random variable. We therefore have
\be
1\geq  \mc{E}_\ell(\tau)  \geq  \mathbb{E}_V [(1 + V\mathbb{E}X)^{-1}] \geq  \mathbb{E}_V [(1 + V M^{1- \nu_\ell /2})^{-1}].
\label{eq:jensen_chain}
\ee
 For any  $\alpha \in [0,1)$, when $\{ V \leq M^{\alpha (\nu_\ell /2 -1)} \}$, we have $(1 + V M^{1- \nu_\ell /2}) \leq 1 +  M^{-(1- \alpha)(\nu_\ell /2-1)}$.  Using this in  \eqref{eq:jensen_chain}, for  $\ell$ such that $\nu_{\ell} > 2$  we have
\begin{equation}
\begin{split}
 \mc{E}_\ell(\tau) & \geq  \frac{P(V \leq M^{\alpha (\nu_\ell /2 -1)})}{1 +  M^{-(1- \alpha)(\nu_\ell /2-1)}} 
 \\  
 & \overset{(a)}{=}  
 \frac{P\Big(U^{\ell}_1 \geq - \frac{\alpha (\nu_\ell /2 -1)}{\sqrt{\nu_{\ell}}} \sqrt{\ln M}\Big)  }{1 +  M^{-(1- \alpha)(\nu_\ell /2-1)}}, 
\end{split}
\label{eq:El_LB}
\end{equation}
where step $(a)$ follows from the definition of $V$ in \eqref{eq:Xrv_def}.

Next consider $\ell$ such that  $2(1- \vem) \leq \nu_\ell \leq  2$, where $\vem:= \frac{\upsilon}{\sqrt{\log M}}$, for some constant $\upsilon >0$. Then, as $(1- \nu_\ell/2) \leq \vem$, from \eqref{eq:jensen_chain} we have
\begin{align}
&  \mc{E}_\ell(\tau)    \geq  \mathbb{E}_V [(1 + V M^{\vem})^{-1}]  \nonumber \\
 & = \expec \Big[ (  1 + e^{\vem \ln M - U_1^{\ell} \sqrt{\nu_\ell \ln M}})^{-1} \Big]  \geq \frac{P(U_1^{\ell} >  2 \upsilon/\sqrt{\nu_\ell})}{1 + \exp(-\upsilon \sqrt{\ln M})}
\end{align}

\textbf{(b)}   For $\delta \in (0,\frac{1}{2})$, first consider $\ell$ such that $\nu_{\ell} >  2 + \delta$. Using the bound  $\mc{Q}(x) \leq\frac{1}{x\sqrt{2 \pi}} \exp\{\frac{-x^2}{2}\}$ for $x > 0$, the relevant term in \eqref{eq:xlb_nonasym} can be bounded from below as
\begin{equation}
\begin{split}
& \frac{ \mc{Q}\Big(-  \frac{\alpha (\nu_\ell /2 -1)}{\sqrt{\nu_{\ell}}} \sqrt{\log M}\Big)  }{1 +  M^{-(1- \alpha)(\nu_\ell /2-1)}} \\
  &   \geq \Big[ 1 - \frac{\sqrt{\nu_{\ell}}M^{-\alpha^2 (\nu_\ell /2 -1)^2/2\nu_{\ell}}}{\sqrt{2 \pi \ln M} (\nu_\ell /2 -1) \alpha} \Big]  \cdot  \frac{1}{1 +  M^{-(1- \alpha)(\nu_\ell /2-1)}} \\
& \geq  \Big[ 1 - \frac{\sqrt{\nu_{\ell}}M^{-\alpha^2 (\nu_\ell /2 -1)^2/2\nu_{\ell}}}{\sqrt{2 \pi \ln M} (\nu_\ell /2 -1) \alpha}\Big] \Big[ 1 -  M^{-(1- \alpha)(\frac{\nu_\ell}{2}-1)}\Big]  \\
 & \geq  1 -\frac{\sqrt{\nu_{\ell}}M^{-\alpha^2 (\nu_\ell /2 -1)^2/2\nu_{\ell}}}{\sqrt{2 \pi \ln M} (\nu_\ell /2 -1) \alpha} -   M^{-(1- \alpha)(\frac{\nu_\ell}{2}-1)}. 
\end{split}
\label{eq:Elt_LB1}
\end{equation}
where the second inequality is obtained using $\frac{1}{1 + x} \geq (1 - x)$, for $x \in [0, 1]$.  Now choose  $\alpha$ as follows:
\[ \alpha = \begin{cases}
1- \delta, \quad &2 + \delta < \nu_{\ell} < 4, \\
\frac{1}{2}, \quad & \nu_{\ell} \geq 4.
\end{cases}
\]
Using this $\alpha$ in \eqref{eq:Elt_LB1}, for $\nu_\ell \geq 2+ \delta$ we obtain
\begin{align*}
& \mc{E}_\ell(\tau) \nonumber \\
&\geq  \Big(1 -\frac{4 M^{-\delta^2(1 - \delta)^2 /32}}{\sqrt{2 \pi \ln M} \delta(1 - \delta)} -   M^{-\frac{\delta^2}{2}}\Big) \mathbf{1}\{2 + \delta < \nu_\ell < 4\} \nonumber  \\ 
& \quad + \Big(1 -\frac{\sqrt{2\nu_{\ell}}M^{-1/8\nu_{\ell}}}{\sqrt{\pi \ln M}} -   M^{-\frac{1}{2}}\Big) \mathbf{1}\{\nu_\ell \geq 4\} \nonumber \\
& \geq  \Big(1 -\frac{ M^{-\kappa_2 \delta^2}}{\delta \sqrt{ \ln M}}\Big) \mathbf{1}\{\nu_{\ell} > 2 + \delta\},
\end{align*}
for a suitably chosen universal positive constant $\kappa_2$.   Next, if $2 < \nu_\ell \leq (2+\delta)$, using $\alpha =0$ in \eqref{eq:xlb_nonasym} yields $\mc{E}_\ell(\tau) \geq 1/4$. Finally  there exists a universal constant $\kappa_3$ such that 
\[ \frac{\mc{Q}( 2 \kappa_3/\sqrt{\nu_\ell})  }{1 +  e^{-\kappa_3 \sqrt{\ln M}} } \geq \frac{1}{4}  \quad \text{ when } \quad \nu_\ell  \geq 2\Big(1- \frac{\kappa_3}{\sqrt{\log M}} \Big). \]

\section{Proof of Lemma \ref{lem:lim_xt_taut}} \label{app:xt_taut}
\renewcommand{\theequation}{B.\arabic{equation}}

Let $x_{t-1} =x < (1-f(M))$. We will  use Lemma \ref{lem:conv_expec}(b) with $\nu_\ell$ determined by \eqref{eq:cell}. 
We only need to consider the case where $\nu_L < (2+\delta)$, because otherwise $ \nu_\ell \geq (2+\delta)$ for $\ell \in [L]$, and \eqref{eq:xlb_asym} guarantees that $x_t \geq (1-f(M))$.  

With $x_{t-1} =x$, we have $\tau_{t-1}^2 = \sigma^2 +P(1-x)$. With $\tau=\tau_{t-1}^2$, we have
\be 
\nu_\ell = \frac{L P_\ell}{R \tau_{t-1}^2} =  \frac{ \tau_0^2}{R \tau_{t-1}^2} L((1 + \snr)^{1/L}-1)  ( 1 +\snr )^{-\ell/L} 
 \label{eq:nuellx}\ee 
for $\ell \in [L]$, where we have used  the expression in \eqref{eq:cell} for $L P _\ell$.
Using \eqref{eq:nuellx} in \eqref{eq:xlb_asym},
\begin{align}
& x_t  \geq (1- f(M)) \sum_{\ell=1}^L   \frac{P_\ell}{P} \, \mathbf{1}\{ \nu_\ell > 2 + \delta \}  
\nonumber \\
& \stackrel{(a)}{=} (1 - f(M))  \nonumber \\
& \qquad  \sum_{\ell=1}^L \frac{P_\ell}{P} \, \mathbf{1}\Big\{\frac{\ell}{L} < \frac{1}{2 \mc{C}} \log \Big(\frac{L((1+\snr)^{1/L} - 1) \tau_0^2}{(2+\delta)  R \tau_{t-1}^2}\Big) \Big\} \nonumber \\
& \stackrel{(b)}{\geq}   (1 - f(M))  \sum_{\ell=1}^L \frac{P_\ell}{P} \, \mathbf{1}\Big\{\frac{\ell}{L} \leq \frac{1}{2 \mc{C}} \log 
\Big(\frac{2 \mc{C} \tau_0^2}{(2+\delta)  R \tau_{t-1}^2}\Big) \Big\} \nonumber \\
& \stackrel{(c)}{\geq}  (1 - f(M))  \frac{P+\sigma^2}{P} \nonumber \\
& \qquad \times \Big[ 1 - \exp\Big\{- \log\Big( \frac{2 \mc{C} \tau_0^2}{(2+\delta)  R \tau_{t-1}^2} \Big) + \frac{2 \mc C}{L}   \Big\} \Big] \nonumber \\
& \stackrel{(d)}{\geq}  (1 - f(M))  \frac{P+\sigma^2}{P}\Big[ 1- \frac{(2 + \delta) R \tau_{t-1}^2}{2 \mc C \tau_0^2} - \frac{5R}{L} \Big]. \label{eq:xtlb0}
\end{align}
In the above, $(a)$ is obtained using the expression for $\nu_\ell$ in \eqref{eq:nuellx}, and inequality $(b)$ by noting that
\[  L((1+\snr)^{1/L} - 1) = L(e^{2 \mc C/L} - 1) \geq 2 \mc{C}.\]
Inequality $(c)$ is obtained by  using the geometric series formula: for any $\xi \in (0,1)$, we have
\ben 
\begin{split}
 & \sum_{\ell=1}^{\lfloor \xi L \rfloor} P_{\ell} = (P+\sigma^2)(1- e^{-2\mc{C}\lfloor \xi L \rfloor/L}) \\
 & \geq  (P+\sigma^2)(1- e^{-2\mc{C}\xi} e^{2 \mc C/L}).    
 \end{split} \een
Inequality $(d)$ uses $e^{2\mc{C}/L} \leq 1 + 4 \mc{C}/L$ for large enough $L$. Substituting $\tau_{t-1}^2 = \sigma^2 +P(1-x)$, \eqref{eq:xtlb0} implies
\begin{align}
& x_t - x  \geq   (1 - f(M))  \frac{P+\sigma^2}{P} \Big(1 - \frac{5R}{L} \Big) \nonumber \\
& \qquad  \qquad - (1-f(M)) \frac{(1+ \delta/2) R}{\mc C}
\Big( \frac{P+ \sigma^2}{P} -x \Big) - x  \nonumber \\
& =  (1 - f(M))  \frac{P+\sigma^2}{P} \Big( 1 - \frac{(1+ \delta/2) R}{\mc C} - \frac{5R}{L}  \Big) \nonumber \\
& \qquad - x \Big(1 -  (1 - f(M)) \frac{(1+ \delta/2) R}{\mc C}   \Big).
\label{eq:xtlb1}
\end{align}
Since $\delta < (\mc C -  R)/\mc C$,  the term $\frac{(1+ \delta/2) R}{\mc C}$ is strictly less than $1$, and the RHS of \eqref{eq:xtlb1} is strictly decreasing in $x$. Using the upper bound of $x < (1- f(M))$ in \eqref{eq:xtlb1} and simplifying, we obtain
\begin{align*}
 x_t - x  &  \geq  (1 - f(M))  \frac{\sigma^2}{P} \Big( 1 - \frac{(1+ \delta/2) R}{\mc C} \Big) \nonumber \\
& \quad - f(M) (1- f(M))\frac{(1+ \delta/2) R}{\mc C}  - \frac{5R(1 +\sigma^2/P)}{L}.
\end{align*}
This completes the proof for $t > 1$. For $t=1$, we start with $x=0$, and we get the slightly stronger lower bound of $\chi_1$ by substituting $x=0$ in \eqref{eq:xtlb1}. 


\section{Concentration lemmas}  \label{app:conc_lemmas}
\renewcommand{\theequation}{C.\arabic{equation}}

In the following $\e >0$ is assumed to be a generic constant, with additional conditions specified whenever needed.  The proofs of  Lemmas \ref{sums}--\ref{powers} can be found in \cite{RushV16}.

\begin{applem}[Hoeffding's inequality  {\cite[Thm.\ 2.8]{BLMconcbook}}]
\label{lem:hoeff_lem}
If $X_1, \ldots, X_n$ are bounded random variables such that $a_i \leq X_i \leq b_i$, then for $\nu = 2[\sum_{i} (b_i -a_i)^2]^{-1}$
\begin{align*}
& P\Big( \frac{1}{n}\sum_{i=1}^n (X_i - \expec X_i) \geq \e \Big) \leq e^{ -\nu n^2 \e^2},  \\
&  P\Big( \Big \lvert \frac{1}{n}\sum_{i=1}^n (X_i - \expec X_i)  \Big \lvert \geq \e \Big) \leq 2e^{ -\nu n^2 \e^2}.
\end{align*}
\end{applem}

\begin{applem}[Concentration of sums]
\label{sums}
If random variables $X_1, \ldots, X_M$ satisfy $P(\abs{X_i} \geq \e) \leq e^{-n\kappa_i \e^2}$ for $1 \leq i \leq M$, then 
\ben
\begin{split}
P\Big(\Big \lvert   \sum_{i=1}^M X_i \Big \lvert  \geq \e\Big) & \leq \sum_{i=1}^M P\Big(|X_i| \geq \frac{\e}{M}\Big) \\
&  \leq M e^{-n (\min_i \kappa_i) \e^2/M^2}.
\end{split}
\een
\end{applem}

\begin{applem}[Concentration of products]
\label{products} 
For random  variables $X,Y$ and non-zero constants $c_X, c_Y$, if
$P( \lvert X- c_X\lvert  \geq \e) \leq K e^{-\kappa n \e^2}$ \text{ and }  $P( | Y- c_Y  |  \geq \e) \leq K e^{-\kappa n \e^2}$,
then
\begin{align*}  
P( | XY - c_Xc_Y |  \geq \e ) & \leq  P\Big(  | X- c_X  |  \geq \min\Big( \sqrt{\frac{\e}{3}}, \frac{\e}{3 c_Y} \Big) \Big) \\
&  \ +   P\Big(  | Y- c_Y |  \geq \min\Big( \sqrt{\frac{\e}{3}}, \frac{\e}{3 c_X} \Big) \Big)  \\ 
 & \leq 2K \exp\Big\{\frac{-\kappa n \e^2}{9\max(1, c_X^2, c_Y^2)}\Big\}.
\end{align*}
\end{applem}


\begin{applem}[Concentration of square roots]
\label{sqroots}
Let $c \neq 0$.  If $ P( \lvert X_n^2 - c^2 \lvert \geq \epsilon) \leq e^{-\kappa n \epsilon^2}$,
then
\ben
P ( \lvert \abs{X_n} - \abs{c} \lvert \geq \epsilon) \leq e^{-\kappa n \abs{c}^2 \epsilon^2}.
\een
\end{applem}
%


\begin{applem}[Concentration of powers]
\label{powers}
Assume $c \neq 0$ and $0 < \e \leq 1$.  If  $P( \lvert X_n - c  \lvert \geq \epsilon) \leq e^{-\kappa n \epsilon^2}$, then for any integer $k \geq 2$
\ben
P( \lvert X_n^k - c^k \lvert \geq \epsilon ) \leq e^{ {-\kappa n \e^2}/[{(1+\abs{c})^k -\abs{c}^k}]^2}.
\een
\end{applem}

\begin{applem}
\label{lem:normalconc}
For a standard Gaussian random variable $Z$ and  $\e > 0$,
$P( \abs{Z} \geq \e) \leq 2e^{-\frac{1}{2}\e^2}$.
\end{applem}


\begin{applem}
\label{lem:max_abs_normals}
Let $Z_1, Z_2, \ldots, Z_N$ and $\tilde{Z}_1, \tilde{Z}_2, \ldots, \tilde{Z}_N$  be i.i.d.\ standard Gaussian random variables and  $0 \leq \e \leq 1$.  Then the following concentration results hold.
\begin{align}
&P\Big(\frac{1}{L}\sum_{\ell = 1}^L \max_{j \in \ind(\ell)} Z_{j}^2 \geq 3 \log M + \e \Big)  \nonumber \\
& \quad \leq \exp\Big\{ \frac{-L }{5} \Big(2\e + \log \frac{M}{70} \Big) \Big\}, \label{res2}  \\
&P\Big(\Big \lvert \frac{1}{N}\sum_{i=1}^N Z_i^2 - 1\Big \lvert \geq \e \Big) \leq 2\exp\Big\{\frac{-N \e^2}{8}\Big\}, \label{res3} \\
& P\Big(\Big \lvert \frac{1}{N}\sum_{i=1}^N Z_i \tilde{Z}_i \Big \lvert \geq \e \Big) \leq 2\exp\Big\{\frac{-N \e^2}{3}\Big\}.  \label{res4}
\end{align} 
\end{applem}

\begin{IEEEproof}
Using a Cram{\'e}r-Chernoff bound \cite[Sec. 2.2]{BLMconcbook}, for any $t >0$ we have
\be
\begin{split}
& P\Big(\frac{1}{L}\sum_{\ell = 1}^L \max_{j \in \ind(\ell)} Z_{j}^2 \geq 3 \log M + \epsilon \Big)  \\
& \leq \exp\Big(- t L (\epsilon +  3\log M) + \sum_{\ell = 1}^L \log \mathbb{E} e^{t \max_{j \in \ind(\ell)} Z_{j}^2} \Big).
\end{split}
\label{eq:ccbound2}
\ee 
Then using the moment-generating function of a Chi-square random variable, we obtain
\ben
\mathbb{E} e^{t \max_{j \in \ind(\ell)} Z_{j}^2} \leq \sum_{j \in \ind(\ell)} \mathbb{E} e^{t Z_{j}^2} = \frac{M}{\sqrt{1 - 2t}}, 
\een
when $0< t < 1/2$.   Using this  bound in \eqref{eq:ccbound2} for  $\ell \in [L]$, we find
\ben
\begin{split}
& P\Big(\frac{1}{L}\sum_{\ell = 1}^L \max_{j \in \ind(\ell)} Z_{j}^2 \geq 3 \log M +\epsilon \Big) \\
&  \leq \exp\Big(\inf_{t \in (0, 1/2)} \Big[ - t L (\epsilon + 3 \log M)+ L \log M  \\
& \qquad - \frac{L}{2} \log (1 - 2t )\Big]\Big).
\end{split}
\een
We choose $t = 2/5$ to obtain the desired bound:
\ben
\begin{split}
& P\Big(\frac{1}{L}\sum_{\ell = 1}^L \max_{j \in \ind(\ell)} Z_{j}^2 \geq 3 \log M +\epsilon \Big) \\
&  \leq \exp\Big( - \frac{2 L \epsilon}{5} - \frac{L}{5}  \log M + \frac{L}{2} \log 5\Big) \\
& \leq \exp\Big(  \frac{-L }{5} \Big(2\e + \log \frac{M}{70} \Big) \Big).
\end{split}
\een

The bounds in \eqref{res3} and \eqref{res4} can be similarly obtained using  Cram{\'e}r-Chernoff bounds. The relevant moment generating functions are 
\ben
\begin{split}
\mathbb{E} e^{t Z_{i}^2}&  = \frac{1}{\sqrt{1 - 2t}},  \  0 <  t < 1/2,  \\
\mathbb{E} e^{t Z_i \tilde{Z}_i} &= \frac{1}{\sqrt{1-t^2}}, \  0 <  t < 1, \quad \text{ for } i \in [N].
\end{split}
\een
The steps to obtain the bounds in \eqref{res3} and \eqref{res4} using these moment generating functions are similar to those for sub-Gamma random variables; see, e.g., \cite[Sec. 2.4]{BLMconcbook}.
\end{IEEEproof}


\section{Other useful lemmas}  \label{app:useful_lemmas}
\renewcommand{\theequation}{D.\arabic{equation}}

\begin{applem}
For any scalars $a_1, ..., a_t$ and positive integer $m$, we have  $(\abs{a _1} + \ldots + \abs{a_t})^m \leq t^{m-1} \sum_{i=1}^t \abs{a_i}^m$.
Consequently, for any vectors $\textbf{u}_1, \ldots, \textbf{u}_t \in \mathbb{R}^N$, $\norm{\sum_{k=1}^t \textbf{u}_k}^2 \leq t \sum_{k=1}^t \norm{\textbf{u}_k}^2$.
\label{lem:squaredsums}
\end{applem}
\begin{IEEEproof} The first result is obtained by applying H{\"o}lder's inequality to the length-$t$ vectors $(\abs{a_1}, \ldots, \abs{a_t})$ and $(1, \ldots, 1)$. The second statement is obtained by applying the result with $m=2$.
\end{IEEEproof}

\begin{applem}[Stein's lemma]
For zero-mean jointly Gaussian random variables $Z_1, Z_2$, and any function $f:\mathbb{R} \to \mathbb{R}$ for which $\expec[Z_1 f(Z_2)]$ and $\expec[f'(Z_2)]$  both exist, we have $\expec[Z_1 f(Z_2)] = \expec[Z_1Z_2] \expec[f'(Z_2)]$.
\label{lem:stein}
\end{applem}

\begin{applem}
Let  $\textbf{u} \in \mathbb{R}^N$ be a deterministic vector, $\tilde{\textbf{A}} \in \mathbb{R}^{n \times N}$ be a matrix with i.i.d.\ 
$\mc{N}(0, \frac{1}{n})$ entries, and  $\mc{W}$  be a $d$-dimensional subspace of $\mathbb{R}^n$ for $d \leq n$. Let $(\textbf{w}_1, ..., \textbf{w}_d)$ be an orthogonal basis of $\mc{W}$ with $\norm{\textbf{w}_{r}}^2 = n$ for
$r \in [d]$, and let  $\proj_\mc{W}^{\parallel}$ denote the orthogonal projection operator onto $\mc{W}$.  Then for $\textbf{D} = [\textbf{w}_1\mid \ldots \mid \textbf{w}_d]$, we have $\proj_{\mc{W}} \tilde{\textbf{A}} \textbf{u} \overset{d}{=} \frac{\norm{\textbf{u}}}{\sqrt{n}}  \proj_\mc{W} \vecZ_u  \overset{d}{=} \frac{\norm{\textbf{u}}}{\sqrt{n}} \textbf{D} \textbf{x}$ where $\textbf{x} \in \mathbb{R}^d$ is a random vector with i.i.d.\ $\mc{N}(0, 1/n)$ entries. \label{lem:gauss_p0}
\end{applem}


\begin{applem}[$\mathcal{H} (d)$ concentration]
\label{lem:Hd_convergence}
Let $\vecZ, \tilde{\vecZ} \in \mathbb{R}^N$ each be standard Gaussian random vectors such that $(Z_i, \tilde{Z}_i)$ are i.i.d. bivariate Gaussian, for $1 \leq i \leq N$.  For $\ell \in [L]$, let $Y^{\ell} = \frac{1}{\log M} \cdot \vecZ_{(\ell)}^*\eta^s_{(\ell)}(\fullb_0 - \tau_s \tilde{\vecZ}_{\ell}),$ where $0 \leq s \leq T$. Then for $\kappa$, a universal positive constant,
\ben
P\Big(\Big \lvert \frac{1}{L} \sum_{\ell = 1}^L (Y^{\ell} - \mathbb{E}[Y^{\ell}])\Big \lvert \geq \e \Big) \leq \exp\{-\kappa L \e^2\}.
\een
\end{applem}

\begin{IEEEproof} We first note that $Y^{\ell}$ is a (scalar) random variable with an $\ell$ dependence.  (It is not an $M$-length vector.)
We represent $Y^{\ell}$ as a sum of a bounded  random variable and an unbounded random variable.
\be
\begin{split}
Y^{\ell} & =  \frac{\vecZ_{(\ell)}^*\eta^s_{(\ell)}(\fullb_0 - \tau_s \vecZ_{s_{(\ell)}})}{\log M} \mathbf{1}\Big\{\max_{i \in \ind(\ell)} \abs{Z_{i}} \leq x\Big\} \\
& + \frac{\vecZ_{(\ell)}^*\eta^s_{(\ell)}(\fullb_0 - \tau_s \tilde{\vecZ}_{(\ell)})}{\log M} \mathbf{1}\Big\{\max_{i \in \ind(\ell)} \abs{Z_{i}} \geq x\Big\},
\end{split}
\label{eq:lemHD1}
\ee
where we specify the value of $x$ later.  Label the first term on the right side of \eqref{eq:lemHD1} as $Y^{\ell}_b$ for `bounded' and the second term as $Y^{\ell}_u$ for `unbounded'.  Therefore, using Lemma \ref{sums},
\be
\begin{split}
& P\Big(\Big \lvert \frac{1}{L} \sum_{\ell = 1}^L (Y^{\ell} - \mathbb{E}[Y^{\ell}])\Big \lvert \geq 2\e \Big) \\
&= P\Big(\Big \lvert \frac{1}{L} \sum_{\ell = 1}^L (Y^{\ell}_b - \mathbb{E}[Y^{\ell}]) + \frac{1}{L} \sum_{\ell = 1}^L Y^{\ell}_u\Big \lvert \geq 2\e \Big) \\
&\leq P\Big(\Big \lvert \frac{1}{L} \sum_{\ell = 1}^L (Y^{\ell}_b - \mathbb{E}[Y^{\ell}])\Big \lvert \geq \e \Big) + P\Big(\Big \lvert \frac{1}{L} \sum_{\ell = 1}^L Y^{\ell}_u\Big \lvert \geq \e \Big).
\label{eq:lemHD2}
\end{split}
\ee
Define $\zeta_{L} = \frac{1}{L}\sum_{\ell = 1}^L \mathbb{E}[Y^{\ell}_u]$.  Noting that $\mathbb{E}[Y^{\ell}] = \mathbb{E}[Y^{\ell}_b] + \mathbb{E}[Y^{\ell}_u]$, we write
\be
\begin{split}
& P\Big(\Big \lvert \frac{1}{L} \sum_{\ell = 1}^L (Y^{\ell}_b - \mathbb{E}[Y^{\ell}])\Big \lvert \geq \e \Big)  \\
& = P\Big(\Big \lvert \frac{1}{L} \sum_{\ell = 1}^L (Y^{\ell}_b - \mathbb{E}[Y^{\ell}_b]) - \zeta_{L} \Big \lvert \geq \e \Big) \\
& \leq P\Big(\frac{1}{L} \sum_{\ell = 1}^L (Y^{\ell}_b - \mathbb{E}[Y^{\ell}_b]) \geq \e + \zeta_{L} \Big) \\
& \qquad  + P\Big(\frac{1}{L} \sum_{\ell = 1}^L (Y^{\ell}_b - \mathbb{E}[Y^{\ell}_b]) \leq -\e + \zeta_{L} \Big).
\end{split}
\label{eq:lemHD3}
\ee
From \eqref{eq:lemHD2} and \eqref{eq:lemHD3} we have
\begin{align}
& P\Big(\Big \lvert \frac{1}{L} \sum_{\ell = 1}^L (Y^{\ell} - \mathbb{E}[Y^{\ell}])\Big \lvert \geq 2\e \Big) 
\nonumber \\
& \leq P\Big(\frac{1}{L} \sum_{\ell = 1}^L (Y^{\ell}_b - \mathbb{E}[Y^{\ell}_b]) \geq \e + \zeta_{L} \Big) + P\Big(\Big \lvert \frac{1}{L} \sum_{\ell = 1}^L Y^{\ell}_u\Big \lvert \geq \e \Big) \nonumber \\
& \quad + P\Big(\frac{1}{L} \sum_{\ell = 1}^L (Y^{\ell}_b - \mathbb{E}[Y^{\ell}_b]) \leq -\e + \zeta_{L} \Big).
\label{eq:lemHD4}
\end{align}
Label the terms on the right side of \eqref{eq:lemHD4} as $T_1$, $T_2$, and $T_3$.  The rest of the proof proceeds as follows:
\begin{enumerate}
\item We show that $- M^{-c_0} \leq \zeta_{L} \leq M^{-c_0}$ for some universal constant $c_0>0$.
\item We apply Hoeffding's inequality to  show that $T_1$ and $T_3$ are bounded by $\exp\{-\kappa_1L \e^2\}$ for some universal constant $\kappa_1>0$.
\item We show that $T_2$ is also exponentially small in $L$.
\end{enumerate}
To show item (1), recalling that $\zeta_{L} = \frac{1}{L}\sum_{\ell = 1}^L \mathbb{E}[Y^{\ell}_u]$, 
we will obtain an upper bound for 
$ \lvert \mathbb{E}[Y^{\ell}_u] \lvert$. 
\be
\begin{split}
\mathbb{E}[Y^{\ell}_u] &= \mathbb{E}\Big[\frac{\vecZ_{(\ell)}^*\eta^s_{(\ell)}(\fullb_0 - \tau_s \tilde{\vecZ}_{(\ell)})}{\log M} \mathbf{1}\Big\{\max_{i \in \ind(\ell)} \abs{Z_{i}} \geq x\Big\}\Big] \\
& \leq \frac{c}{\sqrt{\log M}}\mathbb{E}\Big[ \max_{i \in \ind(\ell)} \abs{Z_{i}} \mathbf{1}\Big\{\max_{i \in \ind(\ell)} \abs{Z_{i}} \geq x\Big\}\Big],
\label{eq:lemHD5}
\end{split}
\ee
where we have used $\sqrt{n P_{\ell}} \leq c \sqrt{\log M}$ for some constant $c > 0$.
Let $W = \max_{i \in \ind(\ell)} \abs{Z_{i}}$, and define $\tilde{W} = W \mathbf{1}\{W \geq x\}$. Then since $\tilde{W}$ is non-negative, we have $\mathbb{E}[\tilde{W}] = \int_0^{\infty} P(\tilde{W} \geq w) dw$.  Note that
\be
P(\tilde{W} \geq w) = \begin{cases}  P(W \geq w) &\mbox{if } w > x \\ 
P(W \geq x)  & \mbox{if } 0 < w \leq x. \end{cases}
\label{eq:lemHD6}
\ee
Then it follows from \eqref{eq:lemHD5} and \eqref{eq:lemHD6},
\begin{align}
& \Big \lvert \mathbb{E}[Y^{\ell}_u] \Big \lvert  \leq \frac{c}{\sqrt{\log M}}
 \Big[\int_{0}^{x} \hspace{-2pt} P(W \geq x) du + \int_{x}^{\infty} \hspace{-2pt} P(W \geq u) du \Big] \nonumber \\
& \leq \frac{c}{\sqrt{\log M}} \Big[x P(W \geq x)+ \int_{x}^{\infty} P(W \geq u) du \Big] \nonumber \\
&\leq \frac{c}{\sqrt{\log M}} \Big[2M e^{-x^2/2} + \int_{x}^{\infty} 2M e^{-u^2/2} du \Big] \label{eq:lemHD7}. \\
& \leq  \frac{c}{\sqrt{\log M}} \Big[2M e^{-x^2/2} + \frac{2M}{x} e^{-x^2/2} \Big].  \label{eq:lemHD7b}
\end{align}
where \eqref{eq:lemHD7} is obtained by noting that for $y >0$,
\ben
\begin{split}
& P \Big( \max_{i \in \ind(\ell)} \abs{Z_{i}} \geq y \Big)  = P\Big( \{ Z_1 \geq y\} \cup \ldots \cup \{ Z_M \geq y \}  \\ 
& \hspace{1.5in}  \cup  \{  Z_1 \leq y \} \cup \ldots \cup 
\{ Z_M \leq y \} \Big)  \\
& \leq 2M e^{-y^2/2} \min \Big\{1, \frac{1}{y \sqrt{2 \pi}} \Big\}.
\end{split}
\een
Inequality \eqref{eq:lemHD7b} also uses the above bound for the Gaussian tail probability.

Now choose $x = k\sqrt{2 \log M}$ for $k>1$ to be fixed later. Then \eqref{eq:lemHD7b} implies 
\ben
\begin{split}
\abs{\mathbb{E}[Y^{\ell}_u]} &\leq \frac{2c}{M^{k^2-1}\sqrt{\log M}}  \Big(1 + \frac{1}{k\sqrt{2 \log M}} \Big). \\
\end{split}
\een
We have therefore shown $\Big \lvert \mathbb{E}[Y^{\ell}_u] \Big \lvert \leq 2cM^{-(k^2-1)}$ for $M$  large enough ($M > e^4$ suffices). So it follows that
\begin{equation}
\begin{split}
& \zeta_{L} \leq \frac{1}{L}\sum_{\ell = 1}^L \abs{\mathbb{E}[Y^{\ell}_u]}  \leq \frac{2c}{M^{k^2-1}} \\
 &  \zeta_{L} \geq \frac{-1}{L}\sum_{\ell = 1}^L \abs{\mathbb{E}[Y^{\ell}_u]} \geq \frac{-2c}{M^{k^2-1}}.
\end{split} 
\label{eq:zetaL_bnd}
\end{equation}

Next we bound the terms $T_1$ and $T_3$ in  \eqref{eq:lemHD4} using Hoeffding's inequality (Lemma \ref{lem:hoeff_lem}).  First  notice that for $\ell \in [L]$, the random variable $Y^{\ell}_b \in [\frac{-x\sqrt{nP_{\ell}}}{\log M}, \frac{x\sqrt{nP_{\ell}}}{\log M}]$, where $x = k \sqrt{2 \log M}$.  To apply Hoeffding's inequality, we also note that 
\ben
\begin{split}
& \sum_{\ell = 1}^L \Big(\Big[\frac{x\sqrt{nP_{\ell}}}{\log M} \Big] - \Big[-\frac{x\sqrt{nP_{\ell}}}{\log M} \Big]\Big)^2 \\
& = \sum_{\ell = 1}^L \frac{4x^2 nP_{\ell}}{(\log M)^2} = \frac{8k^2 nP}{\log M} =  {8k^2 P}{R} L.
\end{split} 
\een
Therefore it follows that
\ben
T_1 = P\Big(\frac{1}{L} \sum_{\ell = 1}^L (Y^{\ell}_b - \mathbb{E}[Y^{\ell}_b]) \geq \e + \zeta_{L} \Big) \leq \exp\Big(\frac{-L(\e + \zeta_{L})^2}{\tilde{c}}\Big)
\een
where  $\tilde{c} = 4k^2P/R$. From \eqref{eq:zetaL_bnd}, $\zeta_L$ can be made arbitrarily small by choosing  $M$ large enough. Therefore, we have $T_1 \leq  \exp\{-\kappa_1 L \e^2\}$ for large enough $M$. Similarly, for sufficiently large $M$ we also have
\ben
T_3 = P\Big(\frac{1}{L} \sum_{\ell = 1}^L (Y^{\ell}_b - \mathbb{E}[Y^{\ell}_b]) \leq -\e + \zeta_{L} \Big) \leq \exp\{-\kappa_1 L \e^2\}.
\een
Finally we bound the second term in  \eqref{eq:lemHD4}.  Using the  Cram{\'e}r-Chernoff bound, for  $t >0$ we have
\be
T_2 = P\Big(\Big \lvert \frac{1}{L} \sum_{\ell = 1}^L Y^{\ell}_u\Big \lvert \geq \e \Big) \leq \exp\Big( \Big[ - t L \epsilon + \sum_{\ell = 1}^L \log \mathbb{E} e^{t \lvert Y^{\ell}_u \lvert}\Big]\Big).
\label{eq:lemHD8a}
\ee
We bound $\mathbb{E} e^{t \lvert Y^{\ell}_u \lvert}$ as follows.  We have
\be
\begin{split}
& \mathbb{E} \exp\{t \lvert Y^{\ell}_u \lvert\}   \\
& = \mathbb{E} \exp\Big\{t\Big \lvert \frac{(\vecZ_{(\ell)})^*\eta^s_{(\ell)}(\fullb_0 - \tau_s \tilde{\vecZ}_{(\ell)})}{\log M} \mathbf{1}\{\max_{i \in \ind(\ell)} \abs{Z_{i}} \geq x\}\Big \lvert \Big\} \\
&\leq \mathbb{E} \exp\Big\{\frac{ct}{\sqrt{\log M}} \max_{i \in \ind(\ell)} \abs{Z_{i}} \mathbf{1}\{\max_{i \in \ind(\ell)} \abs{Z_{i}} \geq x\} \Big\} = \mathbb{E}[U],
\label{eq:lemHD9}
\end{split}
\ee
where we have defined $U := \exp\Big\{\frac{ct}{\sqrt{2 \log M}} \max_{i \in \ind(\ell)} \abs{Z_{i}} \mathbf{1}\{\max_{i \in \ind(\ell)} \abs{Z_{i}} \geq x\}\Big\}$ and used $\sqrt{n P_{\ell}} \leq c \sqrt{\log M}$.  Also recall that $x= k \sqrt{2\log M}$.  As before, let $W = \max_{i \in \ind(\ell)} \abs{Z_{i}}$,  and notice that
\ben
U = \begin{cases} 1 &\mbox{if } W \leq x \\ 
\exp\{\frac{ctW}{\sqrt{\log M}}\} & \mbox{if } W \geq x. \end{cases}
\een 
It follows that
\be
\begin{split}
& P(U \geq u)  \\
& = \begin{cases} 1 &\mbox{if } 0 < u \leq 1 \\ 
P(W \geq x) &\mbox{if } 1 < u \leq \exp\{\frac{ctx}{\sqrt{\log M}}\} \\ 
P(W \geq \frac{(\log u)\sqrt{\log M}}{ct}) & \mbox{if } u > \exp\{\frac{ctx}{\sqrt{\log M}}\}. \end{cases}
\end{split}
\label{eq:lemHD11}
\ee
Let $\tilde{x} = \exp\{\frac{ctx}{\sqrt{\log M}}\} = \exp\{\sqrt{2} kct\}$.  Then using \eqref{eq:lemHD9} and \eqref{eq:lemHD11}, we have
\begin{align}
&  \mathbb{E} e^{t \lvert Y^{\ell}_u \lvert}  \nonumber \\
 & \leq \mathbb{E}[U]  = \int_{0}^{\infty} P(U \geq u) du  \leq 1 + (\tilde{x}  - 1) P(W \geq x)  \nonumber \\
 & \hspace{1in} + \int_{\tilde{x}}^{\infty} P\Big(W \geq \frac{(\log u)\sqrt{\log M}}{ct}\Big)  du \nonumber \\
&\leq 1 + (\tilde{x} - 1)2M e^{-\frac{x^2}{2}} + 2M \hspace{-3pt} \int_{\tilde{x}}^{\infty} \hspace{-5pt} \exp\hspace{-2pt}\Big\{\frac{-(\log u)^2 \log M}{2c^2t^2}\Big\} du
\nonumber  \\
& = 1+ 2 \exp(\sqrt{2} kct) M^{-(k^2-1)} \nonumber \\
& \qquad + 2M \int_{\sqrt{2} k c t}^{\infty} \exp\Big\{\frac{-v^2 \log M}{2c^2t^2} \Big\} e^v dv. \label{eq:lemHD12}
\end{align}
Now by completing the square and simplifying, the integral in \eqref{eq:lemHD12} can be bounded as
\be
\begin{split}
 & \int_{\sqrt{2} k c t}^{\infty} \exp\Big\{\frac{-v^2 \log M}{2c^2t^2} \Big\} e^v dv \\
 & = \exp \Big\{ \frac{c^2t^2}{2 \log M} \Big\} \mc{Q}( k \sqrt{2 \log M} -1) 
 \leq  c_1 M^{-k^2},
 \end{split}
 \label{eq:Qfn_tail}
\ee
for some absolute positive constant $c_1 <2$.  In \eqref{eq:Qfn_tail}, $\mc{Q}(a) = (2 \pi)^{-1/2} \int_{a}^\infty e^{-v^2/2} dv$ is the Gaussian upper tail probability  function. Using \eqref{eq:Qfn_tail} in  \eqref{eq:lemHD12} and taking $t=\frac{1}{\sqrt{2}c}$, we obtain
$\mathbb{E} e^{t \lvert Y^{\ell}_u \lvert} \leq 1 +  c_2 M^{-(k^2-1)},$ where $c_2 >0$ is an absolute positive constant.  Substituting this into \eqref{eq:lemHD8a} we find:
\ben
\begin{split}
& T_2=  P\Big(\Big \lvert \frac{1}{L} \sum_{\ell = 1}^L Y^{\ell}_u\Big \lvert \geq \e \Big) \\
&\leq \exp\Big\{\frac{-L\e}{c \sqrt{2}}\Big\}(1 + c_2 M^{-(k^2-1)})^L \\
& \leq \exp \Big \{\frac{-L\e}{c \sqrt{2}}  \Big\}\exp \{ c_2 L M^{-(k^2-1)}\}  \leq e^{- \kappa L \e },
\end{split}
\een
for some absolute positive constant $\kappa$, when $M$ is sufficiently large. This completes the proof.
\end{IEEEproof}


\begin{applem}{\cite[Lemma 9]{choThesis}}
\label{lem:BC9}
For the function $\eta^t_j:\mathbb{R}^{N} \rightarrow \mathbb{R}$ defined in \eqref{eq:eta_def} for any $j \in [N]$ and $\textbf{s}, \fullD \in \mathbb{R}^N$, the following is true for all $\ell \in [L]$:
\ben
\sum_{i \in \ind(\ell) } \abs{\eta^t_i(\textbf{s}) - \eta^t_i(\textbf{s} + \fullD)} \leq \frac{2nP_{\ell}}{\tau_t^2} \max_{i \in \ind(\ell)} \abs{\Delta_i}.
\een
\end{applem}

\begin{IEEEproof}
From the multivariate version of Taylor's theorem, for any $j \in [N]$ we have 
\be
 \eta^t_j(\textbf{s} + \fullD) =  \eta^t_j(\textbf{s}) + \fullD^T \nabla \eta^t_j(\textbf{s} + c \fullD),
 \label{eq:eta_taylor}
 \ee
for some $c \in (0,1)$. Noting that for $j \in [N]$, $\eta^t_j$ depends only on the subset of its input also belonging to section $\ind(\sect(j))$,  using \eqref{eq:eta_taylor} we have
\ben
\begin{split}
&\sum_{i \in \ind(\ell) } \abs{\eta^t_i(\textbf{s}) - \eta^t_i(\textbf{s}+ \fullD)} \\
&  = \sum_{i \in \ind(\ell) } \Big \lvert \sum_{ j \in \ind(\ell) } \Delta_j \frac{\partial}{\partial s_j} \eta^t_i(\textbf{s}+ c \fullD) \Big \lvert \\
&\overset{(a)}{\leq} \frac{\sqrt{nP_{\ell}}}{\tau_t^2} \sum_{i \in \ind(\ell) } \Big  \lvert\eta^t_i(\textbf{s}+ c \fullD) \Delta_i \Big \lvert \\
 & \quad +  \frac{1}{\tau_t^2} \sum_{i \in \ind(\ell) } \Big \lvert \eta^t_i(\textbf{s}+ c \fullD)  \sum_{j \in \ind(\ell)  } \Delta_j \eta^t_j(\textbf{s}+ c \Delta) \Big \lvert  \\
 & \overset{(b)}{\leq}  \frac{2 nP_{\ell}}{\tau_t^2} \max_{i \in \ind(\ell) } \abs{\Delta_i},
\end{split}
\een
where inequality $(a)$ uses the fact that for $i,j \in \ind(\ell)$,
$\frac{\partial}{\partial s_j} \eta^t_i(\textbf{s}) = \frac{1}{\tau_t^2} \eta^t_i(\textbf{s}) [\sqrt{nP_{\ell}} \, \, \mathbf{1}\{j = i\} - \eta^t_j(\textbf{s})]$. 
Inequality $(b)$ uses the fact that $\sum_{j \in \ind(\ell)} \abs{\eta^t_j(\textbf{s}+ c \fullD)} = \sum_{j \in \ind(\ell)} \eta^t_j(\textbf{s}+ c \fullD) = \sqrt{nP_{\ell}}$. 
\end{IEEEproof}


\begin{applem} \label{lem:expect_etar_etas} Let $\tilde{\vecZ}_r, \tilde{\vecZ}_s \in \mathbb{R}^N$ each be standard Gaussian  random vectors such that the pairs 
$ (\tilde{Z}_{r,i}, \tilde{Z}_{s,i}), \ i \in [N]$, are i.i.d.\  bivariate Gaussian with covariance $\mathbb{E}[\tilde{Z}_{r,i} \tilde{Z}_{s,i}] = (\tau_s/\tau_r)$.   Then for $0 \leq r \leq s \leq T$,
\begin{align}
&  \frac{1}{n}\expec\{ [\eta^r(\fullb - \tau_r \tilde{\vecZ}_r)]^* [\eta^s(\fullb - \tau_s \tilde{\vecZ}_s)] \}  = P x_{r+1}, \label{eq:etar_etas} \\ 
&  \frac{1}{n} \mathbb{E}\{[\eta^r(\fullb - \tau_r \tilde{\vecZ}_r) - \beta]^*  [\eta^s(\fullb - \tau_s \tilde{\vecZ}_s) - \fullb] \} = \sigma^2_{s+1}. \label{eq:full}
\end{align}
\end{applem}

\begin{IEEEproof}
We will use the following fact  \cite[Proposition 1]{RushGVIT17}:
\be
 \expec\{ \fullb^{*}\eta^r(\fullb - \tau_r \tilde{\vecZ}_r) \} = nP x_{r+1}, \quad \text{ for } 0 \leq r \leq t.
 \label{eq:beta_etar}
 \ee
 Let $\textbf{u}^r = \fullb- \tau_r \tilde{\vecZ}_r$ and $\textbf{u}^s = \fullb - \tau_s \tilde{\vecZ}_s$. Recall from \eqref{eq:cond_exp_beta0} that $\eta^r(\fullb - \tau_r \tilde{\vecZ}_r) = \expec[\fullb \mid \textbf{u}^r]$ and $\eta^s(\fullb - \tau_s \tilde{\vecZ}_s) = \expec[\fullb \mid \textbf{u}^s]$. Therefore, for $r \leq s$
\begin{align}
 & \expec\{ [\eta^r(\fullb - \tau_r \tilde{\vecZ}_r)]^* \eta^s(\fullb - \tau_s \tilde{\vecZ}_s) \} \nonumber \\
 &= \expec\{ [\expec[\fullb \mid \textbf{u}^r]]^* \expec[\fullb \mid \textbf{u}^s] \}  \nonumber \\
&= \expec\{ [\expec[\fullb \mid \textbf{u}^r]]^* [\expec[\fullb \mid \textbf{u}^s, \textbf{u}^r] - \fullb + \fullb] \}  \label{eq:usur}\\
& = \expec\{ [\expec[\fullb \mid \textbf{u}^r]]^* \fullb \},  \label{eq:orth_prin}\\
&= nPx_{r+1}. \label{eq:npxr}
\end{align}
which proves \eqref{eq:etar_etas}. In the above, \eqref{eq:usur} holds because $\expec[\fullb \mid \textbf{u}^s, \textbf{u}^r]= \expec[\beta \mid \textbf{u}^s]$ as shown 
below,  \eqref{eq:orth_prin} holds because  $ \expec\{(\expec[ \fullb | \textbf{u}^s , \textbf{u}^r] - \fullb)^* \expec[\fullb | \textbf{u}^r ] \} =0$ due to the orthogonality principle, and \eqref{eq:npxr} follows from \eqref{eq:beta_etar}. 

The result \eqref{eq:full} follows from \eqref{eq:etar_etas} and \eqref{eq:beta_etar}, noting that $\norm{\fullb}^2=nP$. Therefore, the proof is complete once we show that  $\expec[\fullb \mid \textbf{u}^s, \textbf{u}^r]= \expec[\fullb \mid \textbf{u}^s]$. Consider an index $i \in [N]$, and for brevity let $\ell := \sect(i)$.  We then have
\be
\begin{split}
&\expec[\beta_i \mid \fullb - \tau_s \tilde{\vecZ}_s = \textbf{u}^s, \fullb - \tau_r \tilde{\vecZ}_r = \textbf{u}^r ]  \\
& = \expec[\beta_i \mid  \{ \beta_j - \tau_s \tilde{Z}_{s,j} = u^s_j, \  \beta_j - \tau_r \tilde{Z}_{r,j} = u^r_j \}_{j \in \ind(\ell)} ] \\
& = \sqrt{n P_{\ell}} P\big( \beta_i = \sqrt{n P_{\ell}} \mid  \{ \beta_j - \tau_s \tilde{Z}_{s,j} = u^s_j , \\ 
& \hspace{1.6in}  \beta_j - \tau_r \tilde{Z}_{r,j} = u^r_j \}_{j \in \ind(\ell)} \big)\\
&= 
\frac{ \sqrt{n P_{\ell}}   f( \{  u^s_j,  u^r_j \}_{j \in \ind(\ell)} \mid \beta_i = \sqrt{n P_{\ell}})  P(\beta_i = \sqrt{n P_{\ell}})}
{\sum_{k \in \ind(\ell)}  \hspace{-2pt} f( \{  u^s_j,  u^r_j \}_{j \in \ind(\ell)} \mid \beta_k = \sqrt{n P_{\ell}})  P(\beta_k = \sqrt{n P_{\ell}})}
\end{split}
\label{eq:cond_exp_beta}
\ee
where we have used Bayes Theorem with  $$ f( \{  u^s_j,  u^r_j \}_{j \in \ind(\ell)} \mid \beta_k = \sqrt{n P_{\ell}})$$ denoting the joint conditional density function of $\{ \beta_j - \tau_s \tilde{Z}_{s,j}= u^s_j , \, \beta_j - \tau_r \tilde{Z}_{r,j} = u^r_j  \}_{j \in \ind(\ell)}$ given $\beta_k = \sqrt{nP_\ell}$.  Now,  $\fullb$ is independent of $\tilde{\vecZ}_r$ and $\tilde{\vecZ}_s$, and the pairs  $(\tilde{Z}_{r,i}, \tilde{Z}_{s,i})$, $i \in [N]$, are i.i.d.\ bivariate Gaussian with covariance $\mathbb{E}[\tilde{Z}_{r,i} \tilde{Z}_{s,i}] = \frac{\tau_s}{\tau_r}$. We therefore have
\be
\begin{split}
&f( \{ u^s_j, \, u^r_j \}_{j \in \ind(\ell)} \mid \beta_k = \sqrt{n P_{\ell}})  \\
& \propto \exp\Big\{-\frac{\tau_r^2}{2(\tau_r^2 - \tau_s^2)} \Big[\frac{(u^s_k - \sqrt{nP_{\ell}})^2}{\tau_s^2} + \frac{(u^r_k - \sqrt{nP_{\ell}})^2}{\tau_r^2} \\ %
& \hspace{0.5in} - \frac{2(u^s_k - \sqrt{nP_{\ell}})(u^r_k - \sqrt{nP_{\ell}})}{\tau_r^2}\Big]\Big\} \\
&   \, \times \hspace{-4pt}\prod_{j \in \ind(\ell), \, j \neq k}  \hspace{-3pt} \exp\Big\{-\frac{\tau_r^2}{2(\tau_r^2 - \tau_s^2)} \Big[\frac{(u^s_j)^2}{\tau_s^2} + \frac{(u^r_j)^2}{\tau_r^2} - \frac{2u^s_j u^r_j}{\tau_r^2}\Big]\Big\} \\
&  = \exp\Big\{\frac{\tau_r^2 \sqrt{nP_{\ell}}}{\tau_r^2 - \tau_s^2} \Big[\frac{u^s_k }{\tau_s^2} + \frac{u^r_k }{\tau_r^2} - \frac{u^s_k + u^r_k }{\tau_r^2}\Big]\Big\} \\
& \,  \times  \exp\Big\{-\frac{\tau_r^2 n P_{\ell}}{2(\tau_r^2 - \tau_s^2)} \Big[\frac{1}{\tau_s^2} + \frac{1}{\tau_r^2} - \frac{2}{\tau_r^2}\Big]\Big\}\\
& \, \prod_{j \in \ind(\ell)}  \exp\Big\{-\frac{\tau_r^2}{2(\tau_r^2 - \tau_s^2)} \Big[\frac{(u^s_j)^2}{\tau_s^2} + \frac{(u^r_j)^2}{\tau_r^2} - \frac{2u^s_j u^r_j}{\tau_r^2}\Big]\Big\}.
\end{split}
\label{eq:cond_denf}
\ee
Using  \eqref{eq:cond_denf} in \eqref{eq:cond_exp_beta},  together with the fact that $P(\beta_k = \sqrt{n P_{\ell}}) = \frac{1}{M}$ for each $k \in \ind(\ell)$, we obtain
\begin{align*}
 & \expec[\beta_i \mid  \fullb - \tau_s \tilde{\vecZ}_s = \textbf{u}^s, \fullb - \tau_r \tilde{\vecZ}_r = \textbf{u}^r]  \\
 &=  \frac{\sqrt{n P_{\ell}} \exp\Big\{\frac{\tau_r^2 \sqrt{nP_{\ell}}}{\tau_r^2 - \tau_s^2} \Big[\frac{u^s_i }{\tau_s^2} + \frac{u^r_i }{\tau_r^2} - \frac{u^s_i + u^r_i }{\tau_r^2}\Big]\Big\}} 
{\sum_{j \in \ind(\ell)} \, \exp\Big\{\frac{\tau_r^2 \sqrt{n P_{\ell}}}{\tau_r^2 - \tau_s^2} \Big[\frac{u^s_j }{\tau_s^2} + \frac{u^r_j }{\tau_r^2} - \frac{u^s_j + u^r_j }{\tau_r^2}\Big]\Big\}} \\
&= \frac{\sqrt{n P_{\ell}} \exp\Big\{\frac{u^s_i \sqrt{n P_{\ell} }}{\tau_s^2} \Big\}} 
{\sum_{j \in \ind(\ell)} \, \exp\Big\{\frac{u^s_j \sqrt{n P_{\ell}}}{ \tau_s^2}\Big\}} \\
&   =  \expec[\beta_i \mid  \fullb - \tau_s \tilde{\vecZ}_s = \textbf{u}^s],
\label{eq:bit1_def}
\end{align*}
as required.
\end{IEEEproof}

\section*{Acknowledgement}
The authors thank A.\ Greig for the simulations  used in Sec.\ \ref{subsec:code_params}, and A.\ Barron for several helpful discussions. We thank the Associate Editor and the anonymous reviewers for helpful comments and suggestions that led to an improved paper. 

\IEEEtriggeratref{23}

%

\end{document}